\documentclass[twocolumn]{aastex631}
\usepackage{xfrac,graphicx,color,float,hyperref}  
 
\newcommand {\lya}{Ly$\alpha$}
\def\ltsima{$\; \buildrel < \over \sim \;$}
\def\simlt{\lower.5ex\hbox{\ltsima}}
\def\gtsima{$\; \buildrel > \over \sim \;$}
\def\simgt{\lower.5ex\hbox{\gtsima}}
\newcommand {\etal}{et~al.}

\newcommand {\um}{$\mu$m}

\newcommand{\msun}{{\rm\,M$_\odot$}}

\newcommand{\lsun}{{\rm\,L$_\odot$}}

\shorttitle{The COSMOS-Web Survey}
\shortauthors{Casey, Kartaltepe et al.}

\begin{document}

\title{COSMOS-Web: An Overview of the JWST Cosmic Origins Survey}

\correspondingauthor{Caitlin M. Casey, Jeyhan S. Kartaltepe}
\email{cmcasey@utexas.edu, jeyhan@astro.rit.edu}

\author[0000-0002-0930-6466]{Caitlin M. Casey}\altaffiliation{First two authors are co-first-authors} 
\affiliation{The University of Texas at Austin, 2515 Speedway Blvd Stop C1400, Austin, TX 78712, USA}
\affiliation{Cosmic Dawn Center (DAWN), Denmark}

\author[0000-0001-9187-3605]{Jeyhan S. Kartaltepe}\altaffiliation{First two authors are co-first-authors} 
\affiliation{Laboratory for Multiwavelength Astrophysics, School of Physics and Astronomy, Rochester Institute of Technology, 84 Lomb Memorial Drive, Rochester, NY 14623, USA}

\author[0000-0003-4761-2197]{Nicole E. Drakos}
\affiliation{Department of Astronomy and Astrophysics, University of California, Santa Cruz, 1156 High Street, Santa Cruz, CA 95064, USA}

\author[0000-0002-3560-8599]{Maximilien Franco}
\affiliation{The University of Texas at Austin, 2515 Speedway Blvd Stop C1400, Austin, TX 78712, USA}

\author[0000-0003-0129-2079]{Santosh Harish}
\affiliation{Laboratory for Multiwavelength Astrophysics, School of Physics and Astronomy, Rochester Institute of Technology, 84 Lomb Memorial Drive, Rochester, NY 14623, USA}

\author[0000-0003-2397-0360]{Louise Paquereau} 
\affiliation{Institut d’Astrophysique de Paris, UMR 7095, CNRS, and Sorbonne Université, 98 bis boulevard Arago, F-75014 Paris, France}

\author[0000-0002-7303-4397]{Olivier Ilbert}
\affiliation{Aix Marseille Universit\'{e}, CNRS, CNES, LAM, Marseille, France}

\author[0000-0002-8018-3219]{Caitlin Rose}
\affiliation{Laboratory for Multiwavelength Astrophysics, School of Physics and Astronomy, Rochester Institute of Technology, 84 Lomb Memorial Drive, Rochester, NY 14623, USA}

\author[0000-0002-1803-794X]{Isabella G. Cox}
\affiliation{Laboratory for Multiwavelength Astrophysics, School of Physics and Astronomy, Rochester Institute of Technology, 84 Lomb Memorial Drive, Rochester, NY 14623, USA}

\author[0000-0002-8987-7401]{James W. Nightingale}
\affil{Department of Physics, Institute for Computational Cosmology, Durham University, South Road, Durham DH1 3LE, UK}

\author[0000-0002-4271-0364]{Brant E. Robertson}
\affiliation{Department of Astronomy and Astrophysics, University of California, Santa Cruz, 1156 High Street, Santa Cruz, CA 95064, USA}

\author[0000-0002-0000-6977]{John D. Silverman}
\affiliation{Kavli Institute for the Physics and Mathematics of the Universe (WPI), The University of Tokyo, Kashiwa, Chiba 277-8583, Japan}
\affiliation{Department of Astronomy, School of Science, The University of Tokyo, 7-3-1 Hongo, Bunkyo, Tokyo 113-0033, Japan}

\author[0000-0002-6610-2048]{Anton M. Koekemoer}
\affiliation{Space Telescope Science Institute, 3700 San Martin Dr., Baltimore, MD 21218, USA}

\author[0000-0002-6085-3780]{Richard Massey}
\affil{Department of Physics, Centre for Extragalactic Astronomy, Durham University, South Road, Durham DH1 3LE, UK}
\author[0000-0002-9489-7765]{Henry Joy McCracken}
\affiliation{Institut d’Astrophysique de Paris, UMR 7095, CNRS, and Sorbonne Université, 98 bis boulevard Arago, F-75014 Paris, France}

\author[0000-0002-4485-8549]{Jason Rhodes}
\affiliation{Jet Propulsion Laboratory, California Institute of Technology, 4800 Oak Grove Drive, Pasadena, CA 91001, USA}

\author[0000-0003-3596-8794]{Hollis B. Akins}
\affiliation{The University of Texas at Austin, 2515 Speedway Blvd Stop C1400, Austin, TX 78712, USA}

\author[0000-0001-9610-7950]{Natalie Allen}
\affiliation{Cosmic Dawn Center (DAWN), Denmark} 
\affiliation{Niels Bohr Institute, University of Copenhagen, Jagtvej 128, DK-2200, Copenhagen, Denmark}

\author[0000-0002-4465-1564]{Aristeidis Amvrosiadis}
\affil{Department of Physics, Centre for Extragalactic Astronomy, Durham University, South Road, Durham DH1 3LE, UK}
\affil{Department of Physics, Institute for Computational Cosmology, Durham University, South Road, Durham DH1 3LE, UK}

\author[0000-0002-0569-5222]{Rafael C. Arango-Toro}
\affiliation{Aix Marseille Universit\'{e}, CNRS, CNES, LAM, Marseille, France}

\author[0000-0002-9921-9218]{Micaela B. Bagley}
\affiliation{The University of Texas at Austin, 2515 Speedway Blvd Stop C1400, Austin, TX 78712, USA}

\author[0000-0002-0101-6624]{Angela Bongiorno}
\affiliation{INAF-Observatory of Rome, via Frascati 33, 00074 Monteporzio Catone, Italy}

\author[0000-0003-3578-6843]{Peter L. Capak}
\affiliation{Cosmic Dawn Center (DAWN), Denmark} 
\affiliation{Niels Bohr Institute, University of Copenhagen, Jagtvej 128, DK-2200, Copenhagen, Denmark}

\author[0000-0002-6184-9097]{Jaclyn B. Champagne}
\affiliation{Steward Observatory, University of Arizona, 933 N Cherry Ave, Tucson, AZ 85721, USA}

\author[0000-0003-3691-937X]{Nima Chartab}
\affiliation{The Observatories of the Carnegie Institution for Science, 813 Santa Barbara St., Pasadena, CA 91101, USA}

\author[0000-0002-0786-7307]{\'{O}scar A. Ch\'{a}vez Ortiz}
\affiliation{The University of Texas at Austin, 2515 Speedway Blvd Stop C1400, Austin, TX 78712, USA}

\author[0000-0003-4922-0613]{Katherine Chworowsky}
\altaffiliation{NSF Graduate Research Fellow}
\affiliation{The University of Texas at Austin, 2515 Speedway Blvd Stop C1400, Austin, TX 78712, USA}

\author[0000-0002-2200-9845]{Kevin C. Cooke}
\affiliation{Association of Public and Land-grant Universities, 1220 L Street NW, Suite 1000, Washington, DC 20005, USA}

\author[0000-0003-3881-1397]{Olivia R. Cooper}\altaffiliation{NSF Graduate Research Fellow}
\affiliation{The University of Texas at Austin, 2515 Speedway Blvd Stop C1400, Austin, TX 78712, USA}

\author[0000-0003-4919-9017]{Behnam Darvish}
\affiliation{Department of Physics and Astronomy, University of California, Riverside, 900 University Avenue, Riverside, CA 92521, USA}

\author[0000-0002-0786-7307]{Xuheng Ding}
\affiliation{Kavli Institute for the Physics and Mathematics of the Universe (WPI), The University of Tokyo, Kashiwa, Chiba 277-8583, Japan}

\author[0000-0002-9382-9832]{Andreas L. Faisst}
\affiliation{Caltech/IPAC, 1200 E. California Blvd., Pasadena, CA 91125, USA}

\author[0000-0001-8519-1130]{Steven L. Finkelstein}
\affiliation{The University of Texas at Austin, 2515 Speedway Blvd Stop C1400, Austin, TX 78712, USA}

\author[0000-0001-7201-5066]{Seiji Fujimoto}\altaffiliation{NASA Hubble Fellow}
\affiliation{The University of Texas at Austin, 2515 Speedway Blvd Stop C1400, Austin, TX 78712, USA}

\author[0000-0002-8008-9871]{Fabrizio Gentile}
\affiliation{University of Bologna - Department of Physics and Astronomy “Augusto Righi” (DIFA), Via Gobetti 93/2, I-40129 Bologna, Italy}
\affiliation{INAF, Osservatorio di Astrofisica e Scienza dello Spazio, Via Gobetti 93/3, I-40129, Bologna, Italy}

\author[0000-0001-9885-4589]{Steven Gillman}
\affiliation{Cosmic Dawn Center (DAWN), Denmark}
\affiliation{DTU-Space, Technical University of Denmark, Elektrovej 327, DK-2800 Kgs. Lyngby, Denmark}

\author[0000-0003-4196-5960]{Katriona M. L. Gould}
\affiliation{Cosmic Dawn Center (DAWN), Denmark} 
\affiliation{Niels Bohr Institute, University of Copenhagen, Jagtvej 128, DK-2200, Copenhagen, Denmark}

\author[0000-0002-0236-919X]{Ghassem Gozaliasl}
\affiliation{Department of Physics, University of Helsinki, P.O. Box 64, FI-00014 Helsinki, Finland}

\author[0000-0003-4073-3236]{Christopher C. Hayward}
\affiliation{Center for Computational Astrophysics, Flatiron Institute, 162 Fifth Avenue, New York, NY 10010, USA}

\author[0000-0003-3672-9365]{Qiuhan He}
\affiliation{Department of Physics, Institute for Computational Cosmology, Durham University, South Road, Durham DH1 3LE, UK}

\author[0000-0003-2226-5395]{Shoubaneh Hemmati}
\affiliation{Caltech/IPAC, 1200 E. California Blvd., Pasadena, CA 91125, USA}

\author[0000-0002-3301-3321]{Michaela Hirschmann}
\affiliation{Institute of Physics, GalSpec, Ecole Polytechnique Federale de Lausanne, Observatoire de Sauverny, Chemin Pegasi 51, 1290 Versoix, Switzerland}
\affiliation{INAF, Astronomical Observatory of Trieste, Via Tiepolo 11, 34131 Trieste, Italy}

\author[0000-0003-3804-2137]{Knud Jahnke}
\affiliation{Max Planck Institute for Astronomy, K\"onigstuhl 17, D-69117 Heidelberg, Germany}

\author[0000-0002-8412-7951]{Shuowen Jin}
\affiliation{Cosmic Dawn Center (DAWN), Denmark}
\affiliation{DTU-Space, Technical University of Denmark, Elektrovej 327, DK-2800 Kgs. Lyngby, Denmark}

\author[0000-0002-0101-336X]{Ali Ahmad Khostovan}
\affiliation{Laboratory for Multiwavelength Astrophysics, School of Physics and Astronomy, Rochester Institute of Technology, 84 Lomb Memorial Drive, Rochester, NY 14623, USA}

\author[0000-0002-5588-9156]{Vasily Kokorev}
\affiliation{Kapteyn Astronomical Institute, University of Groningen, PO Box 800, 9700 AV Groningen, The Netherlands}

\author[0000-0003-3216-7190]{Erini Lambrides}\altaffiliation{NPP Fellow}
\affiliation{NASA Goddard Space Flight Center, Code 662, Greenbelt, MD, 20771, USA}

\author{Clotilde Laigle}
\affiliation{Institut d’Astrophysique de Paris, UMR 7095, CNRS, and Sorbonne Université, 98 bis boulevard Arago, F-75014 Paris, France}

\author[0000-0003-2366-8858]{Rebecca L. Larson}\altaffiliation{NSF Graduate Research Fellow}
\affiliation{The University of Texas at Austin, 2515 Speedway Blvd Stop C1400, Austin, TX 78712, USA}

\author[0000-0002-9393-6507]{Gene C. K. Leung}
\affiliation{The University of Texas at Austin, 2515 Speedway Blvd Stop C1400, Austin, TX 78712, USA}

\author[0000-0001-9773-7479]{Daizhong Liu}
\affiliation{Max-Planck-Institut f\"ur Extraterrestrische Physik (MPE), Giessenbachstr. 1, D-85748 Garching, Germany}

\author[0000-0002-9104-314X]{Tobias Liaudat}
\affiliation{Universit\'{e} Paris-Saclay, Universit\'{e} Paris Cit\'{e}, CEA, CNRS, AIM, 91191, Gif-sur-Yvette, France}

\author[0000-0002-7530-8857]{Arianna S. Long}\altaffiliation{NASA Hubble Fellow}
\affiliation{The University of Texas at Austin, 2515 Speedway Blvd Stop C1400, Austin, TX 78712, USA}

\author[0000-0002-4872-2294]{Georgios Magdis}
\affiliation{Cosmic Dawn Center (DAWN), Denmark}
\affiliation{DTU-Space, Technical University of Denmark, Elektrovej 327, DK-2800 Kgs. Lyngby, Denmark}
\affiliation{Niels Bohr Institute, University of Copenhagen, Jagtvej 128, DK-2200, Copenhagen, Denmark}

\author[0000-0003-3266-2001]{Guillaume Mahler}
\affil{Department of Physics, Centre for Extragalactic Astronomy, Durham University, South Road, Durham DH1 3LE, UK}
\affil{Department of Physics, Institute for Computational Cosmology, Durham University, South Road, Durham DH1 3LE, UK}

\author[0000-0002-1047-9583]{Vincenzo Mainieri}
\affiliation{European Southern Observatory, Karl-Schwarzschild-Straße 2, D-85748 Garching bei München, Germany}

\author[0000-0003-0415-0121]{Sinclaire M. Manning}\altaffiliation{NASA Hubble Fellow}
\affil{Department of Astronomy, University of Massachusetts Amherst, 710 N Pleasant Street, Amherst, MA 01003, USA}

\author[0000-0001-7711-3677]{Claudia Maraston}
\affiliation{Institute of Cosmology and Gravitation, University of Portsmouth, Dennis Sciama Building, Burnaby Road, Portsmouth, PO13FX, UK}

\author[0000-0001-9189-7818]{Crystal L. Martin}
\affil{Department of Physics, University of California, Santa Barbara, Santa Barbara, CA 93109, USA}

\author[0000-0002-9883-7460]{Jacqueline E. McCleary}
\affiliation{Department of Physics, Northeastern University, 360 Huntington Ave, Boston, MA 02115, USA}

\author[0000-0002-6149-8178]{Jed McKinney}
\affiliation{The University of Texas at Austin, 2515 Speedway Blvd Stop C1400, Austin, TX 78712, USA}

\author[0000-0003-0639-025X]{Conor J. R. McPartland}
\affiliation{Cosmic Dawn Center (DAWN), Denmark}
\affiliation{Niels Bohr Institute, University of Copenhagen, Jagtvej 128, DK-2200, Copenhagen, Denmark}

\author[0000-0001-5846-4404]{Bahram Mobasher}
\affiliation{Department of Physics and Astronomy, University of California, Riverside, 900 University Avenue, Riverside, CA 92521, USA}

\author[0000-0003-3835-9898]{Rohan Pattnaik}
\affiliation{Laboratory for Multiwavelength Astrophysics, School of Physics and Astronomy, Rochester Institute of Technology, 84 Lomb Memorial Drive, Rochester, NY 14623, USA}

\author[0000-0002-7093-7355]{Alvio Renzini}
\affiliation{Istituto Nazionale di Astrofisica (INAF), Osservatorio Astronomico di Padova,
Vicolo dell’Osservatorio 5, 35122, Padova, Italy}

\author[0000-0003-0427-8387]{R. Michael Rich}
\affiliation{Department of Physics and Astronomy, UCLA, PAB 430 Portola Plaza, Box 951547, Los Angeles, CA 90095, USA}

\author[0000-0002-1233-9998]{David B. Sanders}
\affiliation{Institute for Astronomy, University of Hawai’i at Manoa, 2680 Woodlawn Drive, Honolulu, HI 96822, USA}

\author[0000-0002-0364-1159]{Zahra Sattari}
\affiliation{Department of Physics and Astronomy, University of California, Riverside, 900 University Avenue, Riverside, CA 92521, USA}
\affiliation{The Observatories of the Carnegie Institution for Science, 813 Santa Barbara St., Pasadena, CA 91101, USA}

\author[0000-0001-8450-7885]{Diana Scognamiglio}
\affiliation{Argelander-Institut f\"ur Astronomie, Auf dem H\"ugel 71, D-53121, Bonn, Germany}

\author[0000-0002-0438-3323]{Nick Scoville}
\affiliation{Astronomy Department, California Institute of Technology, 1200 E. California Blvd, Pasadena, CA 91125, USA}

\author[0000-0002-5496-4118]{Kartik Sheth}
\affiliation{NASA Headquarters, 300 Hidden Figures Way, SE, Mary W. Jackson NASA HQ Building, Washington, DC 20546, USA}

\author[0000-0002-7087-0701]{Marko Shuntov}
\affiliation{Institut d’Astrophysique de Paris, UMR 7095, CNRS, and Sorbonne Université, 98 bis boulevard Arago, F-75014 Paris, France}

\author[0000-0002-9735-3851]{Martin Sparre}
\affiliation{Institut f\"ur Physik und Astronomie, Universit\"at Potsdam, Karl-Liebknecht-Str.\,24/25, 14476 Golm, Germany}
\affiliation{Leibniz-Institut f\"ur Astrophysik Potsdam (AIP), An der Sternwarte 16, 14482 Potsdam, Germany}

\author[0000-0002-3560-1346]{Tomoko L. Suzuki}
\affiliation{Kavli Institute for the Physics and Mathematics of the Universe (WPI), The University of Tokyo, Kashiwa, Chiba 277-8583, Japan}

\author[0000-0003-4352-2063]{Margherita Talia}
\affiliation{University of Bologna - Department of Physics and Astronomy “Augusto Righi” (DIFA), Via Gobetti 93/2, I-40129 Bologna, Italy}
\affiliation{INAF, Osservatorio di Astrofisica e Scienza dello Spazio, Via Gobetti 93/3, I-40129, Bologna, Italy}

\author[0000-0003-3631-7176]{Sune Toft}
\affiliation{Cosmic Dawn Center (DAWN), Denmark}
\affiliation{Niels Bohr Institute, University of Copenhagen, Jagtvej 128, DK-2200, Copenhagen, Denmark}

\author[0000-0002-3683-7297]{Benny Trakhtenbrot}
\affiliation{School of Physics and Astronomy, Tel Aviv University, Tel Aviv 69978, Israel}

\author[0000-0002-0745-9792]{C. Megan Urry}
\affiliation{Physics Department and Yale Center for Astronomy \&\ Astrophysics, Yale University, PO Box 208120, CT 06520, USA}

\author[0000-0001-6477-4011]{Francesco Valentino}
\affiliation{Cosmic Dawn Center (DAWN), Denmark} 
\affiliation{Niels Bohr Institute, University of Copenhagen, Jagtvej 128, DK-2200, Copenhagen, Denmark}

\author[0000-0002-8163-0172]{Brittany N. Vanderhoof}
\affiliation{Laboratory for Multiwavelength Astrophysics, School of Physics and Astronomy, Rochester Institute of Technology, 84 Lomb Memorial Drive, Rochester, NY 14623, USA}

\author[0000-0002-4437-1773]{Eleni Vardoulaki}
\affiliation{Th\"{u}ringer Landessternwarte, Sternwarte 5, 07778 Tautenburg, Germany}

\author[0000-0003-1614-196X]{John R. Weaver}
\affil{Department of Astronomy, University of Massachusetts Amherst, 710 N Pleasant Street, Amherst, MA 01003, USA}

\author[0000-0001-7160-3632]{Katherine E. Whitaker}
\affil{Department of Astronomy, University of Massachusetts Amherst, 710 N Pleasant Street, Amherst, MA 01003, USA}
\affiliation{Cosmic Dawn Center (DAWN), Denmark}

\author[0000-0003-3903-6935]{Stephen M.~Wilkins}
\affiliation{Astronomy Centre, University of Sussex, Falmer, Brighton BN1 9QH, UK}
\affiliation{Institute of Space Sciences and Astronomy, University of Malta, Msida MSD 2080, Malta}

\author[0000-0002-8434-880X]{Lilan Yang}
\affiliation{Kavli Institute for the Physics and Mathematics of the Universe (WPI), The University of Tokyo, Kashiwa, Chiba 277-8583, Japan}

\author[0000-0002-7051-1100]{Jorge A. Zavala}
\affiliation{National Astronomical Observatory of Japan, 2-21-1 Osawa, Mitaka, Tokyo 181-8588, Japan}

\begin{abstract}

We present the survey design, implementation, and outlook for
COSMOS-Web, a 255\,hour treasury program conducted by the {\it James
  Webb Space Telescope} in its first cycle of observations.
COSMOS-Web is a contiguous 0.54\,deg$^2$ NIRCam imaging survey in four
filters (F115W, F150W, F277W, and F444W) that will reach 5$\sigma$
point source depths ranging $\sim$27.5--28.2\,magnitudes. In parallel,
we will obtain 0.19\,deg$^2$ of MIRI imaging in one filter (F770W)
reaching 5$\sigma$ point source depths of
  $\sim$25.3--26.0\,magnitudes.  COSMOS-Web will build on the rich
heritage of multiwavelength observations and data products available
in the COSMOS field.  The design of COSMOS-Web is motivated by three
primary science goals: (1) to discover thousands of galaxies in the
Epoch of Reionization ($6\simlt z\simlt 11$) and map reionization's
spatial distribution, environments, and drivers on scales sufficiently
large to mitigate cosmic variance, (2) to identify hundreds of rare
quiescent galaxies at $z>4$ and place constraints on the formation of
the Universe's most massive galaxies ($M_\star>10^{10}$\,M$_\odot$),
and (3) directly measure the evolution of the stellar mass to halo
mass relation using weak gravitational lensing out to $z\sim2.5$ and
measure its variance with galaxies' star formation histories and
morphologies.  In addition, we anticipate COSMOS-Web's legacy value to
reach far beyond these scientific goals, touching many other areas of
astrophysics, such as the identification of the first direct collapse
black hole candidates, ultracool sub-dwarf stars in the Galactic halo,
and possibly the identification of $z>10$ pair-instability supernovae.
In this paper we provide an overview of the survey's key measurements,
specifications, goals, and prospects for new discovery.
\end{abstract}


\section{Introduction} \label{sec:intro}

Designed to peer into the abyss, extragalactic deep fields have pushed
the limits of our astronomical observations as far and as faint as 
possible. The first of these deep fields imaged with the {\it Hubble
Space Telescope} \citep[the medium deep survey and the Hubble Deep
Field North, or HDF-N;][]{griffiths96a,williams96a} pushed three
magnitudes fainter than could be reached with ground-based telescopes
at the time. Their data revealed a surprisingly high density of
distant galaxies, well above expectation. This surprise was due to
high-redshift galaxies' elevated surface brightness relative to
nearby galaxies, likely caused by their overall higher star formation
rates. It quickly became clear that ``the Universe at high redshift
looks rather different than it does at the current epoch''
\citep{williams96a}.

This unexpected richness found in these first deep fields marked a
major shift in astronomy's approach to high-redshift extragalactic
science, moving from specialized case studies scattered about the sky
and instead placing more emphasis on statistical studies using
multiwavelength observations in a few deep fields where the density of
information was very high. Such a transformation had a major role in
leveling access to the high-redshift Universe for a wide array of
researchers worldwide, regardless of their individual access to
astronomical observatories. Several other deep fields were pursued in
short order after the HDF-N with {\it Hubble}, the other Great
Observatories, and ancillary observations across the spectrum from the
ground and space \citep[e.g., the HDF-S, CDFN and CDFS, GOODS-N and
  GOODS-S, and the
  HUDF;][]{williams00a,brandt00a,giacconi02a,giavalisco04a,beckwith06a},
complementing each other in depth and area and providing crucial
insight into the diversity of galaxies from the faintest, lowest-mass
systems to the brightest and most rare.

In parallel to the effort to push deep over narrow fields of view,
another experiment with {\it Hubble} transformed our understanding of
large scale structure (LSS) at high redshifts by mapping a contiguous
two square degree area of the sky, $\sim$20 times larger than all
other deep fields of the time combined. Through its large area and
statistical samples (resolving over 2$\times10^6$ galaxies from
$0<z<6$), the Cosmic Evolution Survey \citep[COSMOS;][]{scoville07a}
allowed the first in-depth studies linking the formation and evolution
of galaxies to their larger cosmic environments across 93\%\ of cosmic
time.  By virtue of its large area, COSMOS probed a volume
significantly larger than that of ``pencil-beam'' deep fields and thus
substantially minimized uncertainties of key extragalactic
measurements from cosmic variance.  In addition, the diverse array of
multiwavelength observations gathered in the COSMOS field
\citep{capak07a,ilbert10a,laigle16a,weaver22a} made it possible to
carry out a suite of ambitious survey efforts and understand the
distribution of large scale structure at early cosmic epochs
\citep{scoville13a,darvish15a}.

Deep field images of the distant Universe -- from the deepest, Hubble
Ultra Deep Field (HUDF), to the widest, COSMOS -- have transformed
into rich laboratories for testing hypotheses about the formation and
evolution of galaxies through time. These hypotheses initially
encompassed the first basic cosmological models and ideas regarding
the evolution of galaxy structure. Thanks to the addition of
multiwavelength observations in these deep fields, they expanded to
include hypotheses about the formation of supermassive black holes,
the richness of galaxies' interstellar media, the assembly of gas in
and around galaxies, and the structure of large dark matter haloes.

These deep fields, initially motivated by {\it Hubble} but
substantially enhanced with a rich suite of ancillary ground-based and
space-based data, have deepened our understanding of the evolution of
galaxies across cosmic time. They pushed the horizon of the distant
Universe into the first billion years, a time marking the last major
phase change of the Universe itself from a neutral to an ionized
medium \citep[known as the Epoch of Reionization, or EoR, at
  $z\simgt6$,
  e.g.,][]{stanway03a,bunker03a,bouwens03a,bouwens06a,dickinson04a}. They
also enabled the detailed study of galaxy morphologies
\citep[e.g.,][]{abraham96a,lowenthal97a,conselice00a,lotz06a,Scarlata07a},
stellar mass growth
\citep[e.g.,][]{sawicki98a,brinchmann00a,papovich01a}, the impact of
local environment
\citep[e.g.,][]{balogh04a,kauffmann04a,christlein05a,cooper08a,scoville13a},
the distribution of dark matter across the cosmic web
\citep[e.g.,][]{natarajan98a,mandelbaum06a,Massey07a,leauthaud07a,leauthaud11a},
as well as the discovery of the tight relationship between galaxies
stellar masses and star formation rates \citep[e.g., the galaxies'
  ``star-forming main sequence,''][]{daddi07a,noeske07a,elbaz07a}.

However, due to the expansion of the Universe, the next leap forward
required observations in the near-infrared (NIR) part of the spectrum. That
came with the installation of the WFC3 camera on {\it Hubble} during
the 2009 servicing mission. WFC3 expanded {\it Hubble}'s deep field
capabilities into the NIR at similar depths as was
previously achieved in the optical, enabling a tenfold increase in the
number of candidate galaxies identified beyond $z\simgt6$
\citep{robertson15a,bouwens15a,finkelstein15a,finkelstein16a,stark16a},
from a few hundred to a few thousand as well as the study of galaxies'
rest-frame optical light out to $z\sim3$ (e.g., \citealt{Wuyts11a,Lee13a,
van-der-wel14a,Kartaltepe15b}).
The Cosmic Assembly Near-infrared Deep
Extragalactic Legacy Survey \citep[CANDELS;
][]{grogin11a,koekemoer11a} was particularly pioneering as it imaged
portions of five of the key deep fields (GOODS-N, GOODS-S, UDS, EGS,
and COSMOS) with the F125W and F160W filters over a total area of
$\sim$800 arcmin$^2$.

The successful launch of the {\it James Webb Space Telescope} ({\it JWST})
now marks a new era for studying the infrared Universe and the distant
cosmos. With six times the collecting area of {\it Hubble} and optimized
for observations in the near- and mid-infrared, {\it JWST} is
currently providing images with greater depth and spatial resolution than
previously possible. This is beginning to enable a substantial improvement in our
understanding of galaxy evolution during the first few hundred million
years (the epoch of cosmic dawn, $z\simgt6$) to the peak epoch of galaxy assembly
(known as cosmic noon, $1\leq z \leq 3$). 
Given the tremendous legacy value of the deep
fields imaged by the Great Observatories, several {\it JWST} deep
fields have been planned for the observatory's first year of
observations.
The largest program among these, in both area on the sky and
total prime time allocation, is the COSMOS-Web\footnote{
This survey was originally named COSMOS-{\it Webb}, as a combination
of the telescope name and in reference to the cosmic web, but later
renamed to emphasize the scientific goal of mapping the cosmic web on
large scales as well as to be inclusive and supportive to members of
the LGBTQIA+ community.} 
Survey (PIs: Kartaltepe \& Casey), for which this paper provides an overview.

\begin{figure*}
\centering
\includegraphics[width=\textwidth]{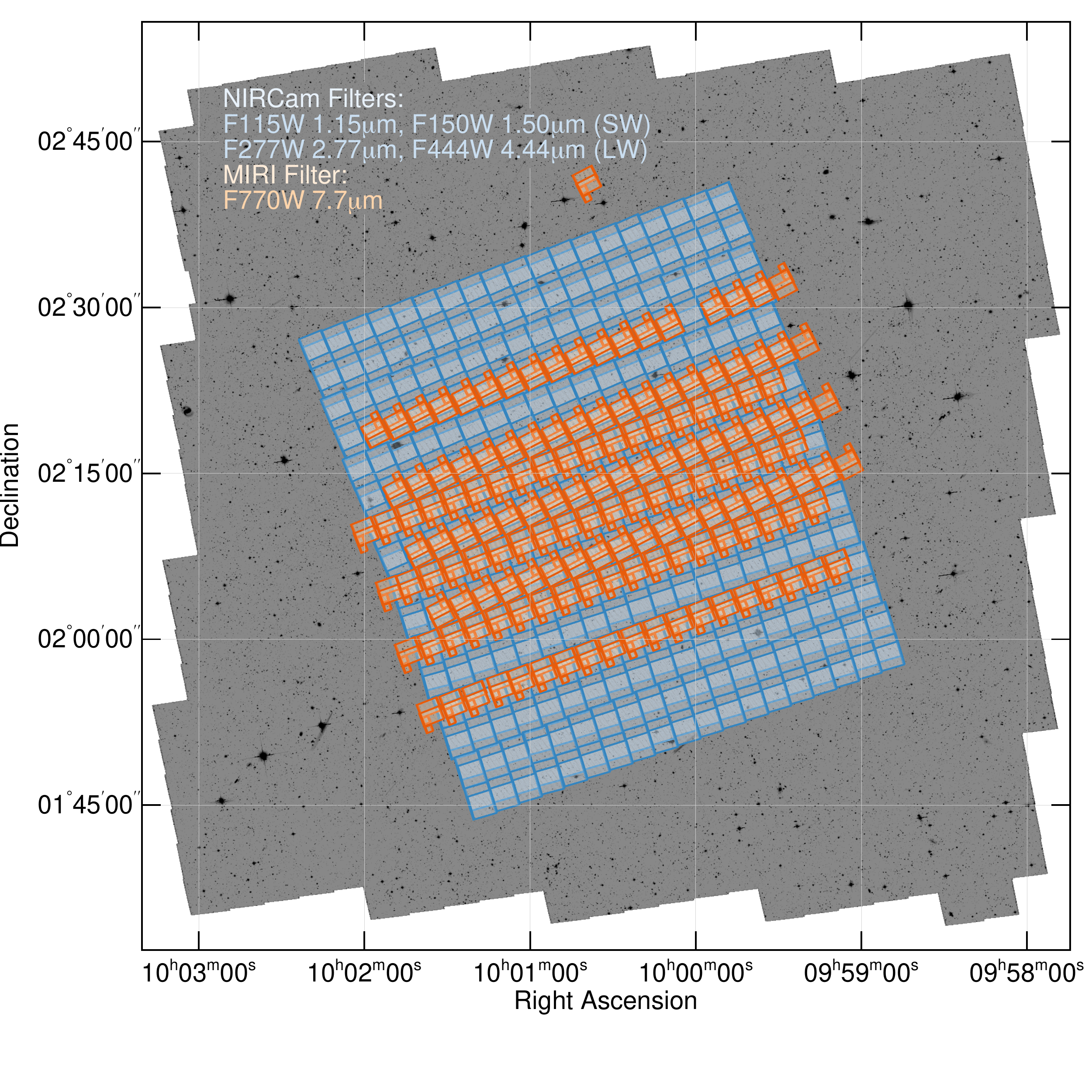}
\caption{A map of the COSMOS-Web tiling pattern embedded within the
  {\it Hubble} ACS F814W mosaic of the COSMOS field
  \citep{scoville07a,koekemoer07a}. The mosaic consists of 152 visits
  where NIRCam serves as the primary instrument (long wavelength
  detector coverage shown in blue) with MIRI in parallel (shown in
  orange).  The entire NIRCam mosaic is centered on the position
  $\alpha$=10:00:27.9, $\delta$=$+$02:12:03.5 and is 41.5 arcminutes
  (in the east-west direction) $\times$ 46.6 arcminutes (in the
  north-south direction) in size. The entire mosaic has an average
  position angle of 110$^{\circ}$, with individual visit PAs equal to
  293$^{o}$ in the northern half and 107$^{\circ}$ in the southern
  half. Three visits required slightly different position angles due
  to availability of guide stars; this includes the lone northern-most
  MIRI tile. The detailed coordinates and position angles of each
  visit are provided in the Appendix, \S~\ref{appendix}.}
\label{fig:jwstmap}
\end{figure*}

COSMOS-Web was designed to bridge deep pencil-beam surveys from {\it
  Hubble} with shallower wide-area surveys, such as those that will be
made possible by facilities like the future {\it Roman Space
  Telescope} \citep{akeson19a} and {\it Euclid}
\citep{euclid-collaboration22a}.  With its unique combination of
contiguous area and depth, COSMOS-Web will enable countless scientific
investigations by the broader community. It will forge the detection
of thousands of galaxies beyond $z>6$, while also mapping the
environments of those discoveries on scales larger than the largest
coherent structures in the cosmic web on $\simgt$10\,Mpc scales. It
will identify hundreds of the rarest quiescent galaxies in the early
Universe ($z>4$) and place constraints on the formation mechanisms of
the most massive galaxies. It will also directly measure the evolution
of the stellar mass to halo mass relation (SMHR) out to $z\sim2.5$ as
a function of various galaxy properties using weak lensing
measurements to estimate halo mass.

This paper describes the motivation for the COSMOS-Web survey as well
as the program's design, providing an initial overview of what is to
come as the data are collected, processed, and
analyzed. Section~\ref{sec:observations} presents the detailed
observational design of the survey and Section~\ref{sec:context}
briefly describes the context of COSMOS-Web among other deep fields
planned for the first year of {\it JWST}
observations. Section~\ref{sec:science} presents the scientific
motivation of the survey as the drivers for the observational
design. In Section~\ref{sec:ancillary}, we share other possible
investigations and predictions for what will be made possible by
COSMOS-Web, beyond the main science goals.  We summarize our outlook
for the survey in section~\ref{sec:summary}.  Throughout this paper,
we use AB magnitudes \citep{oke83a}, assume a Chabrier stellar initial
mass function \citep{chabrier03a}, and a concordance cosmology with
$H_{0} = 70~{\rm km~s^{-1}~Mpc^{-1}}$ and $(\Omega_{tot},
\Omega_{\Lambda}, \Omega_{m}) = (1, 0.7, 0.3)$.

\section{Observational Design}\label{sec:observations}

The observational design of the COSMOS-Web survey is motivated by the
requirements of the primary science drivers described in \S\ref{sec:science} 
while also striving to maximize value for the broader community 
across a wide range of science topics, described in part in 
\S~\ref{sec:ancillary}. Here we describe the detailed layout of the 
COSMOS-Web survey and provide more detailed motivation for 
the design when discussing the science goals in \S~\ref{sec:science}.

\begin{deluxetable*}{c@{ }c@{ }c@{ }c@{ }c@{ }c@{ }c@{ }c}
 \tabletypesize{\small}
 \tablecolumns{7}
 \tablecaption{Summary of COSMOS-Web NIRCam Survey Depth}
 \tablehead{
  \colhead{No. of NIRCam} & \colhead{Total NIRCam} & \colhead{SW Area} & \colhead{F115W Depth} & \colhead{F150W Depth} & \colhead{LW Area} & \colhead{F277W Depth} & \colhead{F444W Depth}\cr
  \colhead{Exposures} & \colhead{Exp. Time (s)} & \colhead{(arcmin$^2$)} & \colhead{ (5$\sigma$)} & \colhead{(5$\sigma$)} & \colhead{(arcmin$^2$)} & \colhead{(5$\sigma$)} & \colhead{(5$\sigma$)}
 }
 \startdata
 1 & 257.68 & 71.3 & 26.87 & 27.14 & 17.8 & 27.71 & 27.61 \\
 2 & 515.36 & 991.6 & 27.13 & 27.35 & 978.0 & 27.99 & 27.83 \\
 3 & 773.05 & 60.0 & 27.26 & 27.50 & 24.4 & 28.12 & 27.94 \\
 4 & 1030.73 & 805.2 & 27.45 & 27.66 & 904.3 & 28.28 & 28.17 \\
 \enddata

 \tablecomments{Depths quoted are average 5$\sigma$ point source
   depths calculated within 0$\farcs$15 radius apertures on data from
   our first epoch of observations without application of aperture
   corrections.}
 \label{tab:obssummarynircam}
\end{deluxetable*}

\begin{deluxetable}{c@{ }c@{ }c@{ }c}
 \label{tab:obssummarymiri}
 \tabletypesize{\small}
 \tablecolumns{4}
 \tablecaption{Summary of COSMOS-Web MIRI Survey Depth}
 \tablehead{
  \colhead{No. of MIRI} & \colhead{Total MIRI} & \colhead{Area Covered} & \colhead{F770W Depth}\cr
  \colhead{Exposures} & \colhead{Exp. Time (s)} & \colhead{(arcmin$^2$)} & \colhead{ (5$\sigma$)} }
 \startdata
 2 & 527.26 & 80.5 & 25.33 \\
 4 & 1054.52 & 430.4 & 25.70 \\
 6 & 1581.77 & 30.8 & 25.76 \\
 8 & 2109.03 & 146.1 & 25.98 \\
 \enddata
 \tablecomments{Depths quoted are average 5$\sigma$ point source
   depths calculated within 0$\farcs$3 radius apertures on data from
   our first epoch of observations, without application of aperture
   corrections. }
\end{deluxetable}

\subsection{Description of Observations}

COSMOS-Web consists of one large contiguous 0.54\,deg$^2$ NIRCam
\citep{rieke22a} mosaic conducted in four filters (F155W, F150W,
F277W, and F444W) with single filter (F770W) MIRI \citep{wright22a}
imaging observations obtained in parallel over a total non-contiguous
area of 0.19\,deg$^2$. The NIRCam mosaic is spatially distributed as a
41.5$'\times$46.6$'$ rectangle at an average position angle of
110$^\circ$; the shorter side of the mosaic is primarily oriented in
the east-west direction. The center of the mosaic is at
$\alpha$=10:00:27.92, $\delta$=$+$02:12:03.5 and is comprised of 152
separate visits (where each visit observes a single tile in the
mosaic\footnote{A single `visit' is a JWST observation acquired in one
block of continuously scheduled time.}) arranged in a 19$\times$8
grid.  The coverage of these visits overlaid on the COSMOS {\it
  Hubble} F814W imaging is shown in Figure~\ref{fig:jwstmap}.

\begin{figure}
\centering
 \includegraphics[width=0.99\columnwidth]{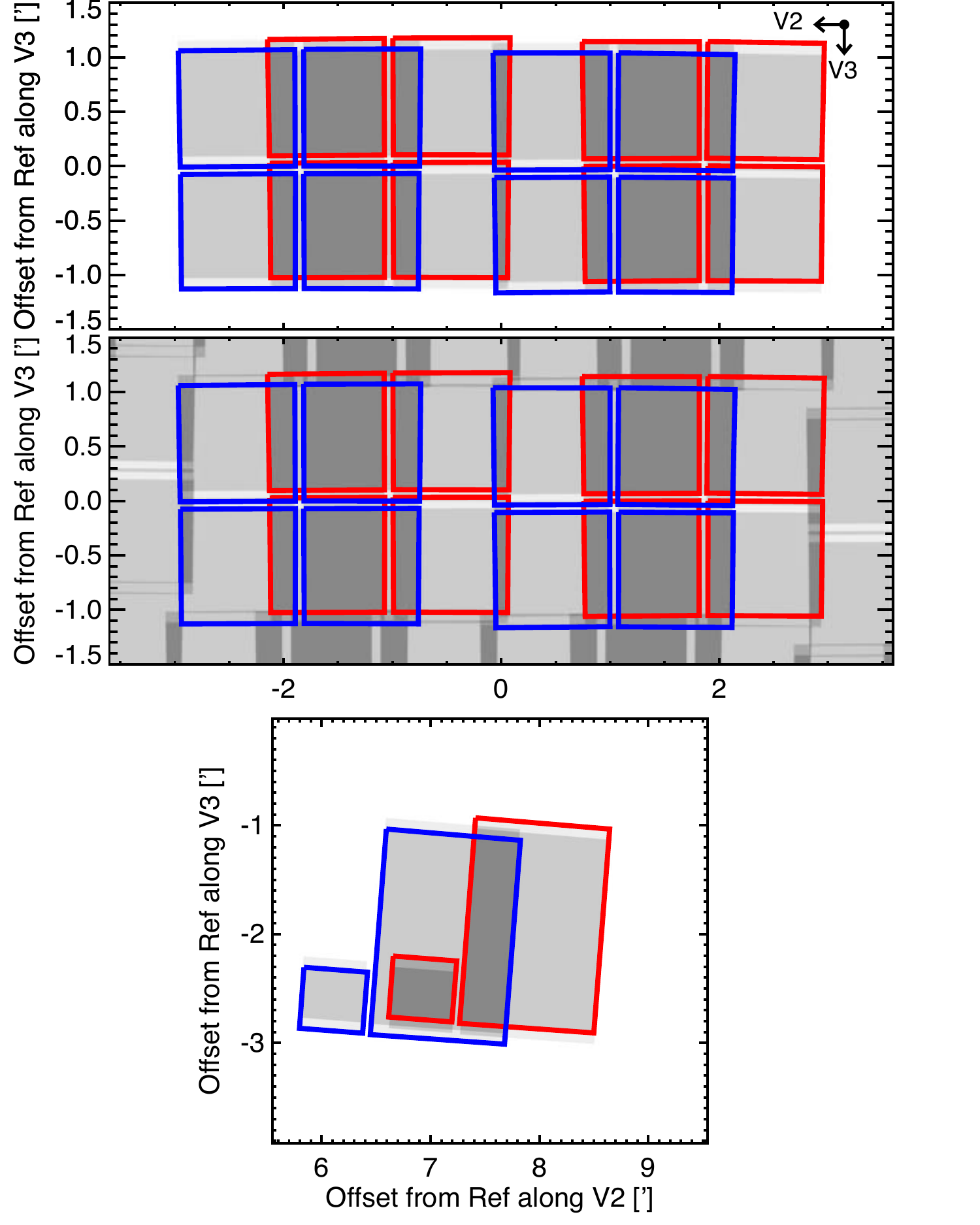}
\caption{An illustration of the 4TIGHT dither pattern for NIRCam prime
  visits (top two panels) and MIRI parallel visits (bottom panel). The
  top panel shows the NIRCam SW exposure map for a single visit with
  coverage ranging from one (lightest) to four (darkest) exposure
  depth. Two of the four dither positions are outlined in color (red
  and blue) for clarity. The middle panel shows the NIRCam SW exposure
  map in the context of the larger COSMOS-Web mosaic. At bottom, the
  MIRI coverage is shown. The axes are positional offsets along the V3
  and V2 angle (i.e., perpendicular and parallel to the PA) relative
  to the reference position, given for each visit in the Appendix,
  \S~\ref{appendix}.}
\label{fig:4tight}
\end{figure}

Each individual visit is comprised of eight separate exposures of
$\sim$257 seconds each, split into two separate executions of the
4TIGHT dither pattern at the same position in the mosaic.  Each 4TIGHT
dither pattern contains four individual integrations; an illustration
of this dither pattern in one standalone visit and embedded in the
larger mosaic is shown in Figure~\ref{fig:4tight}. The first 4TIGHT
dither executes two NIRCam filters -- F115W at short wavelengths (SW)
and F277W at long wavelengths (LW) -- and the MIRI F770W filter in parallel.  
The second execution of the 4TIGHT dither switches NIRCam filters -- to
F150W in SW and F444W in LW -- yet keeps the same MIRI filter, F770W,
for added depth.

The northern half of the mosaic is observed at one position angle,
293$^\circ$, while the southern half of the mosaic is observed at
another, 107$^\circ$. These position angles are relative to the NIRCam
instrument plane and not V3 (which differ by $<$1$^\circ$); they
are also not exactly a 180$^\circ$ flip from one another. Instead they
are staggered by $\pm$3$^\circ$ to make scheduling more flexible while
maintaining a contiguous mosaic using a slight jigsaw pattern to
stitch adjacent visits together. The distribution of half of the
mosaic at one position angle and the other half at another also makes
it possible to fit most of the MIRI parallel exposures fully within
the larger NIRCam mosaic.  A few visits required further position
angle modification due to limitations in guide star catalog
availability at their initially intended angles. The Appendix
(\S~\ref{appendix}) gives detailed information for each individual
visit and a table of all visits.

The depth of the NIRCam observations varies based on the number of
exposures at any position in the mosaic (see
Table~\ref{tab:obssummarynircam}); of the total 1928\,arcmin$^2$
($\approx$0.54\,deg$^2$) area in the NIRCam SW mosaic,
71.3\,arcmin$^2$ ($\sim$3.7\%) will be covered with only a single
exposure per SW filter, 991.6\,arcmin$^2$ ($\sim$51.4\%) will have two
SW exposures, 60.0\,arcmin$^2$ ($\sim$3.1\%) will have three SW
exposures, and 805.2\,arcmin$^2$ ($\sim$41.8\%) will have four SW
exposures. The NIRCam LW mosaic covers a total area of
1924\,arcmin$^2$, of which 17.8\,arcmin$^2$ ($\sim$0.9\%) has single
exposure depth, 978.0\,arcmin$^2$ ($\sim$50.8\%) has two exposure
depth, 24.4\,arcmin$^2$ ($\sim$1.3\%) has three exposure depth, and
904.3\,arcmin$^2$ ($\sim$47.0\%) has four exposure depth. The most
deeply exposed portions of the SW mosaic align with the deepest
portions of the LW mosaic, though the areas differ slightly based on
the differences in detector size and gaps between SW detectors.

Due to the design of the NIRCam mosaic as contiguous, the MIRI
parallel observations are not contiguous but are distributed in 152
distinct regions corresponding to the 152 visits.  MIRI coverage of
each visit has an area of 4.2\,arcmin$^2$ corresponding to the primary
MIRI imager field of view, and 4.5\,arcmin$^2$ when accounting for the
additional area of the Lyot Coronographic Imager\footnote{During MIRI
imaging, the Lyot Coronographic Imager is also exposed using the same
filter and optical path of the imager. Modulo the occulting spot and
its support structure, the Lyot region provides a small amount of
additional survey area for MIRI imaging campaigns.  See the
\href{https://jwst-docs.stsci.edu/jwst-mid-infrared-instrument/miri-features-and-caveats}{JWST
  User Documentation Page\ on MIRI Features and Caveats} for more
details.}. Of that area, 0.55\,arcmin$^2$ (12\%) has two MIRI
exposures, 2.81\,arcmin$^2$ (62\%) has four, 0.21\,arcmin$^2$ (5\%)
has six, and 0.96\,arcmin$^2$ (21\%) has eight MIRI exposures. The
total area covered with MIRI in COSMOS-Web is 688\,arcmin$^2$ or
0.19\,deg$^2$. Of the 152 MIRI visits, 143 (651\,arcmin$^2$, 95\%) are
fully contained within the NIRCam mosaic.  Note that MIRI
observations from PRIMER (GO \#1837) add an additional 53\,arcmin$^2$
of (deeper) 7.7\um\ coverage (see \S~\ref{sec:context} for full
details) contained within the NIRCam footprint, bringing the total
MIRI coverage in COSMOS from these two Cycle 1 surveys to
742\,arcmin$^2$.

\begin{figure*}
 \centering
 \includegraphics[width=\textwidth]{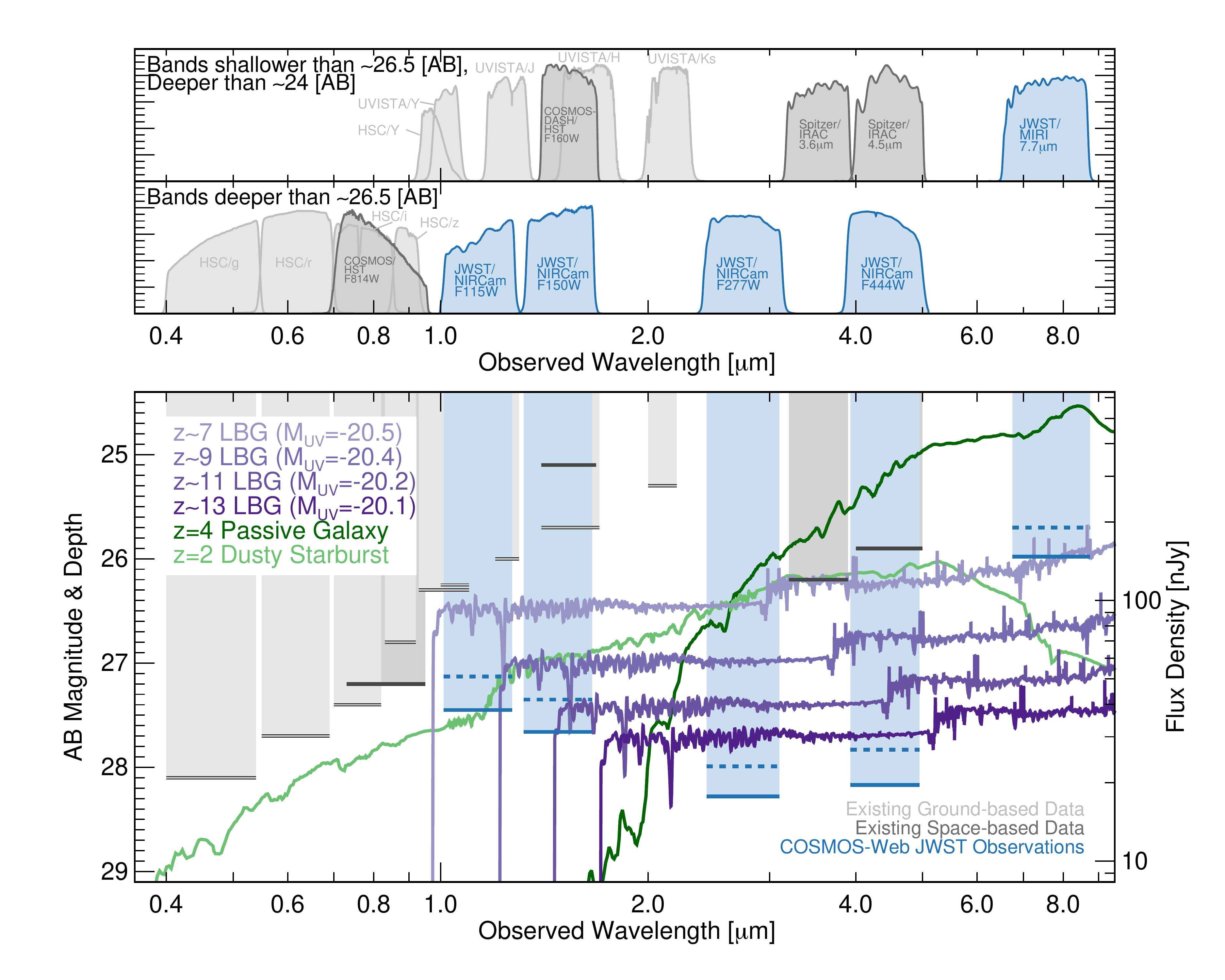}
 \caption{An illustration of the deepest filters available in
   COSMOS-Web and their depths across the spectrum. At top are the
   filter transmission profiles for existing COSMOS datasets that are
   ground-based (light gray), space-based (dark gray), and new
   additions from {\it JWST} for COSMOS-Web (blue).  These filters are
   separated between those that have 5$\sigma$ point source depths
   between 24--26.5 magnitudes (top sub-panel), and those that reach
   depths beyond 26.5 magnitudes \citep[bottom sub-panel;
       see ][for more details]{weaver22a}. We also
   include recent coverage from COSMOS-DASH at
   1.6\um\ \citep{mowla19a,cutler22a}. Note that narrow-band and
   medium-band filters in the field are not shown (as they generally
   have depths shallower than $\sim$24 magnitudes).  At bottom, we
   illustrate the 5$\sigma$ point source depths from these same
   filters, highlighting the depth of COSMOS-Web {\it JWST}
   observations in solid blue at full four-integration depth; dashed
   lines show half two-integration depth (covering approximately half
   the mosaic, as detailed in Tables~\ref{tab:obssummarynircam} and
   ~\ref{tab:obssummarymiri}). Overlaid are several galaxy templates:
   Lyman-break galaxies at $z\sim7-13$ (shades of lavender to dark
   purple), a M$_\star\sim$10$^{10}$\,\msun\ $z=4$ passive galaxy
   (dark green), and a $z=2$ dusty starburst (light green).}
 \label{fig:sedlimits}
\end{figure*}

Table~\ref{tab:obssummarynircam} summarizes the characteristics of the
NIRCam mosaic and the measured depths as a function of number of
exposures.  The NIRCam depths have been measured using data from the
first epoch of COSMOS-Web observations, consisting of six visits (out
of the total 152).  These data are later described in
\S~\ref{sec:scheduling}.  These are broadly consistent with the
expected performance of {\it JWST} in-flight \citep{rigby22a}.  These
depths correspond to 5$\sigma$ point sources extracted within
0$\farcs$15 radius circular apertures in each filter without any
aperture corrections applied. Table~\ref{tab:obssummarymiri} provides
a summary for the MIRI exposures; similarly, these depths are measured
directly using data from the first epoch of observations in COSMOS-Web
using a a 0$\farcs$3 radius circular aperture without aperture
correction.  We note that the measured MIRI depths are significantly
better than expectation from the exposure time calculator.  We
conducted a number of tests to measure this depth accurately,
including a direct comparison of IRAC 8\um\ flux densities with MIRI
7.7\um\ flux densities, measurement of depth within empty apertures in
individual exposures, as well as measurement of the standard deviation
in flux densities for individual sources in individual exposures.  All
tests give consistent results, showing F770W depths nearly a magnitude
deeper than expectation.  The depths of the survey as a function of
wavelength are shown in Figure~\ref{fig:sedlimits} relative to other
existing datasets available in the COSMOS field.

\subsection{Motivation for a Contiguous $\sim$0.5\,deg$^2$ Area}

The contiguous, and roughly square, area of COSMOS-Web is driven by
two of our primary science objectives. The first is to construct large
scale structure density maps at $6<z<10$ to address whether or not the
most UV-luminous systems are embedded in overdense structures (see
\S~\ref{sec:eor} for details). Mapping the large scale environments of
our discoveries and mitigating cosmic variance at these epochs (with
cosmic variance less than 10\%, i.e., $\sigma_v^2<0.10$) requires
contiguous solid angles larger than the expected size of reionization
bubbles at these redshifts \citep{behroozi19a}, $>$\,0.3-0.4\,deg$^2$.
Our 0.54\,deg$^2$ program allows for some uncertainty in the scale of
these reionization bubbles, as some simulations see bubbles extend on
40$'$ scales \citep{daloisio18a,thelie22a}.  Our NIRCam mosaic maps to
$\sim$\,(114\,Mpc)$^2$ between $6<z<8$ and $\sim$\,(122\,Mpc)$^2$
between $8<z<10$ projected on the sky at these epochs.  We describe
more about the expected cosmic variance in COSMOS-Web in
\S~\ref{sec:cosmicvariance}.

The second scientific driver for our contiguous area is the coherence
we can achieve for the weak lensing measurement of galaxies' halo
masses on scales $\simlt$10\,Mpc in order to place constraints on the
SMHR out to $z\sim2.5$ (see \S~\ref{sec:wl} for details). This
requires at least $\sim5$ dark matter halo scale lengths ($\sim3$
proper Mpc across each) of contiguous coverage, for which our survey
will provide $\sim10\times10$ dark matter scale lengths to boost the
signal-to-noise and allow splitting by galaxy type and by mass
\citep{wechsler18a,wang18b,debackere20a,shuntov22a}. Several smaller
non-contiguous areas (of order 0.05\,deg$^2$) would render the SMHR
measurement and calibration of cosmological models severely hindered.

\subsection{Field on the Sky}

\begin{figure*}
 \includegraphics[width=0.99\textwidth]{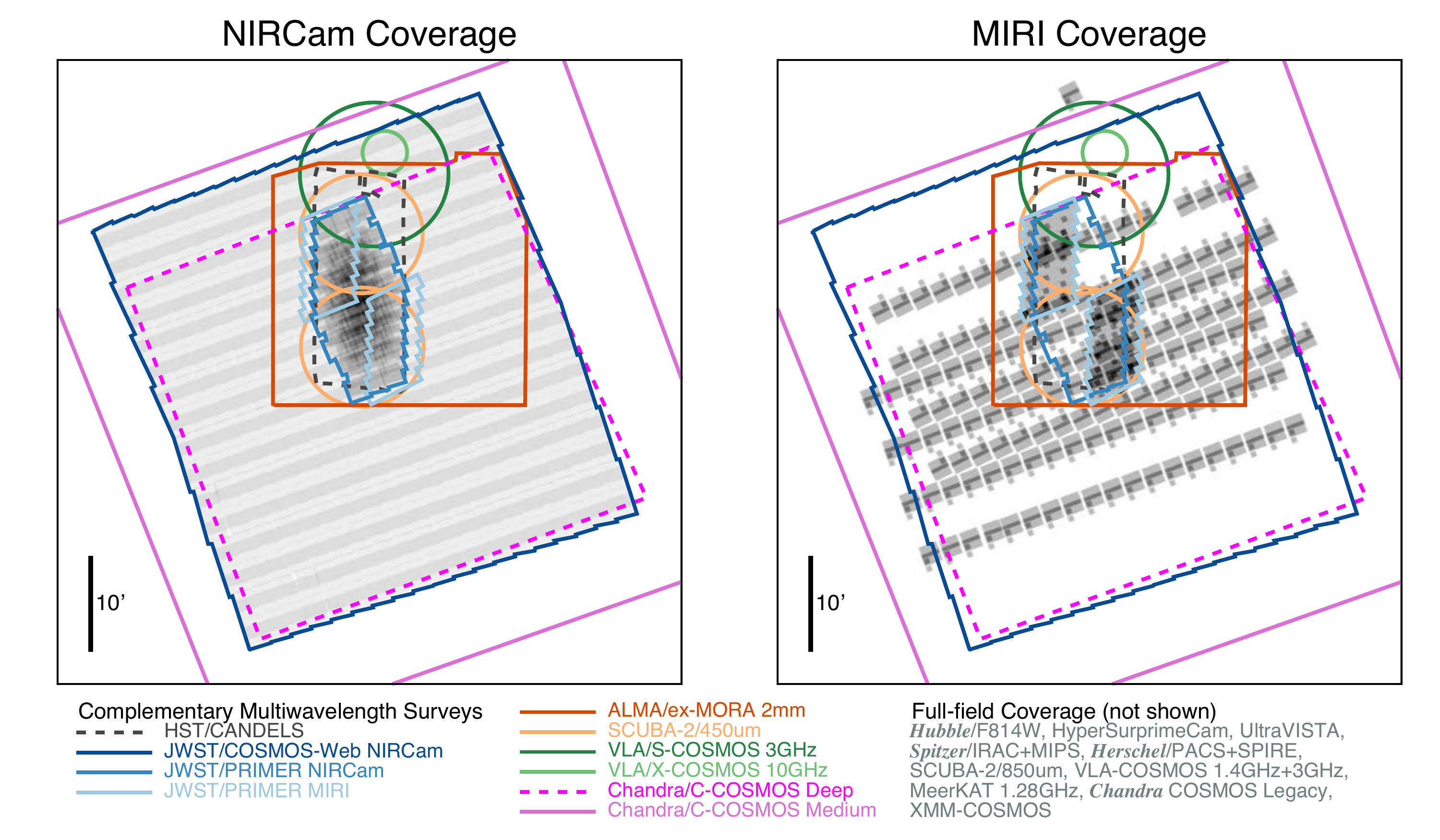}
 \caption{The COSMOS-Web NIRCam (left) and MIRI (right) coverage shown
   together with the PRIMER NIRCam and MIRI coverage in a joint
   exposure map in grayscale. The two instruments' coverage are shown
   separately for clarity. We overlay maps of a number of
   multiwavelength datasets for context. The {\it JWST} coverage from
   COSMOS-Web and PRIMER are shown in shades of blue.  The {\it
     Hubble} CANDELS survey \citep{grogin11a,koekemoer11a} area is
   shown in dashed black, the ALMA extended MORA survey
   \citep[][Long \etal, in preparation]{zavala21a,casey21a} is shown
   in burnt orange and the deep SCUBA-2 450\um+850\um\ coverage area
   of the eS2-COSMOS survey and STUDIES survey \citep{wang17a} is
   shown in light orange \citep[note that the entire COSMOS field is
     covered with 850\um\ SCUBA-2 coverage from the S2COSMOS
     Survey;][]{simpson19a}.  In the radio, we highlight the deep
   continuum coverage at 3\,GHz and 10\,GHz in dark and light green,
   respectively, from the COSMOS-XS survey \citep{van-der-vlugt21a},
   which complements the full-field 3\,GHz VLA COSMOS Survey of
   \citet{smolcic17a}. In the X-ray, we show the area covered by the
         {\it Chandra} C-COSMOS Deep survey (dashed
           magenta) as well as the medium depth
         survey (solid magenta),
         both summarized by \citet{elvis09a}, with full COSMOS field
         coverage extended by \citet{civano16a}. Finally, we note that
         the full COSMOS field has coverage with Subaru's Hyper
         Suprime-Cam \citep{aihara22a}.}
 \label{fig:allmap}
\end{figure*}

The COSMOS field was chosen for these observations for several
reasons. First, the existing HST/ACS F814W coverage
\citep{koekemoer07a} provides crucial value to our science goals of
detecting galaxies beyond $z>6$ using [F814W]-[F115W] colors. Second,
COSMOS has the widest deep ancillary data coverage from X-ray to radio
wavelengths \citep{ilbert13a,laigle16a,weaver22a}. Third, it is an
equatorial field ($\alpha=150^\circ$, $\delta=+2^\circ$), and thus
accessible to all major existing and planned future facilities,
essential for swift and efficient follow-up of {\it JWST}-identified
sources. A sampling of the multiwavelength data already available in
the COSMOS-Web footprint is shown in Figure~\ref{fig:allmap}.

Additionally, COSMOS has been selected or is a likely candidate to be
a deep calibration field for future key projects including {\it
  Euclid}, the {\it Roman Space Telescope}, and the Vera Rubin
Observatory LSST project.  Over 250,000 spectra have been taken of
$>$\,100,000 unique objects in the COSMOS field at $0<z<7$
(A. Khostovan et al.~in preparation), including from large surveys
such as zCOSMOS \citep{lilly07a,lilly09a}, FMOS-COSMOS
\citep{silverman15a,kartaltepe15a,kashino19a}, VUDS \citep{le-fevre15a}, and many
programs using Keck (e.g.,
\citealt{kartaltepe10a,Capak11b,casey12b,Kriek15k,hasinger18a}),
greatly enhancing the accuracy of photometric redshifts for all
sources in the field. Lastly, the quality of photometric redshifts
$\Delta z/(1+z)<0.02$ for galaxies with $i<25$
\citep{ilbert13a,laigle16a,weaver22a} has facilitated the discovery
and analysis of galaxies out to $z\sim7$ and beyond
\citep{bowler17b,bowler20a,stefanon19a,kauffmann22a}.  The photometric
redshifts will be further improved with the addition of COSMOS-Web
(see \S~\ref{sec:photozs}), dramatically improving the accuracy of the
weak lensing measurement of galaxies' halo mass as well as galaxies'
stellar masses and star formation rates across all epochs.

\subsection{Filter Optimization}

We simulated the effectiveness of many filter combinations to deliver
the science objectives described in \S~\ref{sec:science} and
determined that COSMOS-Web should be a four filter NIRCam survey with
MIRI imaging conducted in parallel: F115W+F150W in SW, F277W+F444W in
LW, and F770W with MIRI. Reionization science drives the choice of
F115W and F150W to maximize coverage of the observed
  wavelength of a Lyman break from $6<z<13$; we plan EoR source
  selection using a hybrid photometric redshift and dropout approach
($z\sim\,6-7$ galaxies drop out in {\it HST}-F814W, while
$z\sim\,8-10$ galaxies drop out in the F115W filter, and $z>12$ will
drop out in F150W).
Weak lensing objectives are less sensitive to filter choice but
benefit from tremendous depth in the NIR by increasing the background
source density; we expect $>$10 galaxies per arcmin$^{2}$ at $z>4$
with measurable shapes, in other words, those found above a
15\,$\sigma$ detection threshold.  We calculate the on-sky source
density of galaxies above certain apparent magnitude thresholds from
existing measurements of galaxy luminosity functions from $0<z<10$
\citep[e.g.,][]{arnouts05a,bouwens15a,finkelstein15a}.  Indeed,
preliminary simulations show that galaxies at the 15\,$\sigma$
shape-detection threshold, F277W\,$\sim$\,26.8, with $R_{\rm
  eff}\simlt\,0\farcs3\ (\approx$\,2--3\,kpc), are recovered without
bias introduced from the {\it JWST} point spread function (PSF;
Liaudat \&\ Scognamiglio et al., in preparation).  We find F277W+F444W
to be the most advantageous LW filter combination to improve the
quality of photometric redshifts and mitigate lower redshift
contaminants (more details discussed in \S~\ref{sec:eor}). The F444W
filter is particularly useful for measuring the rest-frame optical
morphologies of galaxies at $z>4$, (e.g., \citealt{Kartaltepe22a}) and
the rest-frame near-infrared morphologies of lower redshift galaxies
(e.g., \citealt{guo22a}).  The LW filters will be useful for the
identification of very red $z=4-6$ quiescent galaxies and measuring
their mass surface densities and morphologies at high signal-to-noise.

The choice of F770W for the MIRI parallel exposures is motivated by
the need to constrain reliable stellar masses for $z>4$ massive
systems. F770W is roughly matched to the {\it Spitzer} 8.0\,\um\ filter
\citep[which has much shallower data in COSMOS;][]{sanders07a}. F770W
data will provide a factor of 50\,$\times$ improvement in depth
relative to {\it Spitzer} 8.0\,\um\ and a factor of 7.6$\times$ improvement
in the beam size, thus opening up detections to the $z>4$
universe. Our MIRI data will cover an area $\sim$3.5$\times$ larger
than all other planned {\it JWST} MIRI deep fields from Cycle 1
combined (see \S~\ref{sec:context}), making it particularly sensitive to rare, bright
objects. F770W optimizes both sensitivity and the uniqueness of longer
rest-frame wavelengths for high redshift galaxies. Longer wavelength
filters would reduce the sensitivity by 10--30$\times$, and F560W does
not provide a sufficient lever arm from F444W to measure high-$z$
galaxy stellar masses.

\subsection{Precision of Photometric Redshifts}\label{sec:photozs}

\begin{figure*}
 \centering
 \includegraphics[width=0.99\columnwidth]{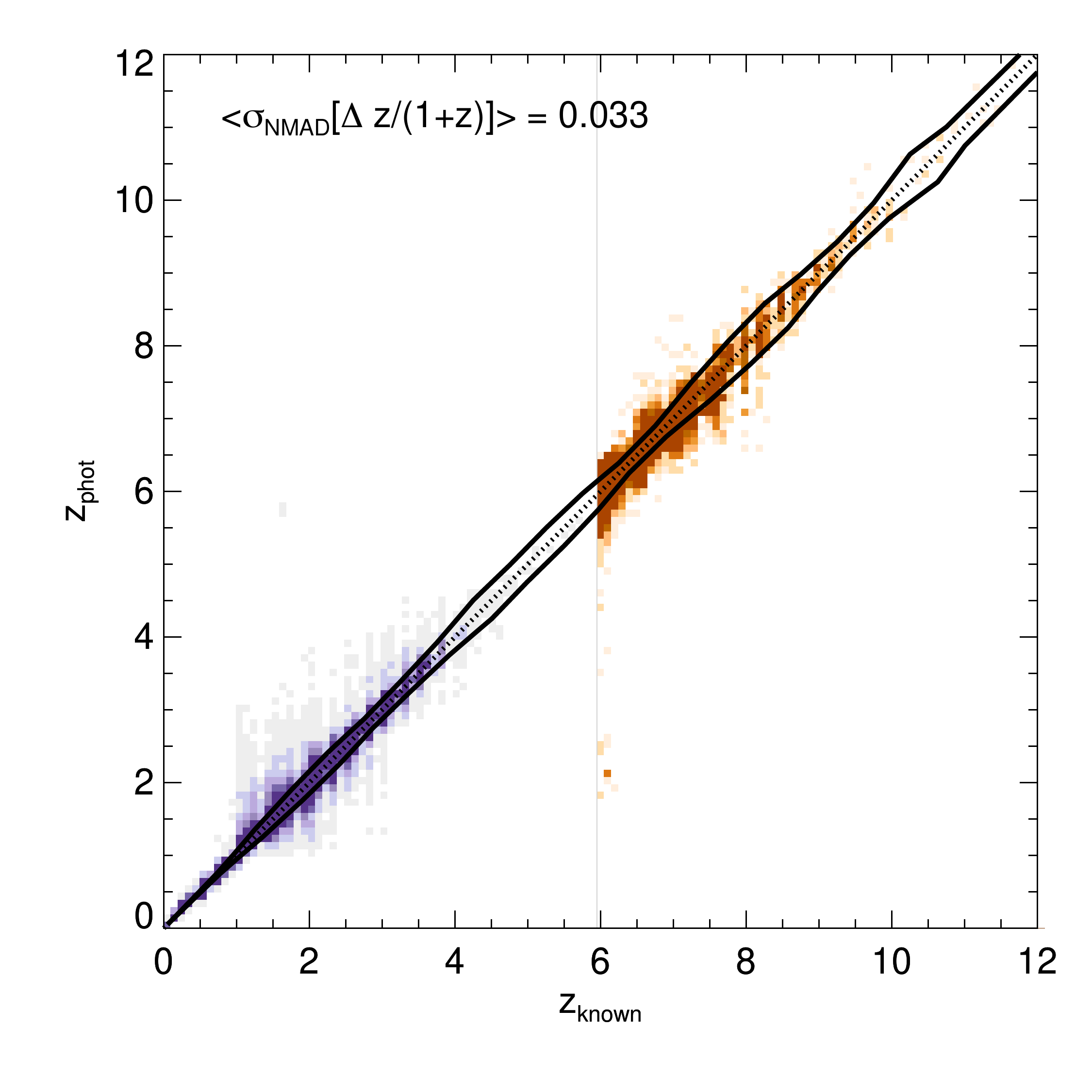}
 \includegraphics[width=0.99\columnwidth]{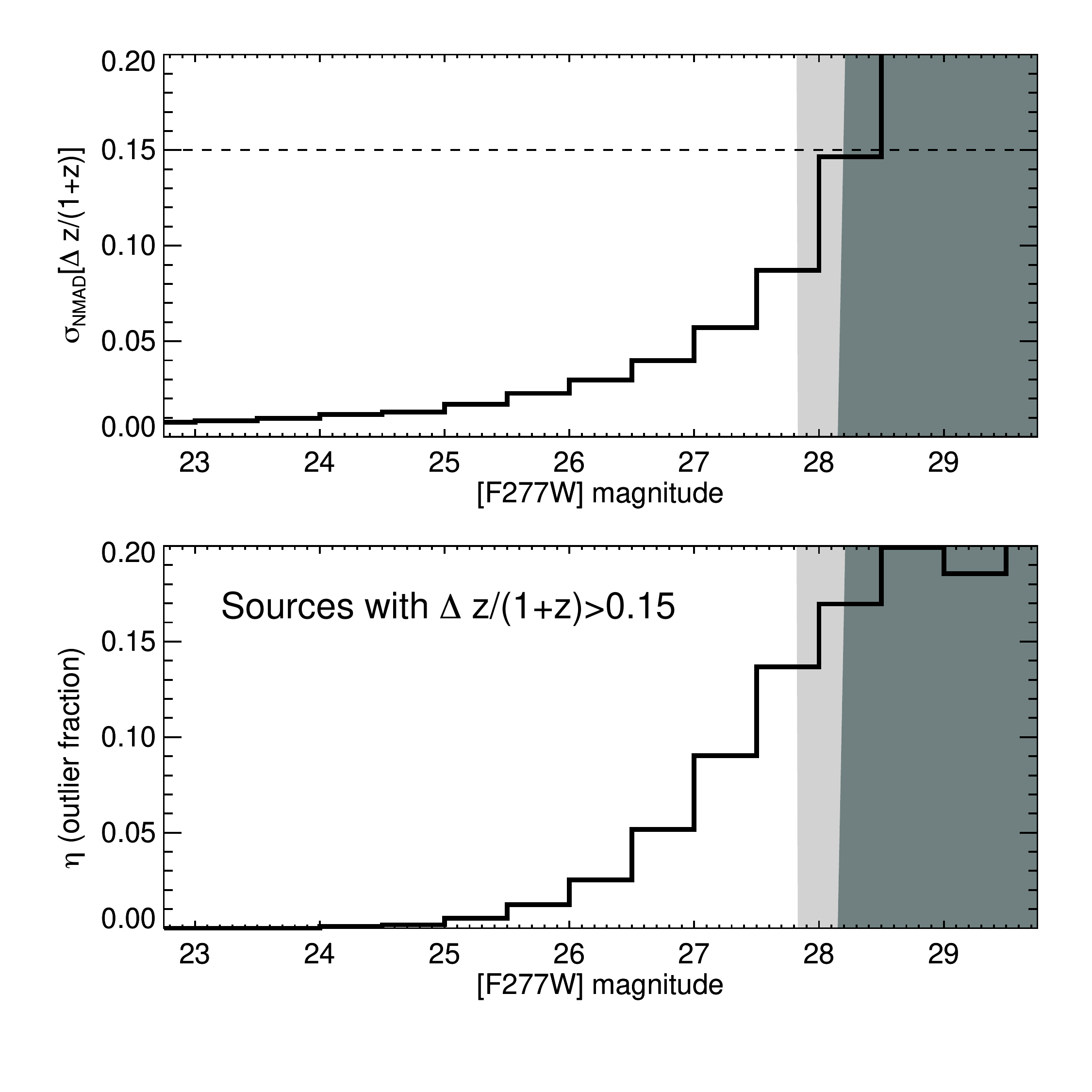}
 \caption{Results of photometric redshift fitting to a set of mock
   galaxies; these mock galaxies have the full set of COSMOS
   photometry, spanning both existing ground and space-based
   photometry \citep[drawn from the limits described in][]{weaver22a},
   as well as model photometry in the {\it JWST} bands corresponding
   to COSMOS-Web. On the left, we show the known redshifts of mock
   galaxies vs.\ the best fit photometric redshifts, which are derived
   by performing spectral energy distribution (SED) fits with {\sc
     LePhare} \citep{arnouts02a,ilbert06a}.  Below $z<6$, the purple
   heat map shows the density of sources down to F277W\,=\,27.5
   magnitudes in 1\%\ of our simulation for clarity; the orange heat
   map shows all simulated sources above $z>6$.  The thick black lines
   enclose the average dispersion about the 1-to-1 line as a function
   of redshift, as measured using the normalized median absolute
   deviation ($\sigma_{\rm NMAD}$). The average precision across all
   redshifts and magnitudes (down to 27.5) is 3.3\%. At the top right,
   we show how the $\sigma_{\rm NMAD}$ statistic varies as a function
   of F277W magnitude; sources down to 27$^{th}$ magnitude will have
   photometric redshifts precise to $<$5\%, and sources closer to the
   detection threshold will have 10--18\%\ precision. At the bottom
   right, we show the anticipated outlier fraction ($\eta$) as a
   function of magnitude, defined as sources with photometric redshift
   precision worse than 15\%. The outlier fraction is less than 5\%
   down to $\sim$26.5, then increases toward 15\%\ near the detection
   cutoff. Neither $\sigma_{\rm NMAD}$ nor $\eta$ show significant
   redshift dependence, other than slight spikes in the range $6<z<7$
   and $9<z<11$. Gray bands show our 5$\sigma$ point source detection
   limits for two-integration and four-integration depths.}
 \label{fig:sim_zphotzspec}
\end{figure*}

A crucial aspect of the design of COSMOS-Web was the selection of
filters, largely driven by finding the most reliable selection of EoR
sources from $6<z<11$.  We generated an empirical light cone of mock
galaxies, populating it with galaxies following the galaxy luminosity
function from the local Universe to $z\sim4$ from \citet{arnouts05a};
from $z\sim4-10$, galaxies are drawn from the UV luminosity functions
of \citet{bouwens15a}.  From $z\sim10-12$, the \citeauthor{bouwens15a}
luminosity functions are extrapolated by fixing $M_\star$ and
extrapolating trends in $\Phi_\star$ and $\alpha$ measured at lower
redshift (i.e., higher redshifts have lower densities and steeper
faint-end slopes).

Once the on-sky density of galaxies (as a function of rest-frame UV
absolute magnitude, $M_{\rm UV}$) is set, we assign a variety of
spectral energy distributions (SEDs) to each galaxy.  Given the focus
on reliability of EoR targets, the SEDs we generate were of somewhat
limited scope, focusing on three families of templates from
\citet{maraston05a} with 61 ages for each.  The primary difference
between templates is the star-forming timescale, with exponentially
declining star-formation histories of 0.25, 1, and 10\,Gyr.  Three
attenuations were used with $E(B-V)=0,\ 0.05$, and $0.1$ (not
including very reddened sources).  Both nebular line emission and IGM
opacity \citep{madau95a} were included.  The choice of SED for a given
galaxy was then assigned using a uniform distribution (with an
allowable star-formation timescale).  While there are clear
limitations to this idealized, empirical calibration sample -- such as
the lack of more diverse SEDs, a mass- or redshift-weighted method of
assigning SEDs, or using a wider set of templates to fit the ensuing
photometric redshifts -- it can still provide a useful first pass at
our photometric redshift precision, particularly for newly-discovered
faint galaxies within the EoR.

Noise is added to the mock observations according to the depth in each
filter (to the greatest depth as quoted in
Tables~\ref{tab:obssummarynircam} and~\ref{tab:obssummarymiri}).
Similarly, known noise characteristics of existing ground-based data
have been added to the galaxies' mock photometry \citep[details of
  those observations are provided by][]{weaver22a}. We use this mock
sample to diagnose the contamination and precision of our photometric
redshifts across all epochs, applying tools we will use for the real
dataset. Specifically, here we use the \texttt{LePhare} SED fitting 
code to derive photometric redshifts
\citep{arnouts02a,ilbert06a}, as implemented for the recent COSMOS2020
compilation by \citet{weaver22a}. Note that in
\S~\ref{sec:eorselection} we explore the specific parameter space of
EoR mock sources from this lightcone in more detail, and here we
present the general characteristics of the expected photometric
redshift quality across all epochs.

Figure~\ref{fig:sim_zphotzspec} shows the input `known' redshift
against the best measured output redshift for all mock sources from
$0<z<12$. The full simulation contains $\sim$3.3M
  sources, 13K of which ($\approx0.4$\%) are at $6<z<12$.  Given the
sheer number of sources in the catalog, we split the simulation into
two regimes: at $z<6$ we only sample a random 1\%\ subset of all
sources for photometric redshift fitting (for computational
ease); in other words, we fit photometric redshifts to
  $\sim$33K sources from $0<z<6$. At $z>6$ we fit all
galaxies so that we adequately sample the full range of
  true EoR source properties.  Thus, in
  Figure~\ref{fig:sim_zphotzspec}, there appears to be a dearth of
  sources at $4\simlt z<6$ due to this differential sampling of
  parameter space, but the apparent differential is simply visual
  (e.g., there are 1K sources modeled in the $4<z<6$ bin). To
understand the improvement in the photometric redshifts provided by
COSMOS-Web data, we compare our inferred mock photometric redshift
quality to those from the COSMOS2020 catalog. Specifically,
\citet{weaver22a} find that sources with $i$-band magnitude between
$25<i<27$ have 5\%\ photometric redshift precision. Over the same
$i$-band magnitude range, we infer that these {\it JWST} data will
improve that statistic to 2.5\%. Both precisions are measured using
the normalized median absolute deviation ($\sigma_{\rm NMAD}$) of
$\Delta z/(1+z)$, a quantity analogous to the standard deviation of a
Gaussian but less sensitive to outliers.

While this direct comparison is useful, we also calculate $\sigma_{\rm
NMAD}$ for intrinsically fainter sources selected at longer
wavelengths. We find that the median precision for sources with
F277W magnitudes ranging 25--26.5 to be 2.3\%\ across all epochs,
and those with F277W magnitudes ranging 26.5--27.5 to be 4.2\%. The
right panel of Figure~\ref{fig:sim_zphotzspec} shows how the
photometric redshift precision is expected to degrade for sources as a function
of F277W magnitude. Similarly, we investigate the outlier fraction,
$\eta$, as a function of magnitude, where outliers are defined as
sources with $\Delta z/(1+z)>0.15$. Outliers are below 10\%\ for sources
brighter than F277W$<27$, increasing steeply to 20\%\ near the 5$\sigma$
detection limit at $\sim$28.2. Note that we analyze both the
photometric redshift precision and outlier fraction as a function of redshift
as well as magnitude; overall, both quantities are somewhat constant
with redshift, with slight spikes in both from $6<z<7$ and $9<z<11$,
which is expected given the lack of complete filter coverage across
expected break wavelengths at those redshifts. We analyze the
efficacy of photometric selection for EoR galaxies further in
\S~\ref{sec:eorselection}.

\subsection{Expected Cosmic Variance}\label{sec:cosmicvariance}

The areal coverage of COSMOS-Web represents a real strength of the
program in the reionization era.  With claims of potential massive
galaxies in the distant universe from smaller surveys that, if
confirmed, may challenge our models of galaxy formation, representative
samples of the distant galaxy population would help establish the true
luminosity function shape and evolution at early times.

Following \citet{robertson10a} \citep[see also][]{trenti08a}, we
estimate the $z\sim9$ cosmic variance of the COSMOS-Web survey.  We
assume the survey area $A=0.54\deg^{2}$, a depth of 27.6 magnitudes,
and the $z\sim9$ luminosity function parameters from
\citet{bouwens21b}.  We perform abundance matching between galaxies
and the halo mass function, assigning the clustering strengths of
halos to their hosted galaxies from the \cite{tinker10a} peak
background split model for the halo bias. We find that the cosmic
sample variance uncertainty of COSMOS-Web at $z\sim9$ is
$\sigma_{v}\approx16\%$, and Poisson uncertainty is
$\sigma_p\approx8\%$, which sum in quadrature to a total expected
uncertainty of $\sigma_{tot}\approx18\%$ (giving a total variance
$\sigma_{v}^2=0.03$).

How does the cosmic variance of COSMOS-Web compare with the collection
of smaller, deeper fields soon available with JWST coverage? Repeating
our calculation for a single 100\,arcmin$^{2}$ field to 29$^{th}$
magnitude, we find such surveys have a $z\sim9$ cosmic sample variance
uncertainty of $\sigma_{v}\approx34\%$.  Five 100\,arcmin$^{2}$ fields
probing independent sight lines have a combined cosmic sample
uncertainty of $\sigma_{v}\approx14\%$, a Poisson uncertainty of
$\sigma_p\approx11\%$ and a total expected variance uncertainty of
$\sigma_{tot}\approx18\%$. Thus COSMOS-Web has comparable statistical
power to the combined power of other JWST Cycle 1 programs conducted
over a smaller area to greater depth.  As discussed in
\cite{robertson10a}, by combining these wide area and pencil beam
surveys the degeneracies in the constraints on luminosity function
parameters, like $M_\star$, $\Phi_\star$, or the faint-end slope
$\alpha$, can be broken or significantly ameliorated.

\subsection{Scheduling of the Observations and First Epoch of Data}\label{sec:scheduling}

COSMOS-Web was awarded a total of 208\,hours, but due to changes in
overhead and the dithering pattern described above, COSMOS-Web will
take a total of 255 hours to execute.  
We requested relatively low zodiacal background
 observations ($<$10-20$^{th}$ percentile) and
 to tile the mosaic at a nearly uniform position angle
on the sky to avoid gaps within the mosaic. COSMOS-Web is observable
in windows in April (PA$\approx$105) and
December/January (PA$\approx$290) of each year. In order
to maximize the amount of overlap between the prime (NIRCam) and
parallel (MIRI) observations, we will observe roughly half of the
mosaic in each window. 

\begin{figure*}
  \centering
  \includegraphics[width=0.99\textwidth]{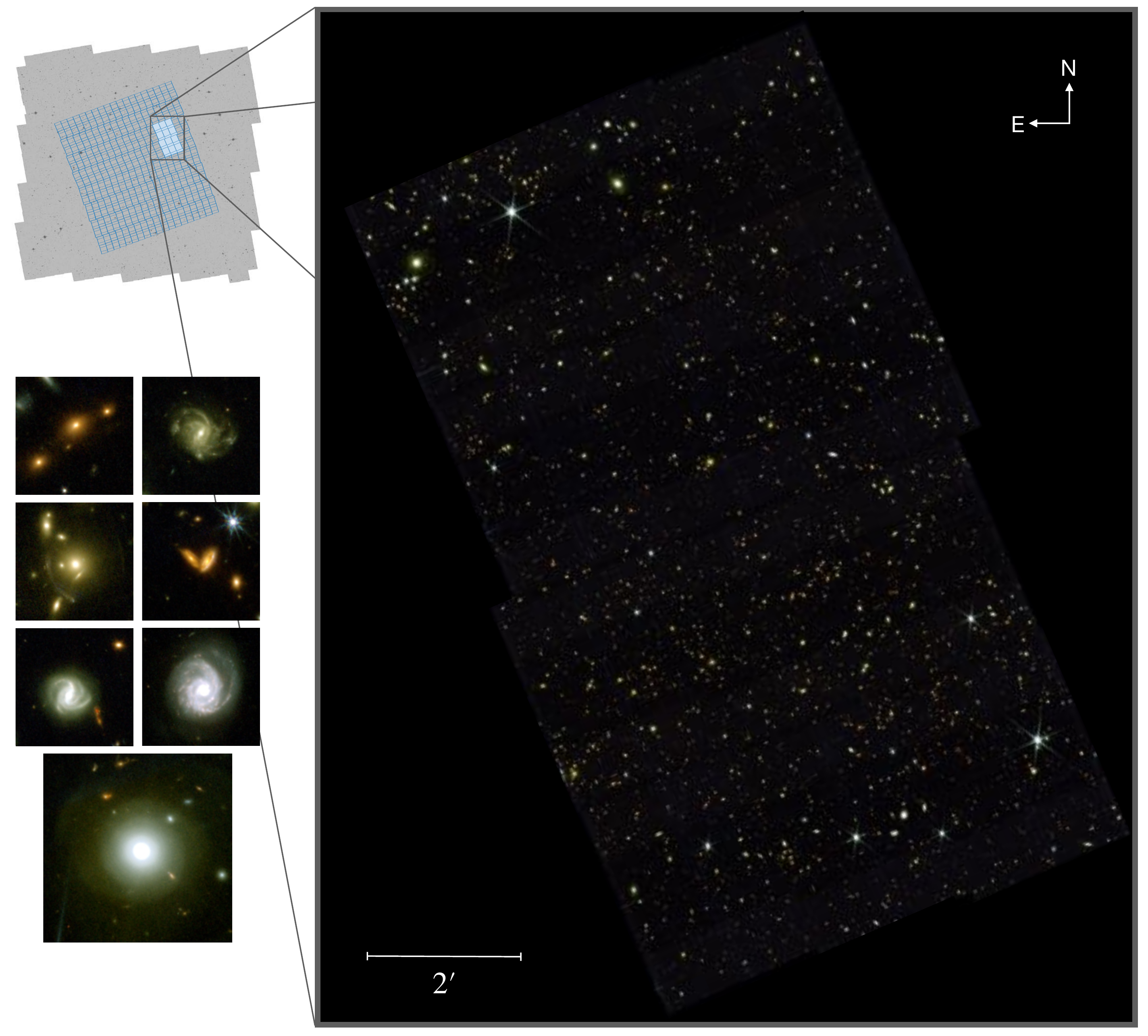}
  \caption{The first epoch of COSMOS-Web NIRCam
      observations obtained on 5-6 January 2023.  These data cover six
      visits (or pointings) out of a total of 152.  The total
      area covered by NIRCam here is $\sim$77\,arcmin$^2$.  The
      relative position of this mosaic in the survey is shown at upper
      left.  At lower left are several zoomed-in $10''\times10''$ cutouts
      (and one $16''\times16''$ cutout) of a handful of interesting 
      objects, highlighting the level of detail revealed by these first
      data.}
  \label{fig:epoch1_nircam}
\end{figure*}

\begin{figure*}
  \centering
  \includegraphics[width=0.99\textwidth]{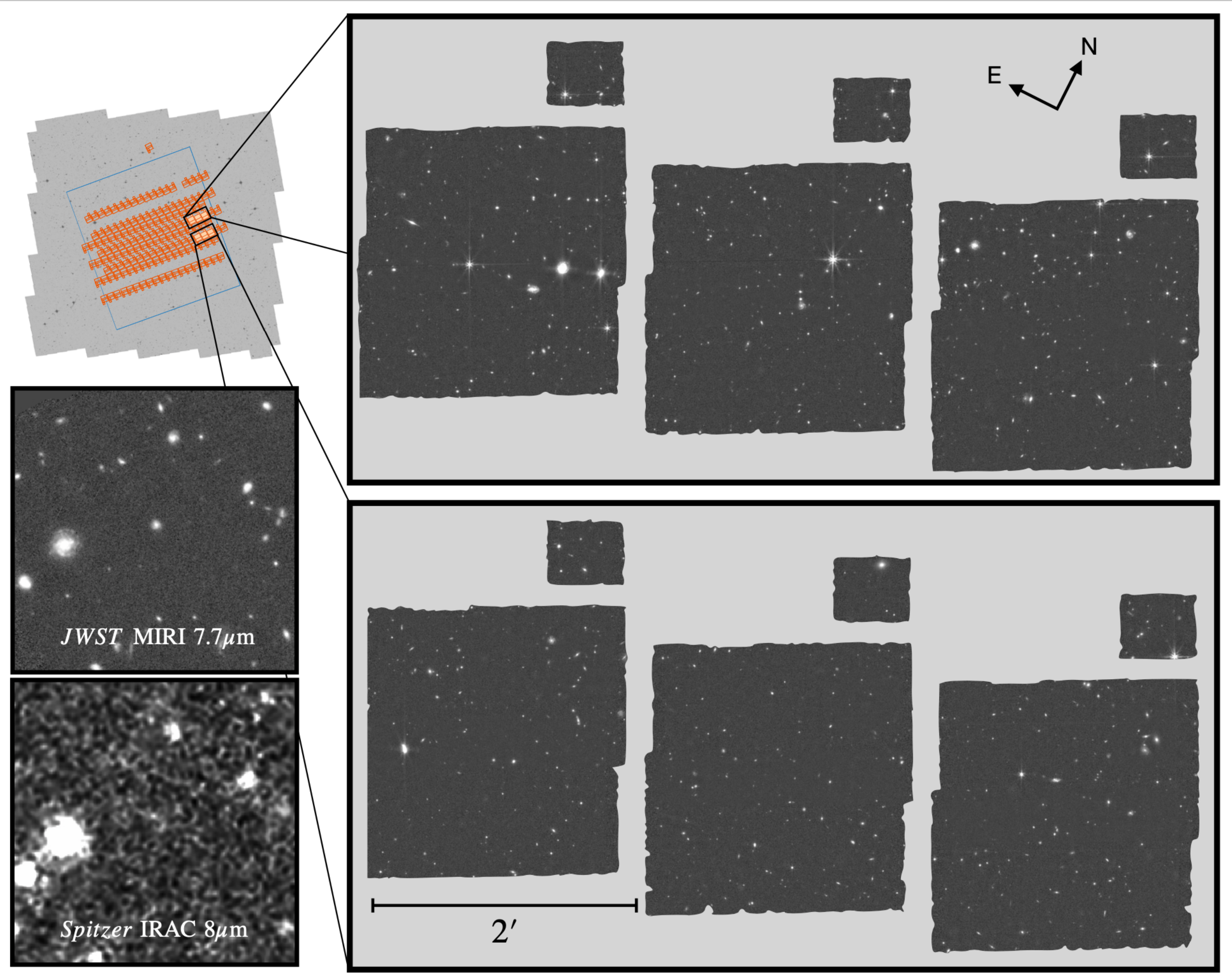}
  \caption{The first epoch of COSMOS-Web MIRI
      observations obtained on 5-6 January 2023.  Covering six visits,
      the MIRI data is distributed in six non-overlapping tiles and
      include data from both the MIRI imager and Lyot Coronograph
      field of view.  At left is a comparison between {\it Spitzer}
      IRAC channel 4 (8\um) data and MIRI 7.7\um\ data in a
      $40''\times40''$ zoom-in panel, highlighting the increased 
      sensitivity and resolution of MIRI observations over those 
      previously obtained with IRAC.}
  \label{fig:epoch1_miri}
\end{figure*}

The first epoch of observations consists of six visits covering
$\sim$77\,arcmin$^2$ with NIRCam and was observed on 5-6 January 2023.
Figure~\ref{fig:epoch1_nircam} shows the NIRCam mosaic of this region
of the field, which is 4\%\ of the final
dataset. Figure~\ref{fig:epoch1_miri} shows the six MIRI tiles from
this epoch.  As of this writing, 77 pointings are scheduled for
April/May 2023 (roughly half the field) and the remaining 69 pointings
in December 2023/January 2024. The COSMOS-Web team will
  release mosaics registered to {\it Gaia} astrometry after each
  subsequent epoch of observations is taken through the Mikulski
  Archive for Space Telescopes (MAST) and the NASA/IPAC Infrared
  Science Archive (IRSA); we will also make these mosaics accessible
  through the IRSA COSMOS cutout
  service\footnote{https://irsa.ipac.caltech.edu/data/COSMOS/index\_cutouts.html}.

\begin{deluxetable*}{lccccc}
 \tabletypesize{\footnotesize}
 \tablecolumns{6}
 \tablecaption{JWST Cycle 1 NIRCam Surveys}
 \tablehead{
  \colhead{Survey} & \colhead{Fields} & \colhead{Area } & \colhead{ SW Filters} & \colhead{ LW Filters} & \colhead{Depth}\cr  
   \colhead{Name} & \colhead{Observed} & \colhead{ arcmin$^{2}$} 
    }
 \startdata
 NGDEEP      & HUDF-Par2 & 10 & F115W, F150W, F200W & F277W, F356W, F444W  & 30.6--30.9 \\
 UDF-Medium & HUDF & 10 & F182M, F210M & F430M, F460M, F480M & 28.0--29.8 \\
 JADES-Deep   & HUDF & 46 & F090W, F115W, F150W, F200W & F277W, F335M, F356W, F410M, F444W & 30.3--30.7   \\ 
 JADES-Medium & GOODS-N, GOODS-S & 190 & F070W$\dagger$, F090W, F115W, F150W, F200W & F277W, F356W, F410M, F444W & 29.1--29.8 \\ 
 CEERS       & EGS & 100 & F115W, F150W, F200W & F277W, F356W, F410M, F444W & 28.4--29.2 \\
 PRIMER      & COSMOS, UDS & 378 & F090W, F115W, F150W, F200W & F277W, F356W, F410M, F444W  & 27.6--29.5 \\
 COSMOS-Web & COSMOS & 1929 & F115W, F150W & F277W, F444W & 26.9--28.3 \\
 \enddata
 \tablecomments{Depths quoted are 5$\sigma$ point source depths.
   NIRCam depths quoted have been drawn from the original proposals
   and pre-flight exposure time calculator estimates within
   0$\farcs$15 radius circular apertures. We have {\it not} adjusted
   for the in-flight calibration \citep{boyer22a} of the instruments;
   however, any differences with these figures is anticipated to be of
   order smaller than a 10\%\ effect, smaller than the typical
   deviation across a mosaic stitched together with non-uniform depth,
   or from variation in depth filter-to-filter. Program IDs for these
   surveys are: NGDEEP (GO \# 2079), UDF-Medium (GO \# 1963),
   JADES-Deep (GTO \# 1180, 1210, 1287), JADES-Medium (GTO \# 1180,
   1181, 1286), CEERS (ERS \# 1345), PRIMER (GO \# 1837), and
   COSMOS-Web (GO \# 1727).  $\dagger$F070W in JADES-Medium imaging is
   only planned for parallel coverage areas currently lacking {\it
     HST}.  }
 \label{tab:cycle1nircam}
\end{deluxetable*}

\begin{deluxetable*}{lccccccc}
 \tabletypesize{\small}
 \tablecolumns{8}
 \tablecaption{JWST Cycle 1 MIRI Surveys}
 \tablehead{
  \colhead{Survey} & \colhead{Fields} & \colhead{Area } & \colhead{Filters} & \colhead{Depth} \\  
   \colhead{Name} & \colhead{Observed} & \colhead{ arcmin$^{2}$} & \colhead{} & \colhead{ } 
 }
 \startdata
 MIRI-HUDF-Deep$^\dagger$ & HUDF & 2.5 & F560W & 28.3--28.5 \\ 
 CEERS       & EGS & 13 & F770W, F1000W, F1280W, F1500W, F1800W, F2100W & 21.6--26.3 \\ 
  JADES-Medium            & HUDF$^*$    & 10   & F770W & 27.1\\
  & GOODS-N, HUDF & 17.5  & F770W, F1280W & 24.7--25.4 \\ 
 MIRI-HUDF-Medium & HUDF & 30 & F560W, F770W, F1000W, F1280W & 23.3--24.8 \\
                         & &  & F1500W, F1800W, F2100W, F2550W & 19.8--23.2 \\
 PRIMER      & COSMOS, UDS & 137 & F770W, F1800W & 22.1--25.4  \\ 
 COSMOS-Web & COSMOS & 697 & F770W & 24.0--25.1$^\ddag$\\ 
 \enddata
 %
 %
 \tablecomments{Depths quoted are 5$\sigma$ point source depths within
   0$\farcs$3 radius circular apertures from the exposure time
   calculator.  Program IDs for these surveys are: CEERS (ERS \#
   1345), MIRI-HUDF-Medium (GO \# 1207), MIRI-HUDF-DEEP (GO \# 1283),
   JADES-Medium (GTO \# 1180 and 1181), PRIMER (GO \# 1837), and
   COSMOS-Web (GO \# 1727). $\dagger$ Note that the MIRI-HUDF-Deep
   Program (GO \# 1283) is nested within the MIRI-HUDF-Medium (GO \#
   1207) program, but both are spatially offset from the JADES-Medium
   HUDF coverage. $^*$ Note that the deeper part of JADES-Medium HUDF
   F770W coverage is nested within the shallower JADES-Medium
   coverage. $^\ddag$ Note that the depth quoted here for
     COSMOS-Web differs from the reported measured depth as given in
     Table~\ref{tab:obssummarymiri}; similarly, PRIMER and CEERS
     measured depths differ from ETC estimates, with measured
     7.7\um\ depths of those programs shown in
     Figure~\ref{fig:context}. What is quoted in this table is from
     the exposure time calculator.  We expect the actual depth of all
     MIRI programs to differ from ETC estimates in a similar manner.}
 \label{tab:cycle1miri}
\end{deluxetable*}

\section{Context of COSMOS-Web Among Other JWST Deep Fields}\label{sec:context}

Several extragalactic deep field surveys will be conducted in the
first year of {\it JWST} observations that span a range of areas,
depths, and filter coverage; their approximate depths and areas are
described in Table~\ref{tab:cycle1nircam} for NIRCam programs and
Table~\ref{tab:cycle1miri} for MIRI programs.  Note that
  the NIRCam depths of other programs quote the pre-flight exposure time
  calculator (ETC) estimates and do not necessarily reflect the actual
  final measured depths of the data. For MIRI, we include the measured
  depths from COSMOS-Web, CEERS, and PRIMER observations along with 
  the updated ETC estimates. The MIRI depth in
  COSMOS-Web is measured to be significantly deeper (by $\sim$1
  magnitude) compared to the ETC estimates.  
  We refer the reader to the recent review by
\citet{robertson22a}, their \S~8.2, as well as their Figure~6, for a
summary of many of the large extragalactic programs, and in particular
their NIRCam coverage. These programs include the Guaranteed Time
Observation (GTO) programs allocated to the instrument teams, the
Director's Discretionary Early Release Science Programs (ERS), as well
as the General Observer (GO) Cycle 1 programs.

\begin{figure*}
 \centering
 \includegraphics[width=0.8\textwidth]{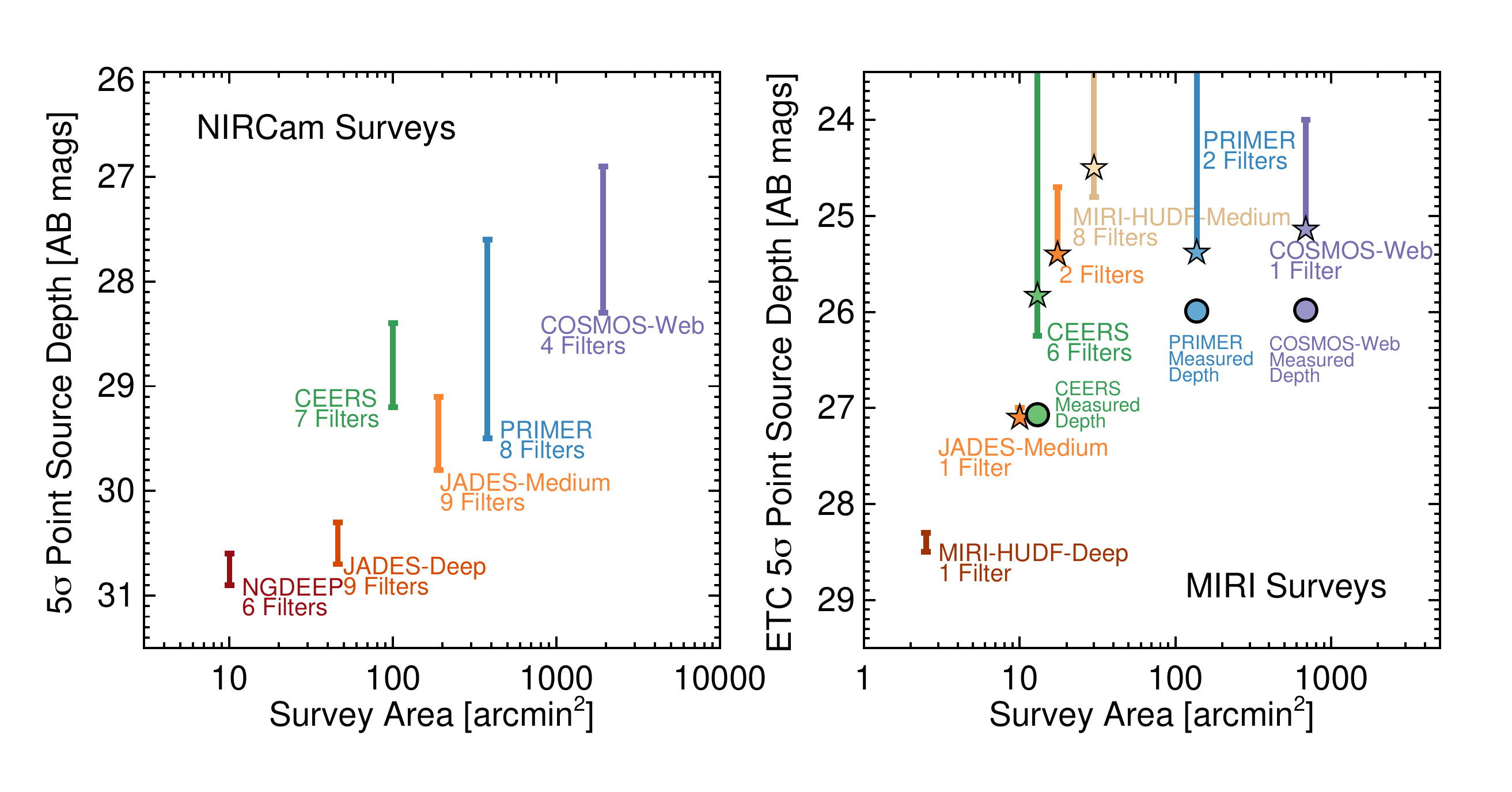}
 \caption{A comparison of several of the {\it JWST} Cycle 1
   extragalactic survey programs in depth and area for NIRCam imaging
   (left) and MIRI imaging (right).  The vertical bars bracket the
   survey depths across all filters.  In the case of MIRI, the dynamic
   range of depths is large due to substantial depth differences by
   filter; the depths at F770W are marked with stars for each
   survey. In the case of COSMOS-Web that has a large
     dither, the vertical bars also capture the range of depths across
     the mosaic.  Note that for MIRI we show the
     exposure time calculator (ETC)-predicted depths, while the
     measured 7.7\,\um\ depths for COSMOS-Web, CEERS, and PRIMER are
     shown with circles.  We expect the depth of all programs to be
     similarly offset between ETC estimates and actual depth
     achieved.  Depths have been converted to approximate 5$\sigma$
   point source depths as detailed in Tables~\ref{tab:cycle1nircam}
   and~\ref{tab:cycle1miri}.}
 \label{fig:context}
\end{figure*}

Figure~\ref{fig:context} shows the relative depth and survey area of
the major broad-band legacy extragalactic programs in Cycle 1, both
for NIRCam imaging and MIRI imaging programs.  To briefly summarize
the relative scope of the NIRCam programs, the deepest surveys are
NGDEEP\footnote{NGDEEP was originally named WDEEP at the time of the
proposal.} (GO \#2079) and the JADES GTO Survey (in particular GTO
\#1180, 1210, \&\ 1287).  These collectively cover about
$\sim$0.05\,deg$^2$ to depths exceeding $\sim$29.5\,mag in several
broad-band filters. The medium depth programs JADES-Medium (including
parts of GTO \#1180, 1181, and 1286), CEERS (ERS \#1345), and PRIMER
(GO \#1837) together cover a total of $\sim$0.18\,deg$^2$ to a depth
$\sim$28--29\,mag.  Note that the UDF Medium Band Survey (GO \# 1963)
achieves similar depths $\sim$28--29.8\,mag in NIRCam medium bands
over 10\,arcmin$^2$ in the HUDF (an area also covered by JADES-Deep in
the broad-bands).  COSMOS-Web (GO \#1727) is the shallowest but
largest program to be observed, covering a total 0.54\,deg$^2$ with
NIRCam to a depth of $\sim$27.5--28.2\,mag across the field.

Planned MIRI programs vary in depth more substantially, as the shorter
wavelength filters achieve much deeper observations per fixed exposure
time.  The MIRI GTO programs adopt two very different approaches: one
(GO \# 1283) goes quite deep in a single MIRI pointing in one filter,
F560W.  The other (GO \# 1207) covers 30\,arcmin$^2$ and uses {\it
  all} 8 broad-band MIRI filters and thus is significantly more
shallow.  MIRI imaging is obtained in parallel to much of the JADES
program (from programs GTO \#1180 and 1181) where a hybrid approach
was adopted, going deep in one filter, F770W, over 10\,arcmin$^2$, and
shallower in two filters, F770W and F1280W, over 15\,arcmin$^2$.
CEERS similarly spans a broad range in depths over 13\,arcmin$^2$
using 6 filters, and PRIMER covers much larger areas over
$\sim$140\,arcmin$^2$ in two filters. Similar to its NIRCam coverage,
COSMOS-Web covers the largest area with MIRI, but with variable depth
(based on the number of exposures) in F770W.  We have shown the F770W
depths of the MIRI surveys using a star in Figure~\ref{fig:context}
for more direct comparisons to the COSMOS-Web depths.

The total area covered by COSMOS-Web in NIRCam is roughly 2.7$\times$
larger than the other planned {\it JWST} extragalactic deep fields
{\it combined}.  For MIRI, COSMOS-Web's coverage is 3.5$\times$ larger
than all other deep field programs {\it combined}.  The extraordinary
range of areas and depths of deep field surveys observed in {\it
  JWST}'s first year will be complementary, and enable a wide range of
scientific studies, spanning the most distant and faintest galaxies
ever detected to the most comprehensive environmental studies of the
distant universe.

\section{Scientific Goals}\label{sec:science}

The scientific breadth of COSMOS-Web has the potential to be
extraordinary, with an estimated $\sim$10$^{6}$ sources to be detected
from $z\sim0.1$ to cosmic dawn. Nevertheless, the survey as proposed
was motivated by three key science areas that ultimately drive the 
design of the survey. The three primary goals of the program are to:

\begin{enumerate}
\item forge the detection of thousands of galaxies in the Epoch of
 Reionization ($6\simlt z \simlt11$) and use their spatial distribution to
 map large scale structure during the Universe's first billion years,
\item identify hundreds of the rarest quiescent galaxies in the first
 two billion years ($z>4$) to place stringent constraints on the
 formation of the Universe's most massive galaxies (with
 $M_\star>10^{10}$\,\msun), and
\item directly measure the evolution of the stellar mass to halo mass
 relation (SMHR) out to $z\sim2.5$ and its variance with galaxies'
 star formation histories and morphologies.
\end{enumerate}
Below we detail the motivation and requirements of each science goal.

\subsection{Mapping the Heart of Reionization}\label{sec:eor}

The first galaxies formed $<\,1$\,Gyr after the Big Bang are thought
to drive the last major phase change of the Universe from a neutral to
ionized intergalactic medium (IGM). This reionization process
\citep{robertson15a} most likely finished around $z\sim6$
\citep{zheng11a,kakiichi16a,castellano16a,ouchi20a} and was halfway completed
by $z\sim7-8$, according to measures of the rest-frame UV galaxy
luminosity function \citep[UVLF;][]{finkelstein16a}. This is in broad
agreement with the {\it Planck} constraint of the instantaneous
reionization redshift $z_{\rm reion}=7.7\pm0.8$
\citep{planck-collaboration20a}. However, neither the start and
duration of reionization, nor the sources responsible --- either
intrinsically luminous galaxies or more intermediate mass galaxies
\citep{naidu20a,hutter21a} --- are well-constrained due to the
relative shortage of both bright and faint $z\sim7-11$ galaxies known
in the pre-{\it JWST} era. Additional complexity is introduced by
potentially significant evolution in the nature of EoR galaxies
themselves: their intrinsic star formation rates (SFR), ionizing power
($\xi_{\rm ion}$), ionizing radiation escape fraction ($f_{\rm esc}$),
number density, physical distribution, and clustering.

Constraining the physics of reionization requires identifying and
characterizing the galaxies that are embedded deep within the
predominantly neutral Universe at $z\simgt$8, though direct detection
of EoR galaxies has been challenging to-date. Pioneering work with
{\it Hubble} led to the discovery of $\sim$\,80 candidate Lyman Break
Galaxies (LBGs) at $z>8$ \citep[see review by][]{finkelstein16a}.
Despite the perceived rapid drop in the UVLF during this epoch
\citep{oesch14a}, there have been a few successful pre-{\it JWST}
detections of surprisingly bright candidate LBGs out to $z\sim11$ (the
most spectacular of which is GNz11 at $z=10.6$,
\citealt{oesch16a,jiang21a,bunker23a}). Although these $z>8$ galaxies
are thought to reside in a predominantly neutral Universe, somehow a
number of them show \lya\ in emission
\citep[e.g.,][]{oesch15a,zitrin15a,hoag18a,hashimoto18a,pentericci18a}. This
is surprising given that those \lya\ photons should have been
resonantly scattered by the mostly neutral IGM
\citep{dijkstra14a,stark16a,garel21a}. Do these \lya\ emitters at
$z>8$ live in special `ionized' bubbles? If they are representative of
the general population, are we missing some fundamental aspect of the
first stage of reionization? These questions can only be answered with
a large sample of bright $z=7-11$ sources across a range of large
scale environments, only possible with a near-IR contiguous wide-area
survey \citep{kauffmann20a}.

The first candidate discoveries of unusually bright
galaxy candidates identified in early {\it JWST} observations
\citep[e.g.,][]{naidu22a,finkelstein22a,finkelstein22b,donnan22a,
  harikane22a,atek22a} suggest that these sources may not be as rare
as our pre-{\it JWST} models of $z>7$ galaxy formation would indicate
\citep[e.g.,][]{mason15a,yung19a,yung20a,behroozi20a,wilkins17a,wilkins22a}.
While we note that these early discoveries are still candidates that
require spectroscopic confirmation \citep[as of this writing only a
  few $z>9$ systems have been spectroscopically-confirmed,]
[]{roberts-borsani22a,williams22a,robertson22c,curtis-lake22a}, the
perceived wealth of {\it bright} candidates 
  may be particularly relevant to understanding the distribution of
galaxies within large scale structure at early times. These bright
candidates theoretically occupy the rarest and most massive dark
matter halos, which are thought to be more highly clustered, and as
such, small area surveys (as has been carried out to-date with {\it
  JWST}) would poorly constrain their volume densities and the
environments in which they live.

The breadth of galaxies' environments at early times is closely
related to how reionization propagated. It is thought that
reionization was predominantly a patchy process, producing ionized
bubbles in the surrounding IGM growing from 5\,--\,20\,Mpc at $z>8$ to
30\,--\,100\,Mpc at $z\sim7$ \citep{furlanetto17a,daloisio18a}. This
corresponds to angular scales of 10\,--\,40\,arcmin across, much
larger than all contiguous NIR deep fields from {\it Hubble}
or other planned deep field areas from {\it JWST} (see Table \ref{tab:cycle1nircam}. 
Furthermore, large
variance in the IGM's opacity from $5<z<7$ quasar sightlines
\citep{becker15a} suggests that the patchiness exceeds
theoretical expectation from the density field alone by factors of a
few, exacerbating uncertainties in reionization constraints from
cosmic variance in existing surveys. Follow-up studies around both
transparent and opaque quasar sightlines indicate a wide variety of
large scale environments \citep{becker18a,davis18a}.

\begin{figure*}
\centering
 \includegraphics[width=0.9\textwidth]{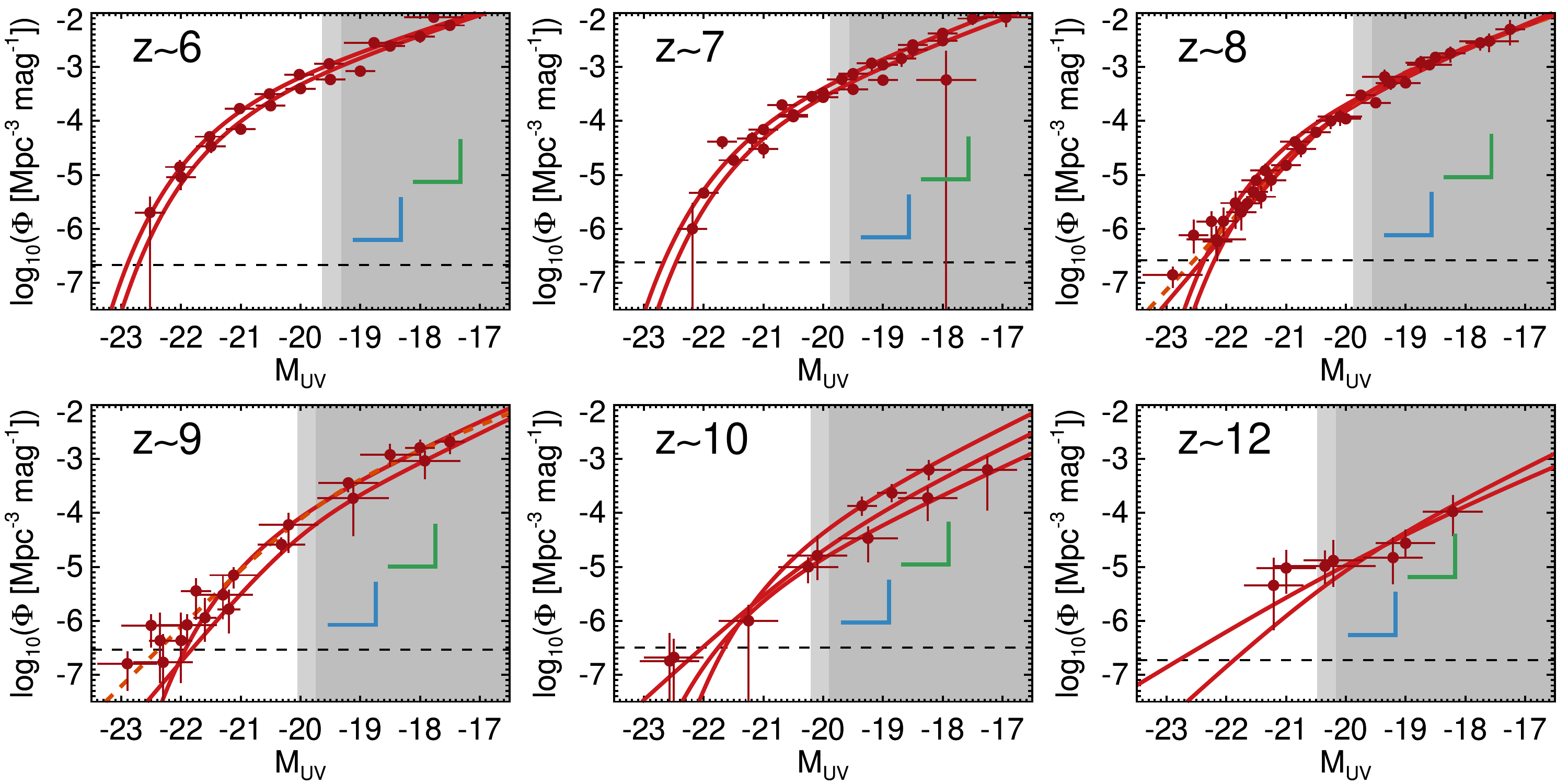}
 \caption{Literature rest-frame UV luminosity functions from
  $6<z<13$; both data points and functional fits are drawn directly
  from the literature to illustrate the range of predictions made
  to-date at each epoch. Data and fits are specifically drawn from
  \citet{mclure13a,oesch13a,bouwens15a,finkelstein15a,finkelstein16a,mcleod16a,stefanon19a,bowler20a,bouwens21b,kauffmann22a,naidu22a,donnan22a,harikane22a}.
  Gray regions mask out rest-frame UV magnitudes where COSMOS-Web
  will not be sensitive; the light gray region marks the limit
  corresponding to our two image depth while the dark gray
  corresponds to four image depth. The horizontal dashed line marks
  the rarity of galaxies at which we would expect only to see one in
  all of COSMOS-Web. The blue and green corners mark the
  sensitivity limits (in depth and source rarity) of all of the
  Cycle 1 medium-depth surveys combined and deep-depth surveys
  combined, respectively.}
 \label{fig:eorlf}
\end{figure*}

COSMOS-Web will grow the census of EoR ($z>6$) galaxies beyond what is
known from {\it Hubble} surveys by a factor of $\sim$5 and quantify
the evolution of the UVLF, stellar mass function (SMF), and star
formation histories of galaxies across the Universe's first billion
years. By observing a large contiguous area, COSMOS-Web will detect a
factor of $\sim$6-7 times more sources at or above the knee of the
luminosity function, $L_\star$, than expected from all other {\it
  JWST} deep field efforts combined. Figure~\ref{fig:eorlf} shows the
aggregate UVLF measurements from the literature to-date from $6<z<13$,
combining {\it Hubble} samples with the most recent results from {\it
  JWST}. Table~\ref{tab:eorsources} gives statistics on the predicted
number of EoR galaxies to be found in COSMOS-Web, calculated directly
from the compiled UV luminosity functions, relative to other Cycle~1
medium and deep programs.

Massive galaxies above $L_\star$ are most likely to trace the
highest-density peaks from which the reionization process was likely
to begin. In particular, the 0.54\,deg$^2$ survey area of COSMOS-Web
is sufficiently large to capture reionization on scales larger than
its expected patchiness, minimizing the effect of cosmic variance. As
a contiguous survey, COSMOS-Web will sample the full range of
environments at this epoch, provided large scale structure is
clustered on scales within an order of magnitude of their predicted
scales \citep{gnedin14a}. This contrasts with, for example, the
innovative {\it Hubble} and {\it JWST} pure-parallel surveys (e.g.,
BoRG, \citealt{schmidt14a,calvi16a} and PANORAMIC, GO \#2514) that, by
design, will sample a wide variety of environments but cannot
directly map the large scale environments of their discoveries.

\begin{deluxetable*}{lcccccc}
 \tabletypesize{\footnotesize}
 \tablecolumns{7}
 \tablecaption{Number of Sources Expected between $6<z<13$ in Cycle 1 Programs}
 \tablehead{
  \colhead{Survey} & \colhead{$z\sim6$} & \colhead{$z\sim7$} & \colhead{$z\sim8$} & \colhead{$z\sim9$} & \colhead{$z\sim10$} & \colhead{$z\sim12$}\cr
  \colhead{} & \colhead{($\Delta z = 1$)} & \colhead{($\Delta z = 1$)} & \colhead{($\Delta z = 1$)} & \colhead{($\Delta z = 1$)} & \colhead{($\Delta z = 1.5^*$)} & \colhead{($\Delta z = 2$)}
 }
 \startdata
 COSMOS-Web & 2900--4000 & 1000--1500 & 500--680 & 150--160 & 30--70 & 12--25 \\
\hline
 All Medium Cy\,1 Programs$^\dagger$ & 3800--5000 & 1600--2400 & 1000--1300 & 230--450 & 70--260 & 37--44 \\
 All Deep Cy\,1 Programs$^\ddag$ & 900--1100 & 450--640 & 310--350 & 90--150 & 20--100 & 13--15 \\
\hline
Medium Cy\,1 at COSMOS-Web Depth & 1300--1700 & 460--680 & 230--300 & 30--80 & 14--36 & 5--11 \\
Deep Cy\,1 at COSMOS-Web Depth & 110--150 & 40--60 & 19--26 & 3--7 & 1--3 & 0--1 \\
\hline
\hline
COSMOS-Web Depth in $M_{\rm UV}$ & --19.3 & --19.6 & --19.6 & --19.7 & --19.9 & --20.2 \\
\enddata
%
 \tablecomments{Here we refer to all `medium' depth Cycle 1 programs
  ($\dagger$) as surveys reaching $\sim$28.5--29.5\,mags in broad-band filters from
  Table~\ref{tab:cycle1nircam}, including JADES-Medium, CEERS and
  PRIMER. The `deep' Cycle 1 programs ($\ddag$) refer to JADES-deep
  and NGDEEP together, which will reach depths exceeding 29.5\,mags. 
  $^*$ Note that the $z\sim10$ bin spans $9.5<z<11$.}
 \label{tab:eorsources}
\end{deluxetable*}

\subsubsection{Impact beyond $z>8$}

Beyond the halfway point of reionization, COSMOS-Web is likely to
detect hundreds of intrinsically bright galaxies at $8<z<11$ embedded
deep in the predominantly neutral IGM. This will increase the number
of known $z>8$ galaxies from the pre-{\it JWST} era by a factor of 10
above a luminosity of $L_\star$. Through such a transformative sample
of luminous $z>8$ candidates, these discoveries will allow the first
constraints on the bright-end of the UVLF and SMF at $z\simgt8$ with
minimal uncertainty from cosmic variance, minimized to
$\simlt$10\%\ on scales of 0.5\,deg$^2$ at our detection threshold of
$\sim$27.5\,magnitudes \citep{behroozi19a}. Table~\ref{tab:eorsources}
shows the expected total number of sources COSMOS-Web will find,
totaling to $\sim$600--900 above $z>8$ and 12--25 from $11<z<13$.

Our NIRCam filter combination is specifically optimized for $8<z<11$
galaxy selection above the F115W detection limit of $\sim$27.4 mag as
shown in Figure~\ref{fig:sedlimits}. Such systems are expected to see
a significant drop in the F115W filter. If we account for a possible
deviation from a Schechter UVLF as measured by wide/shallow
ground-based UVLF estimates at $z>8$ (shown as double power laws in
Figure~\ref{fig:eorlf}), our detections will likely exceed 1000
sources above $z>8$, sufficient to map their spatial distribution and
trace large scale structure at such early times. Even with our
fiducial expectations in 0.54\,deg$^2$ coverage, we expect to see a
factor of $\sim$7 improvement in the number of $z>8$ candidate
galaxies above $L_\star$ over all previous {\it Hubble} work and a
factor 2$\times$ larger samples at those luminosities than all other
planned Cycle~1 programs combined.

\subsubsection{Inferring the bright-end shape of the UVLF and SMF}

While CANDELS found only $\sim$\,2\,--\,10 galaxies at $z>6$ with
$M_{\star}>10^{10.5}$\,\msun, and none above
10$^{11}$\,\msun\ \citep{grazian15a,song16a}, the wider Ultra-VISTA
survey \citep{bowler14a,bowler20a} found a larger number of massive
galaxies than expected based on an extrapolation of a Schechter
function fit to the CANDELS-measured SMF. The recent 
  candidate discovery of intrinsically bright $z>10$ galaxies in
small-area early release {\it JWST} observations
\citep[e.g.,][]{naidu22a,castellano22a,finkelstein22a,finkelstein22b,donnan22a,atek22a}
also hint at a possible overabundance of massive galaxies compared to
a Schechter function expectation. This excess of bright sources could
indicate that the most massive galaxies are highly clustered and/or
that the SMF at $z>6$ departs from Schechter
\citep{bowler17b,davidzon17a}.  COSMOS-Web will greatly improve the
dynamic range of luminosities (and thus masses) probed beyond all
other NIR surveys, detecting $\sim$\,280--500 bright $M_{\rm UV}<-21$ galaxies at $z\sim6-8$
and $\sim$\,30--80 at $8<z<13$, corresponding
to stellar masses $\simgt4\times$10$^{9}$\,\msun.  We calculate these
estimates using the literature parameterized luminosity functions
shown in Figure~\ref{fig:eorlf} integrated down to $M_{\rm UV}=-21$,
significantly above our detection threshold as detailed in
Table~\ref{tab:eorsources}. Given these statistics, a Schechter UVLF
will be distinguishable from a double power-law in this dataset at a
minimum of $\sim$4$\sigma$ out to $z=9$; this estimate is based on the
Poisson uncertainties in the expected number of sources
to-be-discovered in COSMOS-Web given a Schechter function and a
conservative estimate on the bright-end slope of the UVLF in the case
of a double power-law (using $\beta=-4$). Such a deviation could be
indicative of a primordial galaxy formation stage with different star
formation timescales \citep{finkelstein15a,yung19a}, a lack of dust,
or before the onset of feedback from Active Galactic Nuclei (AGN).

\subsubsection{Selection of EoR Sources}\label{sec:eorselection}

As discussed in \S~\ref{sec:photozs}, we generate a mock lightcone of
the COSMOS-Web field containing an idealized sample of $0<z<12$
galaxies, and here we use that simulated photometric catalog to
diagnose contamination and precision of our EoR photometric redshifts,
applying tools we will use for the real dataset.

\begin{figure}
\includegraphics[width=0.99\columnwidth]{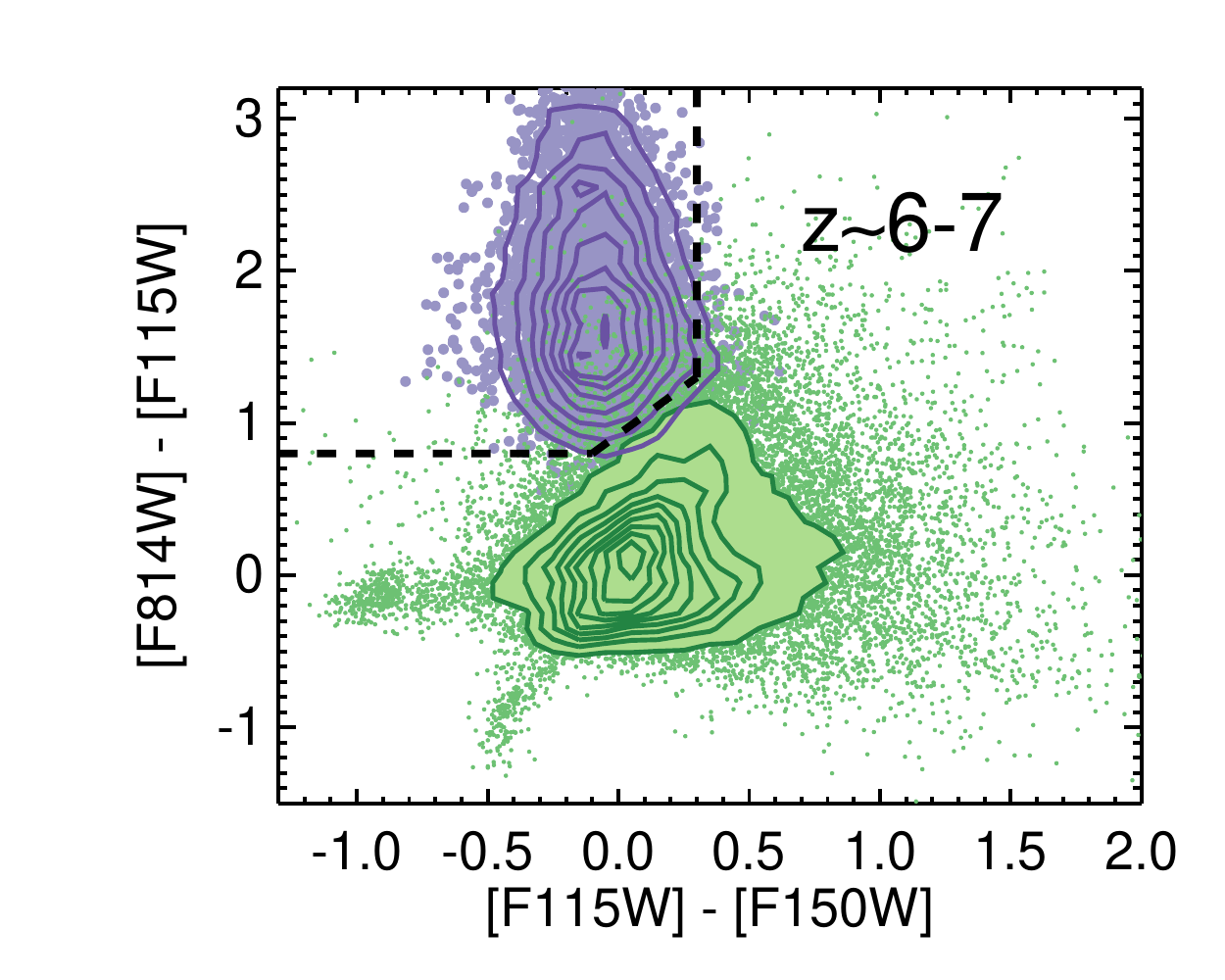}
\includegraphics[width=0.99\columnwidth]{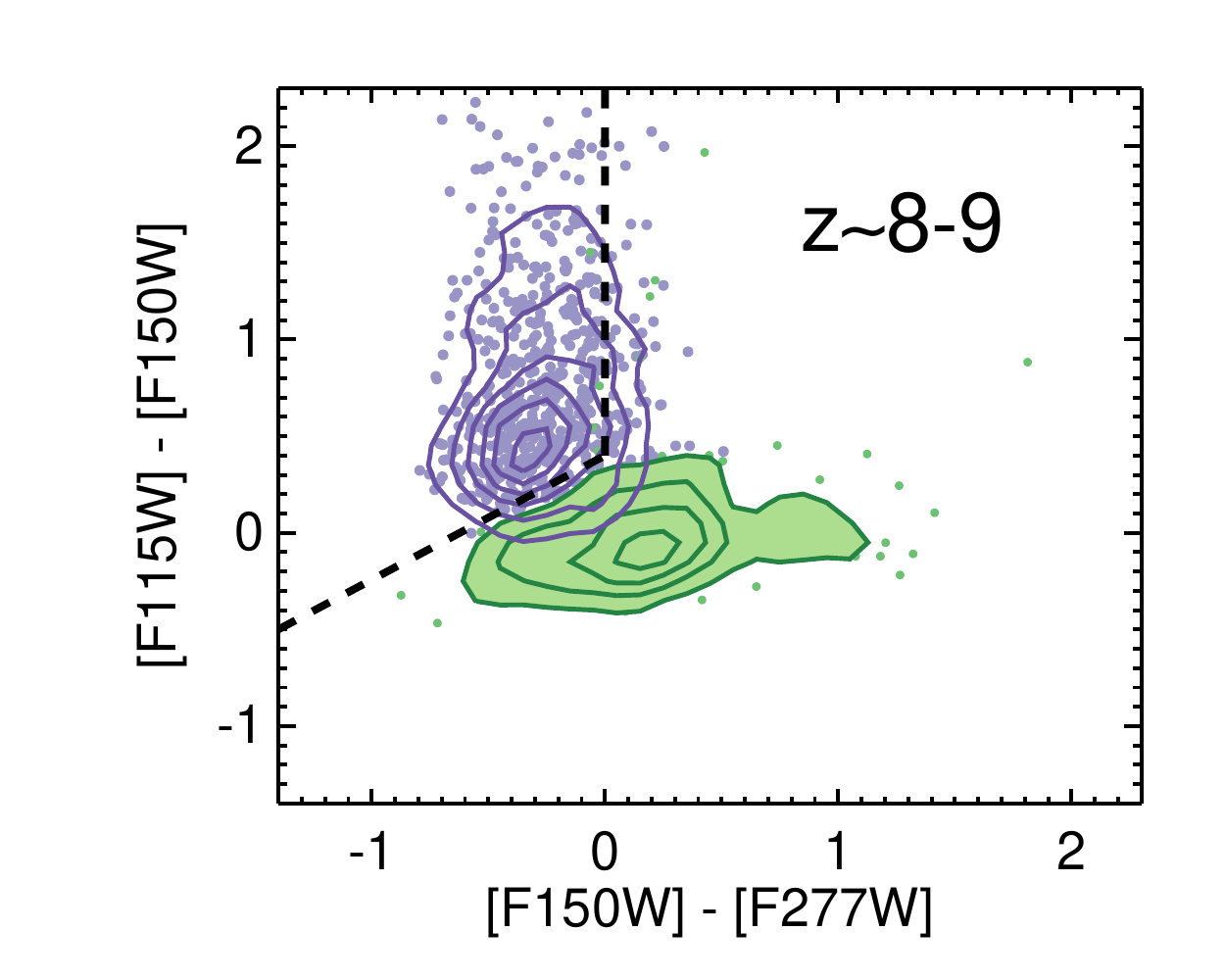}
\caption{Color-color diagrams of mock galaxies drawn from a
  semi-analytic model illustrating the selection of $z\sim6-7$ (top
  panel) and $z\sim8-9$ galaxies (bottom panel) using the COSMOS-Web
  filter-set. In the $z\sim6-7$ panel, green points and contours
  illustrate the distribution of mock galaxies at all redshifts
  relative to those at $6<z<7.5$, shown in purple points and contours.
  A strong drop in the [F814W]--[F115W] color and a blue
  [F115W]--[F150W] correlates strongly with galaxies at $z\sim6-7$;
  $z\sim1$ sources serve as the major contaminant due to degeneracy
  with the Balmer break.  In the $z\sim8-9$ panel, green points and
  contours show galaxies with photometric redshift solutions above
  $z=5$ and purple points highlight sources with known redshifts
  $8<z<9.5$. At these redshifts, we expect the drop to migrate to the
  [F115W]--[F150W] color yet sources are still expected to be
  relatively blue in [F150W]--[F277W].}
\label{fig:eorcolors}
\end{figure}

Figure~\ref{fig:eorcolors} highlights the distribution of mock
galaxies in color-color space for $z\sim6-7$ and $z\sim8-9$ galaxies
against potential contaminating populations. The primary contaminants
in both redshift regimes are $1<z<4$ faint galaxies ($\sim$27$^{th}$
mag).  The F814W and F115W filters are effective drop out filters for
the two redshift regimes, though small gaps in wavelength coverage
between filters imply that photometric redshift precision in
COSMOS-Web will be somewhat less accurate than in fields with more
complete filter coverage. We find that
contamination rates are most significant (up to $\sim$20\%) within
0.5\,magnitudes of our 5$\sigma$ point source detection limit, where
the constraint on drop filters is slightly weaker. We also anticipate
relatively elevated contamination ($\sim$20\%) in the redshift range
$5.5<z<6.5$ due to both the gap between F814W and F115W as well as the
relative depth difference between the filters (where F814W is
shallower but also serves as the drop out filter). For $6.5<z<9.5$, we
anticipate contamination rates below $\sim$10\%\ with a photometric
redshift precision of $\Delta z/(1+z)\approx0.02-0.04$. Above
$z\sim9.5$, the precision of photometric redshifts is degraded
substantially by the lack of coverage at 2\um\ (see
Figure~\ref{fig:sedlimits}); while some candidate $z>12$ sources may
be identified, they would require spectroscopic follow-up to confirm
their redshifts, as NIRCam photometry would not constrain them very
precisely. We will present further analysis of photometric redshift
precision, as well as EoR sample contamination and completeness, in a
forthcoming COSMOS-Web paper on the rest-frame UV luminosity function
(Franco \etal, in preparation).

An important consideration for the selection of EoR galaxy candidates
will also be contamination of samples with lower redshift strong
nebular emission line sources.  For example, an underlying dust
obscured (and reddened) rest-frame optical continuum superimposed with
strong emission lines can masquerade as a bluer rest-frame UV
continuum in {\it JWST}'s broadband filters, as shown by \citet{zavala22a}
and \citet{naidu22b}.  In that case one might expect dust continuum
emission at millimeter wavelengths, representing reprocessed emission
from hot stars. However, \citet{fujimoto22a} demonstrates that even a
lack of dust continuum emission cannot rule out possible contamination
from type-II quasars or AGN \citep[in this case, the area covered with
  MIRI in F770W could lead to the detection of AGN that satisfy LBG
  selection criteria;][]{fujimoto22b}.  While these possible foreground
contaminants have come to light with the
identification of ultra high-redshift sources ($z>12$), it is
nevertheless an important consideration in the identification of {\it
  all} EoR candidates, given the relatively sparse broad-band sampling
available to select such sources.  Follow-up spectroscopy of many EoR
candidates in the next year will elucidate the level of contamination
present in such samples and play a crucial role in informing
statistics about large samples selected in COSMOS-Web.

\begin{figure*}
 \centering
 \includegraphics[width=0.9\textwidth]{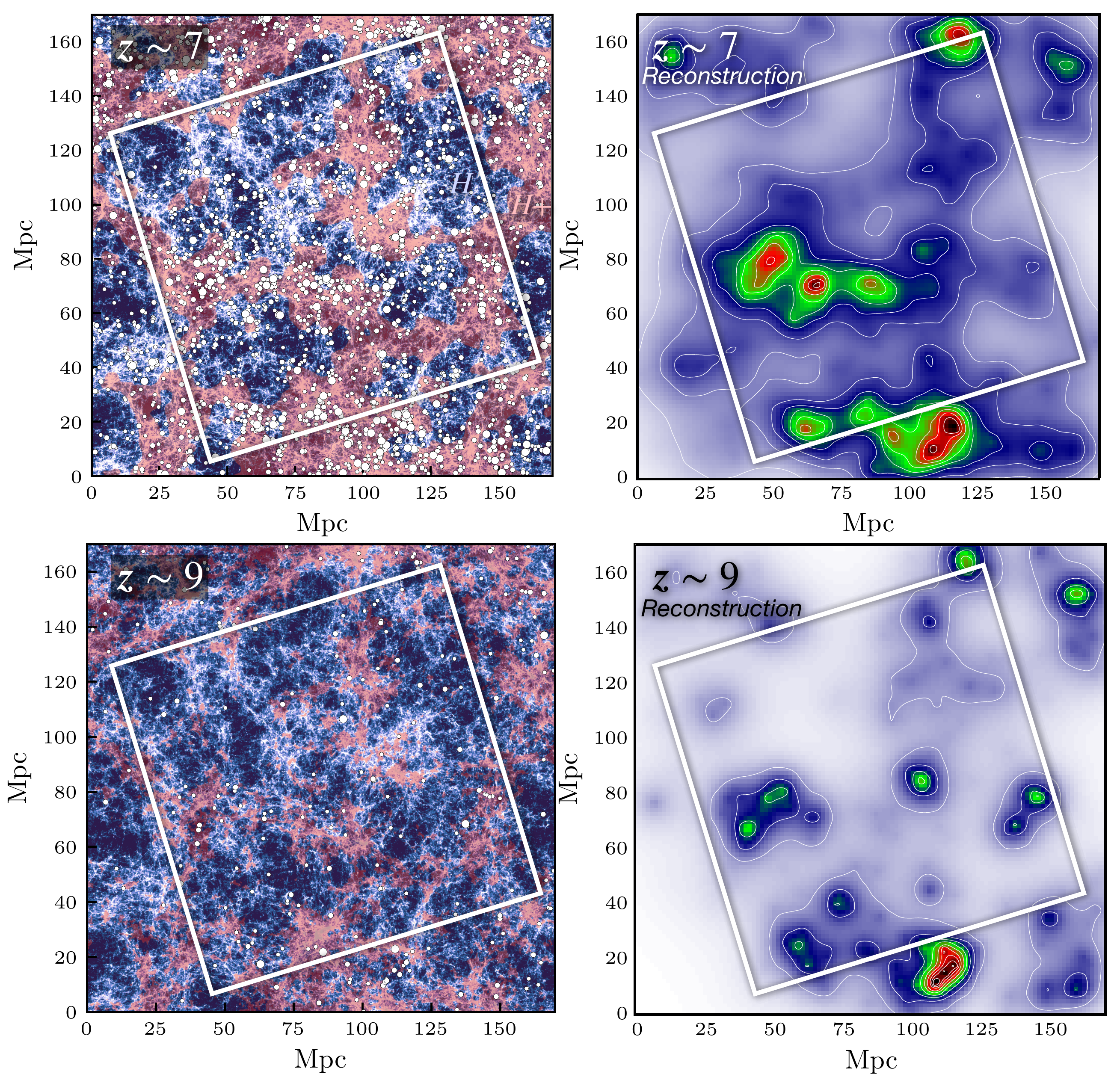}
 \caption{At left are two snapshots of a cosmological n-body
   simulation performed using GADGET-2 \citep{springel05a} spanning a
   (100\,h$^{-1}$\,Mpc)$^3$ volume (not the real COSMOS survey
   field). The cube is here projected from one side and has a
   thickness equivalent to $\delta z=0.5$ at $z=7$ and $\delta z=0.7$
   at $z=9$.  The underlying dark matter distribution is shown in blue
   (void indicated by darker blue) and the distribution of ionized
   hydrogen gas (H$+$) is shown in pink, whereas regions of neutral
   IGM have the underlying blue dark matter distribution
   visible. Galaxies that are detectable in COSMOS-Web are shown as
   white points (having F150W\,$<$\,27.5). Larger points represent
   more luminous (with F150W\,$<$\,26.5) galaxies. At right we show
   the recovered galaxy density maps inferred from the same simulation
   snapshots using the observational limits of our survey. The
   recovered maps use an adaptive kernel smoothing on a global 5\,Mpc
   kernel scale \citep{darvish15a} and include a modeling of sources'
   incompleteness and photometric redshift uncertainties,
   demonstrating our ability to recover large scale structure at these
   redshifts.
  The galaxies responsible for reionization may be expected to be
  strongly clustered on 30\,--\,100\,cMpc (10\,--\,40\,arcmin)
  scales, much larger than all existing contiguous near-infrared
  {\it Hubble} deep surveys and other planned Cycle 1 {\it JWST}
  surveys. COSMOS-Web will span an area the size of the white box,
  about (46\,arcmin)$^2$, and will cover a mix of 4\,--\,16
  independent reionization bubbles or neutral gas regions per
  $dz=0.3$ slice across 12 independent slices. With $\sim$5000
  sources detectable across $6<z<8$ and $\sim$600 sources at
  $8<z<10$, COSMOS-Web will be uniquely situated to gathering
  statistical samples of EoR density environments. Further
  follow-up observations in Ly$\alpha$ may then reveal the
  relationship between mass overdensities and ionized bubbles. }
 \label{fig:eorsim}
\end{figure*}

\subsubsection{The first maps of LSS during the EoR}

The full 0.54\,deg$^2$ COSMOS-Web survey will allow the direct
construction of large scale structure density maps of galaxies
spanning $z\sim6-10$. Such snapshots of the density field will provide
a direct test as to whether or not the brightest, most massive
galaxies are indeed highly clustered, as suggested by cosmological
simulations \citep[e.g.,][]{mcquinn07a,behroozi13a,chiang17a}. Though
some massive galaxies have been identified at this epoch from
Ultra-VISTA data \citep{bowler20a,endsley22b,kauffmann22a}, it remains
to be seen whether or not they sit in overdense environments. Existing
{\it Hubble} and other planned {\it JWST} surveys are insufficient to
answer this question due to their limited areas, however, COSMOS-Web
will have both the depth and area to enable this measurement.

Figure~\ref{fig:eorsim} illustrates our approach by using a mock
catalog from a cosmological simulation \citep[GADGET2; ][]{springel05a}
at $z\sim7$ with width $\Delta z \approx 0.5$ (and $z\sim9$ with width
$\Delta z\approx0.7$) to reconstruct the underlying density map from
simulations using the weighted adaptive kernel smoothing technique
\citep{darvish15a} on 5\,Mpc scales. We have used this simulation to
directly test our ability to reconstruct the density field of galaxies
from detectable sources. We infer that we will be able to reconstruct
$\sim$12 independent mappings of the full density field between
$6<z<10$ based on our simulated photometric precision ($\Delta
z/(1+z)\sim0.02-0.04$) and low contamination rates using a combination
of color cuts and photometric redshift fitting (see
\S~\ref{sec:eorselection}). The smoothing scale of 5\,Mpc is an ideal
scale to achieve a S/N in the overdensity measurement of $>5\,\sigma$
per beam and S/N\,$>20\,\sigma$ per overdense structure (with $>25$
galaxies per beam in the highest density regions). COSMOS-Web will
provide the first direct measurement of the physical scale and
strength of overdensities at these epochs for direct comparison to
the hypothesized scale of reionization-era bubbles that theoretically
emanate from them.

\subsubsection{Masses of EoR Galaxies}\label{sec:eormass}

COSMOS-Web will enable crucial stellar mass constraints for the most
massive EoR galaxies via the detection of rest-frame optical light
(e.g., \citealt{faisst16a}). In particular, with the MIRI F770W
observations covering 0.19\,deg$^2$, we expect to detect 90-130
galaxies at $6<z<8$ and 2--3 galaxies at $8<z<10$ based on estimates
from the UVLF.  This would double the expected number of EoR galaxies
with rest-frame optical detections from the other Cycle 1 {\it JWST}
programs.  At these redshift regimes this corresponds to rest-frame
1\um\ and 7700\,\AA\ light, respectively.  This will place unique
constraints on the physical characteristics of extremely rare
M$_\star\simgt10^{10}$\,\msun\ galaxies in the Universe's first few
hundred Myr for the subset of EoR sources that we expect to detect
  with MIRI, as no galaxies with these high masses are
expected to be found in the other Cycle 1 {\it JWST} surveys that
cover smaller areas \citep[cf. the sources identified by][with extreme
  stellar masses established at $z\sim7-9$; though those sources'
  stellar masses may yet be highly uncertain, e.g.,
  \citealt{endsley22a}]{labbe22a}. The detection of $\sim$100 galaxies
in this mass regime will provide important clues to the star formation
histories of the Universe's most massive halos in the first Gyr after
the Big Bang, which are currently unconstrained.

\subsubsection{Follow-up of EoR Sources}

Beyond the direct EoR discoveries that COSMOS-Web will make through
its {\it JWST} imaging, follow-up observations will further enhance
the impact of this program and shed light on key unknowns. These
include (1) rest-frame UV diagnostics with {\it JWST} NIRSpec that
will constrain ionizing photon production in $z\simgt6$ sources
\citep[i.e., constraints on $f_{\rm esc}$ and $\xi_{\rm ion}$,
  e.g.,][]{chisholm20a}, (2) deep rest-frame UV observations of
Ly$\alpha$ to infer local variations in the IGM neutral fraction with
Keck, Subaru, VLT (which can typically reach line
  sensitivities of $\sim$10$^{-18}$\,erg\,s$^{-1}$\,cm$^{-2}$), and
future 30\,m-class telescopes (the extremely large
  telescopes, ELTs, that will push fainter), and (3)
obscured star-formation and cold ISM content of dust and metals from
ALMA detections of the FIR continuum and the FIR fine-structure atomic
cooling lines \citep{laporte17a,hashimoto18a,bakx22a,fujimoto22a},
which will inform stellar population synthesis models of galaxies'
first light, metals, and dust.  COSMOS-Web, as a wide and shallow
survey, will be particularly useful for the detection of bright, rare
candidates that are well optimized for ground-based follow-up.  These
future observations will be crucial for detailed characterization of
EoR overdensities, unlocking direct comparisons between mapped
reionization bubbles (measured via Ly$\alpha$ follow-up) and {\it
  JWST}-measured density maps, as shown from a simulation in
Figure~\ref{fig:eorsim}.

\subsection{The Buildup of the Massive Galaxy Population}\label{sec:mg}

The wide-area coverage of COSMOS-Web, in particular the combination of
the NIRCam LW (2.8\,\um\ and 4.4\,\um) and MIRI (7.7\,\um) observations 
with the already existing wealth of optical to NIR data in COSMOS, will 
allow us to take the first census of massive galaxies from the end of the 
EoR to the peak of galaxy assembly. Within the footprint of COSMOS-Web, 
we expect firm identification of half a million galaxies at all redshifts,
$\sim$32,000 of which will be detected in MIRI F770W imaging,
allowing us to constrain stellar masses, sizes, morphologies, star 
formation rates, and AGN activity for galaxies across a wide swath of 
cosmic time.

\subsubsection{The First Quiescent Galaxies}\label{sec:quiescent}

\begin{figure}
 \centering
 \includegraphics[width=0.4\textwidth]{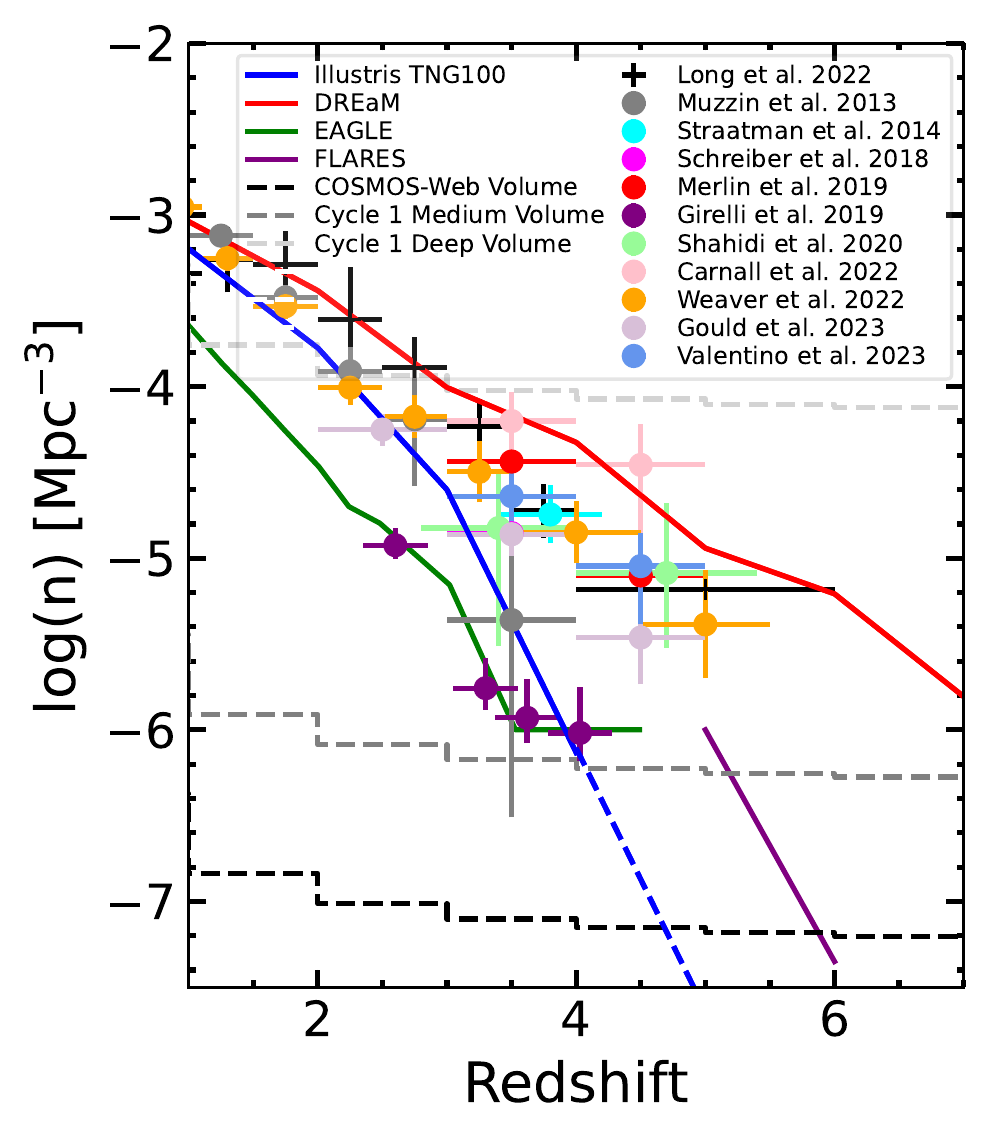}
 \caption{The number density of quiescent galaxies (specific SFR,
   SFR/M$_{\star}$ $<10^{-11}$\,yr$^{-1}$) as a function
   of redshift for galaxies with M$_{\star}>10^{10}$\,M$_{\odot}$
   selected from Illustris TNG100 (blue) and extrapolated beyond $z>4$
   (blue dashed) where no quiescent galaxies are found in the
   Illustris TNG100 volume.  Other simulation predictions for this
   population are shown from EAGLE (green), FLARES (purple), and the
   DREaM (red) semi-empirical model and predictions from the empirical
   model of \cite{long22a} in black. Overplotted is a collection of
   number densities of quiescent galaxies from the literature,
   illustrating the wide range that both the observations and
   simulations span.  The dashed lines correspond to the number
   density of one object in the NIRCam volume of COSMOS-Web (black),
   medium-volume JWST surveys such as CEERS and PRIMER (dark gray),
   and deep volume surveys such as JADES-Deep and NGDEEP (light
   gray). Only COSMOS-Web has the volume necessary to place strong
   constraints on the number densities of quiescent galaxies at $z>4$
   if they are indeed as rare as expected. }
 \label{fig:quiescent}
\end{figure}

The growing census of massive quiescent galaxies at early epochs
\citep[$M_\star\simgt10^{10}\,$\,M$_\odot$ out to $z\sim3-5$,
e.g.,][]{straatman14a,Glazebrook17a,schreiber18b,merlin19a,
tanaka19a,Girelli19a,Valentino20a,Forrest20b,Carnall22a,carnall23a,Rodighiero22a} 
has presented a strong challenge to theoretical models of early massive galaxy
formation \citep[e.g.,][see Figure \ref{fig:quiescent}]{feldmann16a,
Steinhardt16a,Cecchi19a}. In order to build up their significant stellar 
masses and quench their star formation so early in the Universe's 
history, these galaxies must have formed their stars at exceptionally high 
rates ($\gg$100\,\msun\,yr$^{-1}$, comparable to luminous infrared 
galaxies; \citealt{sanders96a}, and dusty star-forming galaxies, 
DSFGs; \citealt*{casey14a}) at very early times and then abruptly 
shut down the production of stars well within the Universe's first Gyr. 
The existence of these sources and their relative abundance provide 
important tests of the galaxy assembly process and the physical
processes driving the quenching of star formation at this early epoch.

The quiescent galaxy mass function beyond $z\sim4$ is currently
unconstrained, partly because of the difficulty of detecting these
rare galaxies in existing deep field observations (with volume
densities $\simlt10^{-5}$\,Mpc$^{-3}$) and partly because such
galaxies are particularly difficult to separate from DSFGs and 
post-starburst galaxies that can
mimic the same red colors (see Figure \ref{fig:sedlimits}). Detecting
them requires deep rest-frame optical observations over wide areas of
the sky. COSMOS-Web will provide the ideal dataset for identifying
candidate quiescent galaxies and measuring (or placing constraints) on
their number densities and relative abundances. Figure
\ref{fig:quiescent} highlights the expected number density of massive
(M$_{\star}>10^{10}$\,M$_{\odot}$) quiescent (specific SFR,
SFR/M$_\star<10^{-11}$\,yr$^{-1}$) galaxies from the cosmological
hydrodynamical simulations IllustrisTNG100 \citep{pillepich18a}, EAGLE
\citep{McAlpine16a}, and FLARES \citep{Lovell22a}, as
well as the DREaM semi-empirical model \citep{Drakos22a} and 
predictions from the empirical model of \cite{long22a}, in
comparison to some of the currently identified quiescent galaxy
candidates in the literature
(\citealt{muzzin13a,straatman14a,merlin19a,schreiber18b,Girelli19a,
  Shahidi20a,Carnall22a, Weaver22b,Gould23a,Valentino23a}). 
Note that each study selects quiescent galaxies slightly differently, 
and the resulting samples span a range of stellar mass cuts, with 
the vast majority of candidates having M$_{\star}>10^{10}$\,M$_{\odot}$. 
When multiple mass cuts are quoted by a given study, we show 
number densities above this mass limit for consistency.

The quiescent galaxy sample from IllustrisTNG100 was selected using
the publicly
available\footnote{\url{https://www.tng-project.org/data/}} star
formation rate \citep{pillepich18a, Donnari19a} and stellar mass value
that corresponds to the mass within twice the half-mass radius of each
object \citep{Rodriguez-Gomez16a}.  Note that there are no quiescent
galaxies in the IllustrisTNG100 volume beyond $z>4$ using this
definition. We similarly selected the quiescent galaxy sample from the
public EAGLE galaxy
database\footnote{\url{https://icc.dur.ac.uk/Eagle/database.php}}
\citep{McAlpine16a} using the recommended aperture size of 30 physical
kpc.  These hydrodynamical simulations are calibrated to reproduce
physical properties in the local Universe and predict the the SFR and
M$_{\star}$ values at high redshift.  For FLARES, we use the number
densities measured by \cite{Lovell22a}.  These
simulations generally underpredict the observed number densities of
quiescent galaxies in the literature (though the observations span a
wide range of values).  On the other hand, semi-analytic models like
DREaM are calibrated to match scaling relations at all redshifts. The
DREaM number densities in Figure \ref{fig:quiescent} are based on the
SMF of \cite{williams18a} and are a close match to the high end of the
observed number densities.

Even though true quiescent galaxies are expected to be rare at $z>4$,
with the large area of COSMOS-Web we will be able to identify
massive quiescent galaxy candidates and place robust constraints on their
abundances as a function of redshift if they are brighter than our
detection limit with number densities $\geq$10$^{-7}$\,Mpc$^{-3}$.
This measurement will also be less impacted by the effects of cosmic
variance than similar measurements from smaller area surveys (e.g.,
\citealt{Carnall22a}).

\begin{figure*}
 \centering
 \includegraphics[width=0.9\textwidth]{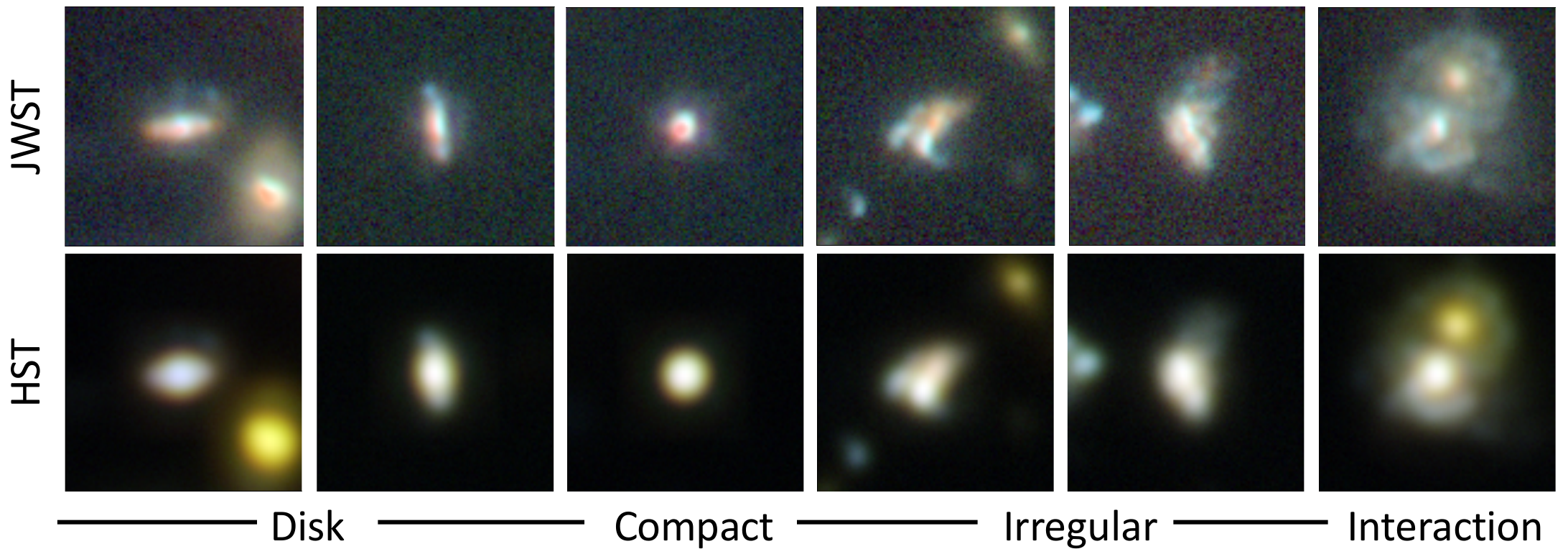}
 \caption{A selection of three color NIRCam (F115W+F150W+F277W)
   cutouts of galaxies at $z>3$ with varied morphologies selected from
   IllustrisTNG mock images \citep{rose22a} with noise added
   consistent with the COSMOS-Web depth. Each cutout is 3$''$ on a
   side.  These illustrate COSMOS-Web's ability to distinguish between
   different morphological types and detect a diversity of
   morphologies thanks to {\it JWST}'s sensitivity and resolution at
   these redshifts. }
 \label{fig:morph}
\end{figure*}

The combination of NIRCam and MIRI filters over 0.20\,deg$^2$
(including the COSMOS-Web and PRIMER MIRI imaging that fall within the
NIRCam footprint) will enable quiescent galaxies to be distinguishable
from dusty star-forming interlopers via color-selection and SED
analysis using the well-sampled rest-frame optical
photometry. Additionally, the complementary (sub)millimeter
observations over the COSMOS field (see Figure \ref{fig:allmap}) will
enable the direct identification of dusty galaxies at $z>4$ and
therefore disentangle them from quiescent and EoR galaxy candidates
(see \citealt{zavala22a,naidu22b,fujimoto22a} for detailed discussion
of the difficulty in separating these populations using NIRCam colors
alone).  Specifically, the existing SCUBA-2 and future TolTEC
observations will cover COSMOS to a depth of
SFR\,$\simgt$\,50\,\msun\,yr$^{-1}$, while the ALMA MORA survey
\citep{zavala21a,casey21a,manning22a} and its continuation (ex-MORA;
Long \etal, in preparation) will cover 0.2\,deg$^2$ ($\sim$1/3 of
COSMOS-Web), in addition to the public ALMA archival pointings from
A3COSMOS \citep[totalying 0.12\,deg$^2$ across all of
  COSMOS;][]{liu19a} and will directly detect DSFGs at $z>4$ in excess
of SFR\,$\simgt$\,100\,\msun\,yr$^{-1}$.

In addition to identifying the highest redshift quiescent galaxies,
COSMOS-Web observations will allow us to study their properties in
detail. MIRI 7.7\,\um\ observations (rest-frame
1.1--1.5\,\um\ at $4<z<6$)  for a subset will provide a long wavelength 
lever arm to accurately determine their masses.  The full multiwavelength SED will 
enable us to measure their SFRs and constrain their star formation 
histories (SFHs) and dust attenuation, with improved uncertainties on the SFHs
with constraints from {\it JWST} (e.g., \citealt{whitler22a}). With
the high resolution NIRCam and MIRI imaging we will be able to study
their morphologies in great detail (see Figure \ref{fig:morph}) and
robustly measure their rest-frame optical sizes as well as constrain
the physical distribution of their SFR, mass, and dust content, giving
insight into how these galaxies may have quenched. This will enable a
detailed investigation of the galaxy size-mass relation for quiescent
systems $<$2\,Gyr after the Big Bang, extending our understanding of
size growth out to higher redshifts and less extreme massive galaxies
than has been possible before (e.g.,
\citealt{Toft07a,van-der-wel14a,Straatman15a,Shibuya15a,faisst17a,kubo18a,Whitney19a})
and a statistically robust study of their progenitors.

Within the COSMOS-Web footprint, we expect to detect $\sim$\,13,000
massive galaxies (M$_{\star}>10^{10}$\, M$_{\odot}$) between $4<z<6$
($\sim$\,2,300 with MIRI coverage) of which we estimate there will be
at least $\sim$\,350 quiescent candidates (in the NIRCam
  mosaic, and 120 with MIRI coverage, selected to have
  sSFR$<10^{-11}$\,yr$^{-1}$) scaling the COSMOS2020 estimates of
  source counts at these redshifts \citep{weaver22a}; this will be
  $\sim$10$\times$ improvement over current $z>4$ quiescent galaxy
  candidate samples. Follow-up spectroscopic observations for
  subsamples of these quiescent galaxies (e.g., such as those by
  \citealt{schreiber18b,Valentino20a}) will be able to confirm their
  redshifts, measure their velocity dispersions, and more fully
  characterize their ages and star formation histories, enabling us to
  separate true quiescent galaxies from post-starburst systems.

\subsubsection{Dusty Star Forming Galaxies}\label{sec:dsfg}

DSFGs are an intrinsically rare population (with number densities
$\simlt$10$^{-5}$\,Mpc$^{-3}$) whose individual discoveries,
particularly at $z>5$ test the limits of galaxy formation models
\citep*[see reviews by][]{casey14a,hodge20a}. They are largely
regarded as the dominant progenitor population of high-redshift
quiescent galaxies, given their prodigious rates of star formation
($\simgt$100--1000\,M$_\odot$\,yr$^{-1}$) and similar volume densities
(though both are quite uncertain). While DSFGs are typically easily
identified directly via FIR emission or their (sub)mm emission in
single-dish or interferometric maps, often their more detailed
physical characterization remains elusive. This may include the
measurement of their redshifts or masses.  It is difficult due to
significant degeneracies in their submm emission with redshift and
significant dust obscuration of the rest-frame UV and optical
emission.  Radio continuum emission can also be a vital tool in
detecting DSFGs, and often facilitates quick multiwavelength
identification via precise astrometric constraints
\citep[e.g.,][]{algera20a,talia21a,enia22a}.  From ancillary
FIR/submm data already in hand covering COSMOS-Web, we know of
$\sim$1100 DSFGs at all redshifts detected by SCUBA-2 and {\it
  Herschel} that will be covered by the NIRCam mosaic with
luminosities $\simgt$10$^{12}$\,\lsun; many of these do not yet have
spectroscopic redshifts and confirmed counterparts, though a
significant fraction ($\simgt$50\%) have follow-up continuum ALMA
observations providing precise astrometric constraints
\citep{liu19a,simpson20a}.

COSMOS-Web will transform our understanding of the stellar content in
DSFGs at all redshifts, but in particular shed light on the rarest
DSFGs found at $z>5$ (of which there are fewer than two dozen with
spectroscopic redshifts). Based on recent models of the obscured
galaxy luminosity function \citep{zavala21a}, we estimate that
$\sim$40--70 of the $>$10$^{12}$\,\lsun\ DSFGs in the NIRCam mosaic
will lie above $z=4$, and $\sim$3--10 above $z=6$. Including those
with an order-of-magnitude lower luminosity ($>10^{11}$\,\lsun), the
statistics inflate by an order of magnitude. Roughly a third of DSFG
samples, and especially those selected at longer wavelengths, are
invisible even in deep {\it Hubble} imaging
\citep{franco18a,Gruppioni20a,casey21a,manning22a}.  In contrast, {\it
  JWST} imaging (both with NIRCam and MIRI) pushes to depths
sufficient to capture DSFGs' highly obscured stellar emission,
enabling measurement of more precise photometric redshifts than are
currently accessible, in addition to constraints on their morphologies
and stellar masses.  For example, the vast majority of
  DSFGs are detected in deep {\it Spitzer}/IRAC imaging (with
  [4.5\,\um]\,$<\,26$); thus, we expect detection of all DSFGs in the NIRCam LW
  filters particularly because the median stellar mass of the
  population is expected to be high, $\sim7\times10^{10}$\,\msun
  \citep{hainline11a}, roughly a factor of $\sim$150--200$\times$
  larger than the stellar masses of galaxies at the NIRCam LW
  detection limit at $z\sim5$.  Through more reliable optical-IR
photometric redshifts, combined with additional (sub)mm constraints on
their redshifts \citep{cooper22a}, these data will unlock many
unknowns about the evolution of and buildup of mass in such extreme
star-forming galaxies at early times.

\subsection{Linking Dark Matter with the Visible}\label{sec:wl}

The link between galaxies' dark matter halos and their baryonic
content is of fundamental importance to cosmology. Yet directly
observable tracers of halo mass are not available for the vast
majority of galaxies, and in their place, either halo occupation
distribution (HOD) modeling \citep{seljak00a,cowley18a} or abundance
matching \citep{kravtsov04a,conroy09a,behroozi19a} are used to infer
halo mass from galaxies' stellar masses \citep[via the
  stellar-mass-to-halo mass relation,
  SMHR;][]{croton06a,somerville08a}. However, the evolution of
galaxies is direct evidence for the complexity of the halo-baryon
relationship \citep{legrand19a,shuntov22a}.  Halos provide the
potential well for accretion of fresh gas, which in turn fuels stellar
mass growth through star formation. Merging also substantially boosts
stellar mass growth and relates directly to the physical interactions
of halos which occurs on scales larger than individual galaxies.
Indeed, it is thought that on such large scales, galaxies' halo mass
growth should be independent of the baryonic processes within
galaxies. If measurable, they could provide a direct path to
constraining galaxy growth and their relationship to quenching
mechanisms. Obtaining direct measurements of halo masses not only
helps us to constrain the astrophysics of galaxies
\citep{mandelbaum06a,mandelbaum14a} but also gives independent
measurements on cosmological parameters
\citep{zheng07a,yoo06a,yoo09a}.

\begin{figure*}
 \includegraphics[width=0.99\textwidth]{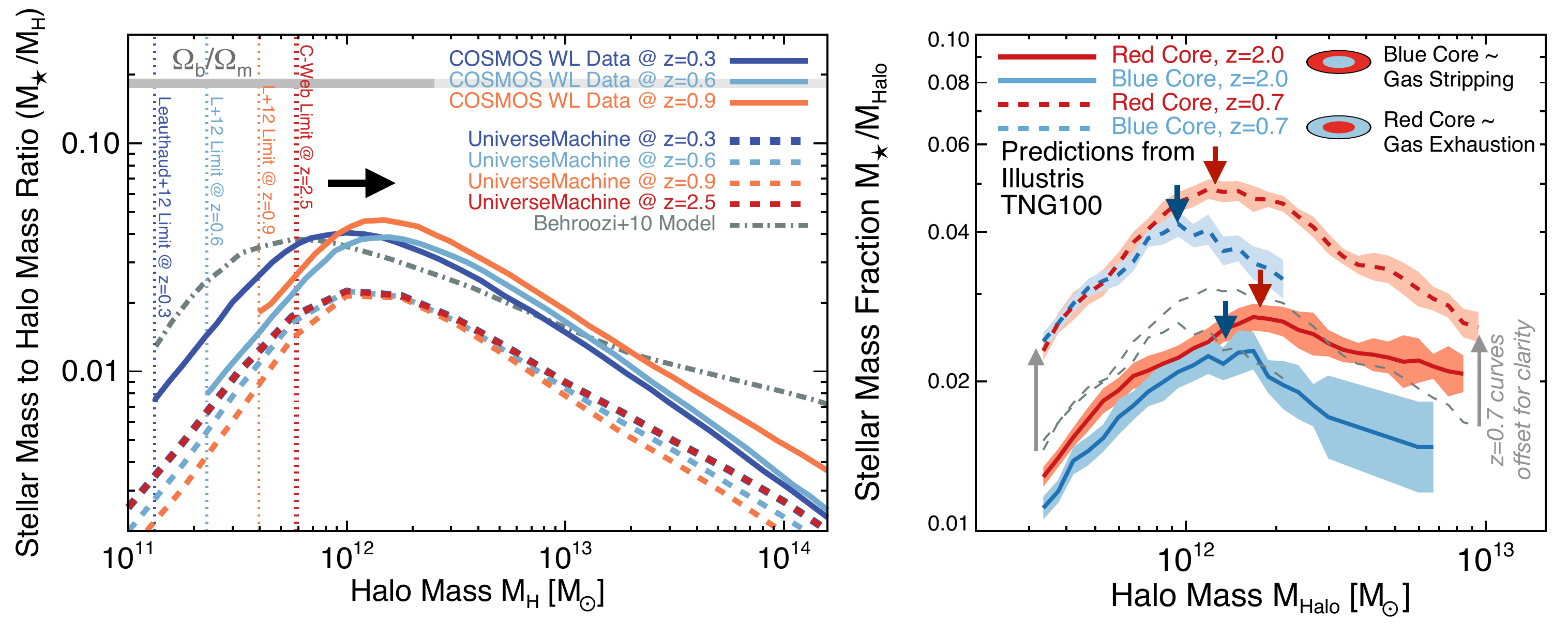}
 \caption{At left, the stellar mass to halo mass ratio as a function
   of halo mass. Curves are shown for central galaxies only, not
   including satellites, though both relations may be constrainable
   with our dataset. Solid lines are measurements from COSMOS ACS weak
   lensing data \citep{leauthaud12a} at $z=0.3$ (dark blue), $z=0.6$
   (light blue), and $z=0.9$ (orange). In comparison, the SMHR from
   cosmological simulations is overplotted (gray dot-dashed line from
   \citealt{behroozi10a} and dashed line colored dashed lines from
   UniverseMachine, \citealt{behroozi19a}) at matched redshifts
   extending out to $z=2.5$ (red). The existing weak lensing
   measurements show evolution out to $z\sim1$ (black arrow noting
   SMHR peak evolving); COSMOS-Web weak lensing measurements will have
   the power to extend this analysis to $z=2.5$, where the vertical
   red line will represent the lower mass limit at that redshift. At
   right, predictions from hydrodynamic simulations, specifically
   Illustris TNG100 \citep{pillepich18a}, suggest that ellipticals
   with blue or red cores experience different quenching mechanisms
   (gas stripping shown in blue, and gas exhaustion in red).  These
   differences are reflected in how their $M_\star/M_{\rm H}$ evolves
   with time ($z\sim2$ in solid; $z\sim0.7$ in dashed). COSMOS-Web
   data will have the potential to provide powerful constraints on the
   evolution of the SMHR for different galaxy populations and test
   fundamental galaxy quenching models.}
 \label{fig:wl}
\end{figure*}

Directly measuring halo masses out to large galactocentric radii
($\sim$1\,Mpc, needed to probe the underlying dark matter) can be done
either with galaxy-galaxy lensing \citep{brainerd96a} or using
kinematic tracers like satellite galaxies \citep{mckay02a}. Given the
sparsity of bright satellites beyond the local Universe and rarity of
strongly-lensed galaxies, weak lensing (WL) is the only tool that can
be used as a direct probe of halo masses for a large sample of
galaxies across cosmic time \citep{sonnenfeld21a}. An innovative
method combining galaxy clustering measures with HOD modeling and weak
lensing was demonstrated by \citet{leauthaud11a} and
\citet{leauthaud12a} using the COSMOS single band F814W {\it Hubble}
imaging to measure SMHR evolution from $0.2<z<1.0$ at
$M_{\star}>10^{10}$\,\msun. These measurements are shown in the left
panel of Figure~\ref{fig:wl}.

COSMOS-Web's 4-band NIRCam imaging spanning $>$0.5\,deg$^2$, joined
with the high quality 40$+$ band imaging constraining galaxies' masses
and photometric redshifts in COSMOS \citep[][see also
  Figure~\ref{fig:sim_zphotzspec}]{weaver22a}, will be the best
available dataset from which high resolution weak lensing mass mapping
measurements can be done. This will involve a careful reconstruction
of the PSF for each exposure in each filter and measurements of source
centroids, shapes, and orientations. These measurements will then be
combined with the best possible photometric redshifts to infer
evolution in the SMHR \citep{leauthaud07a,leauthaud11a}.  Extending
from $z\sim1$ to $z\sim2.5$ and to depths an order of magnitude deeper
in halo mass at fixed redshift is enabled by the significant boost in
spatially-resolved background and foreground sources where the weak
lensing signal goes roughly as the square-root of the foreground
source density multiplied by the background source density,
$\propto\sqrt{N_{\rm fg}(z)}\sqrt{N_{\rm bg}(>z)}$.  The density of
background sources will exceed 10\,arcmin$^{-2}$ out to $z=4$, with
$\sim$110,000 sources at $z>2.5$ above 15$\sigma$, which is the
necessary detection threshold for adequate shape recovery
\citep{jee17a}.  Furthermore, these data will push that deep in {\it
  each independent filter}, thus will be the first wide, {\it deep}
multi-band survey from space; simultaneous weak lensing measurement in
multiple bands will both let us go even deeper and provide independent
cross-checks of instrumental effects like the PSF calibration.  {\it
  Euclid} and {\it Hubble} cannot observe at such long wavelengths
\citep{lee18a} and {\it Roman} will not achieve such high
resolution. Contiguous, high resolution NIR imaging from {\it JWST} in
  COSMOS-Web will thus serve as a much needed absolute calibration of
  the SMHR relation out to $z\sim2.5$ that can be leveraged by other
  weak lensing surveys conducted on larger scales.

\subsubsection{Evolution in the SMHR}

\citet{leauthaud12a} found unexpected evolution in the characteristic
mass at which the SMHR is maximized (downsizing), in other words where
the peak efficiency (around $\sim$10$^{12}$\,\msun) evolves downward
from $z\,\sim\,0.9$ to $z\,\sim\,0.3$. \citet{shuntov22a} similarly
finds this downsizing trend with the COSMOS2020 catalog using
clustering and constraints on the stellar mass function out to
$z\sim3$.  COSMOS-Web will significantly strengthen measurements to
$z\sim2.5$ with the important addition of weak lensing constraints,
facilitating a re-calibration of hydrodynamical simulations and
semi-analytic models that produce mock observables essential for much
of cosmology and extragalactic astrophysics. This has important
implications for how HOD modeling or abundance matching is used in the
literature and how semi-analytic models and cosmological
hydrodynamical simulations generate observables, on which much of
extragalactic astrophysics relies.

Such weak lensing measurements rely on contiguous coverage over a
large ($\simgt$0.5\,deg$^2$) area, otherwise they are substantially
affected by edge effects \citep{mandelbaum05a,massey07b,han15a}.
COSMOS-Web, with its large, deep, and contiguous coverage, will enable
direct measurements of galaxies' halo masses out to $z\sim2.5$ down to
$M_\star>$\,few$\,\times10^{9}$\,\msun\ (down to 10$^8$\,\msun\ at
$z\sim1$), well beyond current data limitations (above
$10^{10}$\,\msun\ at $z\simlt1$) and future planned weak lensing
measurements (e.g., from {\it Euclid} or {\it Roman}). Extending weak
lensing measurements to $z\sim2.5$ is essential for simulation
calibration due to the significant evolution in galaxies' properties
\citep[e.g., SFRs;][]{noeske07a,Whitaker14a} in the past 11\,Gyr from
$z=0-2.5$.

The potential to extend SMHR constraints to higher redshifts is also
possible using similar techniques to \citet{shuntov22a}. Such high
redshifts and great mass depths can be reached due to the dramatic
increase in the number of background sources for weak lensing and
sources at all epochs that will have high-quality photometric
redshifts.

\subsubsection{Constraining the Dependency of the SMHR on Resolved Baryonic Observables}

Given that COSMOS-Web data will be obtained in multiple filters, it
will be the first sufficiently large dataset to test for alternate
dependencies of resolved baryonic observables (e.g., color as a tracer
of quenching mechanisms) on halo mass.  This type of differential
measurement with galaxy type is demonstrated by \citet{tinker13a} out
to $z\approx1$, who find that star-forming galaxies grow in lock-step
with their dark matter halos, while quiescent galaxies have baryonic
growth that is outpaced by dark matter growth. Higher redshifts can be
reached by conducting the same experiment at longer wavelengths,
boosting observed densities of high-$z$ sources. COSMOS-Web will push
the limits of weak lensing's direct measurement of halo masses to
$z\sim2.5$ with $M_\star>10^{10}$\,\msun\ such that halo masses can be
independently constrained as a function of galaxy type over a
significant portion of the Universe's history.

Resolved color gradients in galaxies are thought to be the hallmark
tracer of the quenching process.  Bluer cores likely trace systems
where cold gas has been stripped from the periphery
\citep[e.g.,][]{meschin14a}, while redder cores trace gas exhaustion,
where gas at the galaxy core is not replenished
\citep[e.g.,][]{kawata08a,tacchella15a}. Flat color gradients are
expected for galaxy collisions, in which the gas supply is consumed
quickly with no preferred radial distribution
\citep[e.g.,][]{springel05a,sparre15a}.  While differential dust
attenuation may complicate the interpretation of galaxies' color
gradients, some independent observations at long wavelengths could
break the degeneracy (see the discussion later in
\S~\ref{sec:dustattenuation}).

The right panel of Figure~\ref{fig:wl} illustrates the expected
difference between the SMHR of bluer-cored galaxies and redder-cored
galaxies. Do galaxies with different gradients show different
evolution in their SMHRs? Cosmological simulations predict that
similar SMHRs may point to the significant role of major galaxy
mergers in the quenching process, while different SMHRs would point to
feedback quenching mechanisms. At $z\simlt2.5$, our weak lensing measurements
of halo masses for large samples can be directly compared with
assertions that massive $>10^{10}$\,\msun\ galaxies evolve from
star-forming to quenched in $\sim$100\,Myr \citep[e.g.,][]{barro13a}.

\section{COSMOS-Web's Impact on Other Topics}\label{sec:ancillary}

The breadth of scientific studies that COSMOS-Web may advance is
extraordinary and impossible to anticipate in full. Below we describe
some key ancillary science cases that could make significant strides
given the layout and plans for the COSMOS-Web Treasury program.
 We emphasize that COSMOS-Web's contribution to
  these areas will be powerful, though not made in isolation; much
  of the progress will be significantly aided by, if not fully dependent
  on, the legacy of data obtained in the COSMOS field from other
  observatories.

\subsection{Galaxy Morphologies and Sizes to $z\sim 8$}
 
Over the age of the Universe, galaxies have undergone dramatic
morphological transformations. Today's galaxies are a mix of
well-formed spiral disk galaxies, ellipticals, and irregular galaxies
and deep rest-frame optical images from {\it Hubble} have shown that
the basis for what is known as the Hubble sequence was already in
place by $z\sim3$
\citep[e.g.,][]{Wuyts11a,van-der-wel14a,kartaltepe15a}. Early {\it
  JWST} studies (e.g., \citealt{robertson22b,ferreira22a,Kartaltepe22a}) 
  are finding that galaxies at even higher redshifts
have a wide diversity of morphologies, and a significant fraction
already show evidence for disks and spheroids. However, a large
fraction of galaxies at high redshift also have irregular
morphologies, some of which may be signatures of galaxy mergers and
interactions (e.g., \citealt{Kartaltepe12a}) and some may be
indicative of other physical processes such as disk instabilities
(e.g., \citealt{Keres09a,genzel11a}). In order to quantify the
morphological transformation of galaxies from very early epochs to
today, and understand the physical drivers responsible, large samples
at high redshift are required.

COSMOS-Web will spatially resolve the rest-frame optical emission of 
tens of thousands of galaxies from $z=3-8$, enabling a detailed morphological 
classification into spheroids/disks/irregulars and identification of 
interaction and merger signatures. These measurements will enable studies
of morphological transformation as a function of environment and the relative 
roles of different physical processes responsible for enhanced star formation
and black hole growth in the early universe. These large samples of morphology
measurements will be essential training samples for machine learning 
algorithms (e.g., \citealt{Snyder19a, Pearson19a,Hausen20a, 
Ciprijanovic20a,rose22a}) to classify galaxies, identify merger signatures,
and identify unique morphologies that may otherwise be missed.

The evolution of galaxy sizes is also a useful tool for investigating
the evolutionary history of galaxies and connecting the properties of
today's galaxies to their progenitors in the early universe. Over the
past decade, a number of studies have found evidence for strong
evolution in the optical/UV sizes of galaxies, with effective radii
growing by a factor of 2--7 since $z\sim2$ (e.g.,
\citealt{van-der-Wel08a,Buitrago08a}), suggesting that these massive
galaxies have evolved through minor mergers in this time period (e.g.,
\citealt{Naab09a, Bluck12a,Furlong17a}).  Both star-forming and
quiescent populations of galaxies have been found to evolve in size,
with samples of compact star forming \citep{Barro14a,Barro14b} and
compact quiescent (e.g., \citealt{Toft07a, van-Dokkum08a, Bezanson09a,
  barro13a}) galaxies identified at cosmic noon. At even higher
redshifts, $z=3-7$, significant, though less steep, evolution has been
found by a number of studies (e.g.,
\citealt{van-der-wel14a,Straatman15a, Shibuya15a,Whitney19a}), and a
range of physical mechanisms driving this evolution have been
suggested, including major and minor mergers (e.g., \citealt{Bluck12a,
  Wellons16a}), rejuvenated star formation in the galaxy's outer
regions due to gas accretion \citep{Conselice13a,
  Ownsworth16a,dekel20a}, quasar feedback (e.g.,
\citealt{Fan08a,dubois16a}), and progenitor bias
\citep{van-Dokkum01a}.

\cite{Wellons16a} used Illustris to track the evolution of a sample of
compact quiescent galaxies at $z\sim2$ and found a diverse range of
properties among their descendants, with very few remaining compact in
the present day, in agreement with observations \citep{trujillo09a,
  tortora18a,scognamiglio20a}. Most growth appears to be driven by the
delivery of ex-situ mass and the impact of galaxy mergers and both are
closely linked to a galaxy's environment
\citep{trujillo07a,song21a}. The progenitors of these compact
quiescent galaxies themselves could have formed through gas-rich major
mergers (e.g., \citealt{Hopkins09a,barro13a,wellons15a}) or through
clump migration (e.g., \citealt{Dekel14a}).  Additionally, the
extension of the size-mass relation into the EoR is currently not
well-constrained.  UV measurements from the HUDF and CANDELS have
found a range of sizes for these early galaxies (0.3--1\, kpc,
\citealt{Oesch10a, Ono13a, Curtis-Lake16a}) where mass is not very
well measured given the limited scope of detection bands to the
rest-frame UV.

Robust size measurements free from redshift bias are needed to
adequately trace the evolution of galaxy sizes from the early
Universe, which require deep rest-frame optical imaging of galaxies
out into the EoR. In addition, the combination of analysis of
galaxies' rest-frame optical sizes can be compared to the sizes of
their dust and gas reservoirs \citep[e.g.,][]{hodge20a} to further 
place constraints on the morphological transformation of galaxies
across cosmic time.

\subsection{Spatially Resolved Galaxy SEDs}

The high resolution and deep images provided by COSMOS-Web NIRCam
images will enable detailed pixel-by-pixel SED fitting of galaxies
across a wide redshift range and down to lower stellar masses than has
been possible to date (e.g.,
\citealt{Wuyts12a,Jafariyazani19a,Abdurrouf21a,Abdurrouf23a}).  The
resulting mass, star formation rate, and dust attenuation maps can be
used to study the star formation and quenching process in galaxies
\citep[e.g.,][]{tacchella15a}.  The large number of sources in
COSMOS-Web will enable the study of trends as a function of redshift,
environment, and position relative to the star forming main sequence.

Mass maps that represent the overall resolved stellar mass of galaxies
can be used for the morphological measurements described above.
For example, \cite{Cibinel15a} show that morphological measurements
using the mass maps of galaxies are better able to pick out features
indicative of galaxy mergers than similar measures using standard light
images. Similarly, precise size measurements can be made using mass
maps in comparison with standard measurements 
(e.g., \citealt{Suess19a,Mosleh20a}). Clumps can be 
more easily identified using stellar mass maps and star formation rate maps
can be used to quantify the growth of stellar mass in galaxies as a 
function of their morphology and environment.

\subsection{Constraints on the Dust Attenuation Law}\label{sec:dustattenuation}

The dust attenuation law plays a crucial role in SED modeling for
galaxies at all redshifts \citep{salim20a} but is heavily dependent
on dust grain properties, total dust content, and dust geometry within
galaxies' interstellar media. Without direct constraints, most SED
fitting routines blindly adopt one of a few common dust attenuation
curves, for example that of the Milky Way galaxy \citep{cardelli89a}
or the `Calzetti' curve \citep{calzetti00a}. Such blind adoption of
an attenuation law that may or may not be applicable can result in
substantial systemic biases introduced to extrapolated dust emission
models, mass estimates, and star-formation rates \citep{mitchell13a,laigle19a}. 

Well-sampled SEDs -- from the rest-frame UV through the near-infrared
-- facilitate a direct measurement of the dust attenuation law
\citep[e.g.,][]{kriek13a}. This is done by construction of broad SEDs
that can then be fit to stellar population synthesis models with a
range of dust attenuation law prescriptions to infer the best-fit
solutions. The broader COSMOS survey includes 40$+$ bands of coverage
from the far-UV through the mid-infrared spanning both narrow and
broad-band filters; such well-sampled SED coverage is sufficient to
constrain some variation in the dust attenuation law, as has been
measured in similar datasets
\citep{pannella15a,salmon16a,reddy18a}. However, a key limitation in
constraining any possible evolution in the dust attenuation law comes
from limited samples at higher redshifts, particularly at epochs where
one might expect sufficiently different content and distribution of
galaxies' dust reservoirs. The added near-infrared (and mid-infrared)
depth brought by COSMOS-Web will be crucial to dramatically increase
the number of known, well-characterized galaxies out to $z\sim4$ whose
photometry can then be extracted across all COSMOS datasets to piece
together large statistical samples of SEDs. These SEDs can, in turn,
be used to infer redshift evolution in dust attenuation. A presumption
of energy balance -- where absorbed rest-frame UV emission is
re-emitted at long wavelengths -- can then be directly tested against
deep submillimeter observations in the field, stacked using
single-dish datasets \citep[e.g.,][]{oliver12a,simpson19a} or
individual constraints from galaxies observed by ALMA to much greater
depths \citep[e.g., the A3COSMOS project;][]{liu19a}.

In addition to broad SED constraints, the ability to spatially resolve
colors on kpc scales using NIRCam and {\it Hubble}/ACS imaging will allow
direct measurement of the impact of dust geometry on galaxies'
integrated SEDs. This will be particularly useful for galaxies already
detected by ALMA, of which we estimate there are $\sim$1000 (from ALMA
Cycles 0-9) across the COSMOS-Web mosaic footprint. Dust geometry in
complex ISM environments has long been a nuisance to SED fitting, as
it often results in decoupling of the stellar and dust SEDs
\citep{lower22a}. COSMOS-Web will allow direct constraints on the
relative degree of decoupling and its correlation to galaxy morphology
as a function of color.

\subsection{Finding \&\ Characterizing Protoclusters}

Galaxy clusters represent the most massive gravitationally bound
structures in the Universe, and yet the history of their assembly is
observationally uncertain.  Galaxy clusters are typically found at
$z\simlt1.5$ thanks to thermal Bremsstrahlung radiation in the X-ray
\citep{kravtsov12a} or via the Sunyaev-Zel'dovich effect in the
millimeter \citep{menanteau09a,vanderlinde10a} due to a hot
$\sim$10$^7$\,K intracluster medium.  A complete catalog of X-ray
groups identified in COSMOS is compiled by
\citet{gozaliasl19a}. However, the progenitors of galaxy clusters --
called protoclusters -- are observationally more elusive
\citep{overzier16a}.  They have not yet virialized, thus their
intracluster medium is not yet substantially heated to be
distinguishable from the surrounding IGM.  Before virialization, at
$z\simgt2$, overdense environments are extended in large filaments
that may span up to $\simgt$10\,comoving Mpc scales
\citep{muldrew15a,chiang17a}.  At $z\simgt2$, these physical scales
span 10--30\,arcminutes across, thus wide field-of-view surveys are
needed to detect and characterize their spatial distribution.

Due to its large solid angle and sufficient depth to detect structures
at $z>2$, COSMOS has served as a primary observational field used to
detect and analyze protoclusters at high redshifts
\citep[][]{yuan14a,casey15a,diener15a,chiang15a,hung16a}.  Such works
have highlighted some of the challenges in constraining the forward
evolution of such diffuse structures, where it is particularly
difficult to constrain protoclusters' halo masses, and yet total halo
mass is crucial to the interpretation of their long-term evolutionary
path \citep{sillassen22a}.  One particular structure, now dubbed
``Hyperion,'' lies in the center of the COSMOS field at $z\sim2.5$
with an estimated $z=0$ halo mass exceeding 10$^{15}$\,\msun; a
subcomponent of Hyperion has been discussed in the literature as a
possible proto-virialized cluster core through the detection of
associated extended X-ray emission
\citep{wang16a,wang18a,champagne21a}.  Its filamentary structures
extend half a degree across and coincide well with the coverage of
COSMOS-Web, which will allow a much richer mapping of its constituent
galaxies at fainter luminosities.  While spectroscopic follow-up will
solve an essential piece of the puzzle in spatially mapping the full
extent of known structures like Hyperion in COSMOS, the precise
photometric redshifts provided by COSMOS-Web will dramatically improve
the efficiency of follow-up.  For example, reducing
  $\sigma_{\rm NMAD}(\Delta z/(1+z))$ from $\sim$0.06 to 0.03 for
  $\sim$27$^{th}$ magnitude sources reduces the uncertainty in
  line-of-sight projected distance by a factor of $\sim$2 to
  $\sim$100\,Mpc from $z\sim2-5$.  While still significantly larger
  than the expected line-of-sight distances within protocluster
  environments, the increased precision will significantly improve the
  efficiency of follow-up spectroscopic campaigns targetting sources
  with photometric redshifts consistent with an overdensity of
  spectroscopic redshifts.

At higher redshifts, the prospect for discovering new protoclusters
in COSMOS-Web is significant.  Based on the $z\sim0$ cluster mass
function \citep[e.g.,][]{bahcall93a}, we expect $\sim$30 structures
between $2<z<8$, $\sim$20 of which will be $4<z<8$, that eventually
collapse into $>5\times10^{14}$\,\msun\ clusters at $z=0$.  Some of
these we may have already found the first hints of based on
ground-based data \citep[e.g.,][]{brinch22a}, and the added depth and
photometric redshift precision of COSMOS-Web will push the potential
discovery space significantly. A more efficient mapping of such
structures over an unbiased area will then allow more detailed
investigations of the assembly history of protoclusters themselves
\citep{casey16a}.

\subsection{Strong Lensing}

The past three decades have seen the discovery of hundreds of
galaxy-scale strong lenses
\citep[e.g.,][]{bolton08a,gavazzi12a,rojas21a}.  COSMOS-Web will
better resolve the 40$+$ candidate strong lenses currently known in
the COSMOS field from existing {\it Hubble} data and ground-based
observations \citep{faure08a,jackson08a}, and has the potential to
discover many more previously unknown galaxy-galaxy lenses
 due to the survey's added depth in the near-infrared,
  sensitive to fainter, higher redshift background sources.

We perform a simple estimate of how many lenses will be in COSMOS-Web
by first estimating the total number of galaxies acting as potential
lenses.  As intrinsically massive galaxies are needed to cleanly
resolve the lensed system, we assume lensing galaxies are already part
of the current COSMOS2020 catalog \citep{weaver22a}.  We use the
criterion $0.2<z_{\rm lens}<1.5$, M$_{\star}>10^{9}\,$\msun, and
SFR$<10^{-1}\,$\msun\,yr$^{-1}$, resulting in a selection of
$\sim$2$\times$10$^5$ galaxies.  We take the median of the
distribution of the stellar-to-halo mass relation from
\citet{shuntov22a} to define four mass bins: $9.0<\log M_\star<9.5$
($M_{\rm halo}\sim2\times10^{11}$\,\msun), $9.5<\log M_\star<10.0$
($M_{\rm halo}\sim4\times10^{11}$\,\msun), $10.0<\log M_\star<10.5$
($M_{\rm halo}\sim7\times10^{11}$\,\msun), $10.5<\log M_\star<11.0$
($M_{\rm halo}\sim1.2\times10^{12}$\,\msun).  We estimate the surface
on the sky where multiple imaging occurs ($2<z_{\rm source}<13$)
assuming lens mass profiles are isothermal spheres.  We use the
estimated number of $\sim$78 galaxies per arcmin$^2$ at $2<z<13$ drawn
from UV luminosity functions \citep[compiled 
  by][]{behroozi19a}.  This calculation does not account for factors
hindering the confirmation of lens candidates (e.g., confusion with
spiral arms) and therefore may overestimate the number of lenses
confirmed.  To bring the number of lenses closer to realistic numbers,
we perform the same calculation in the larger COSMOS field and rescale
our results to the size of COSMOS-Web, assuming that 70 lenses are
confirmed.  This yields an expected $\sim$90 {\it new} lenses will be found
in COSMOS-Web.

{\it JWST}'s unprecedented depth and resolution will lead to the discovery
of the highest density of strong lenses per square degree, making it
ideal for inferring line-of-sight shear with strong lenses
\citep{fleury21a,hogg22a} and complementing COSMOS-Web's weak lensing
analysis \citep[see also][]{kuhn21a}. Uniform multiband imaging of
every strong lens will be available, overcoming challenges with
deblending the lens and source light \citep{etherington22a}.  The
highly magnified source galaxy population will allow for studies of
high redshift galaxy formation \citep[e.g.,][]{swinbank15a} as well as
detailed studies of the central mass profile of lensing galaxies
\citep[e.g.,][]{koopmans09a,nightingale19a,shajib21a,etherington22a}
and dark
matter contents \citep[e.g.,][]{vegetti14a,he22a}.

\subsection{Identifying Candidate Direct Collapse Black Holes}\label{sec:dcbh}

Direct Collapse Black Holes (DCBHs) have been proposed to resolve the
mysterious quick growth of the Universe's first supermassive black
holes with $M_{\rm BH}\sim10^{9}$\,M$_\odot$
\citep{volonteri10a,volonteri12a,natarajan11a}, found out to redshifts
$z\sim7.5$ \citep[e.g.,][]{wang21a}. DCBHs are hypothesized to form
black hole seeds of significant mass
\citep[$\sim10^{4-6}$\,M$_\odot$;][]{shang10a,johnson12a} from the
primordial collapse of an atomic-cooling halo whereby strong
Lyman-Werner photons could dissociate $H_2$ and prevent gas
fragmentation, allowing the formation of DCBHs with significant mass
and growth rates possibly exceeding Eddington rates
\citep[e.g.,][]{volonteri05a,alexander14a}. Though no DCBH candidates
have been directly confirmed, they are thought to have significant
infrared through submillimeter emission, resulting in a steep, red
near-infrared spectrum; they are also expected to have X-ray emission
\citep{natarajan17a}.

\citet{pacucci16a} present two possible candidate DCBHs from CANDELS
data in the GOODS-S area, both of which have photometric redshifts
larger than $z\sim6$ and robust X-ray detections as well as very steep
infrared spectra. The improved depth of COSMOS-Web compared to CANDELS
(out to 4.4\,\um\ or 7.7\um\ in the NIRCam mosaic or MIRI-covered
subset) will allow more robust identification of fainter DCBH
candidates with more robust photometric redshifts.  The number density
of the CANDELS-identified sources extrapolated to COSMOS-Web implies
that we may find $\sim$20 such candidates in our full survey volume at
$z>6$. Such sources will then require spectroscopic follow-up with
{\it JWST} to confirm their nature, assess their black hole masses,
and to inform predictions for future deeper X-ray observations that
may provide further confirmation.

\subsection{Supermassive Black Hole - Galaxy Coevolution}

COSMOS-Web will open new avenues to study AGN and quasars at high
redshift. At $z > 6$, the black hole population with masses down to
$10^{6}$ M$_{\odot}$ can be revealed through color-selection (see, for
example, \citealt{goulding22a}) some of which may have resulted from
an earlier DCBH event (see \S~\ref{sec:dcbh}). 
Using a semi-analytic model for the formation of the first galaxies and 
black holes \citep{trinca22a}, we expect $\sim50$ black holes within
the COSMOS-Web volume that have $7<z<10$ and masses of
10$^{6}-10^8$\,M$_{\odot}$.
With the spatial resolving
power of {\it JWST}, morphologies can be decomposed into unresolved
AGN emission and more diffuse host galaxy emission
\citep{Kocevski22a,Ding22a} on spatial scales of $\sim$1\,kpc. With
AGN-free host galaxy images, we can measure the mass relation between
black holes and their hosts (M$_{\rm BH}$ vs.\ M$_{\rm host}$) beyond
$z\sim3$ \citep{trakhtenbrot15a,suh20a}, carry out spatially-resolved studies of the
stellar populations up to $z\sim2$, perform a quasar-galaxy cross
correlation analysis \citep{garcia-vergara17a}, and assess the
influence of mergers
\citep[e.g.,][]{mechtley16a,shah20a}. Furthermore, MIRI will aid in
our ability to determine the demographics of the AGN population
including the obscured population through detection of steep infrared
(unresolved) sources with dominant torus emission.

\subsection{Searches for $z>10$ Pair Instability Supernovae}

\begin{figure}
 \includegraphics[width=0.9\columnwidth]{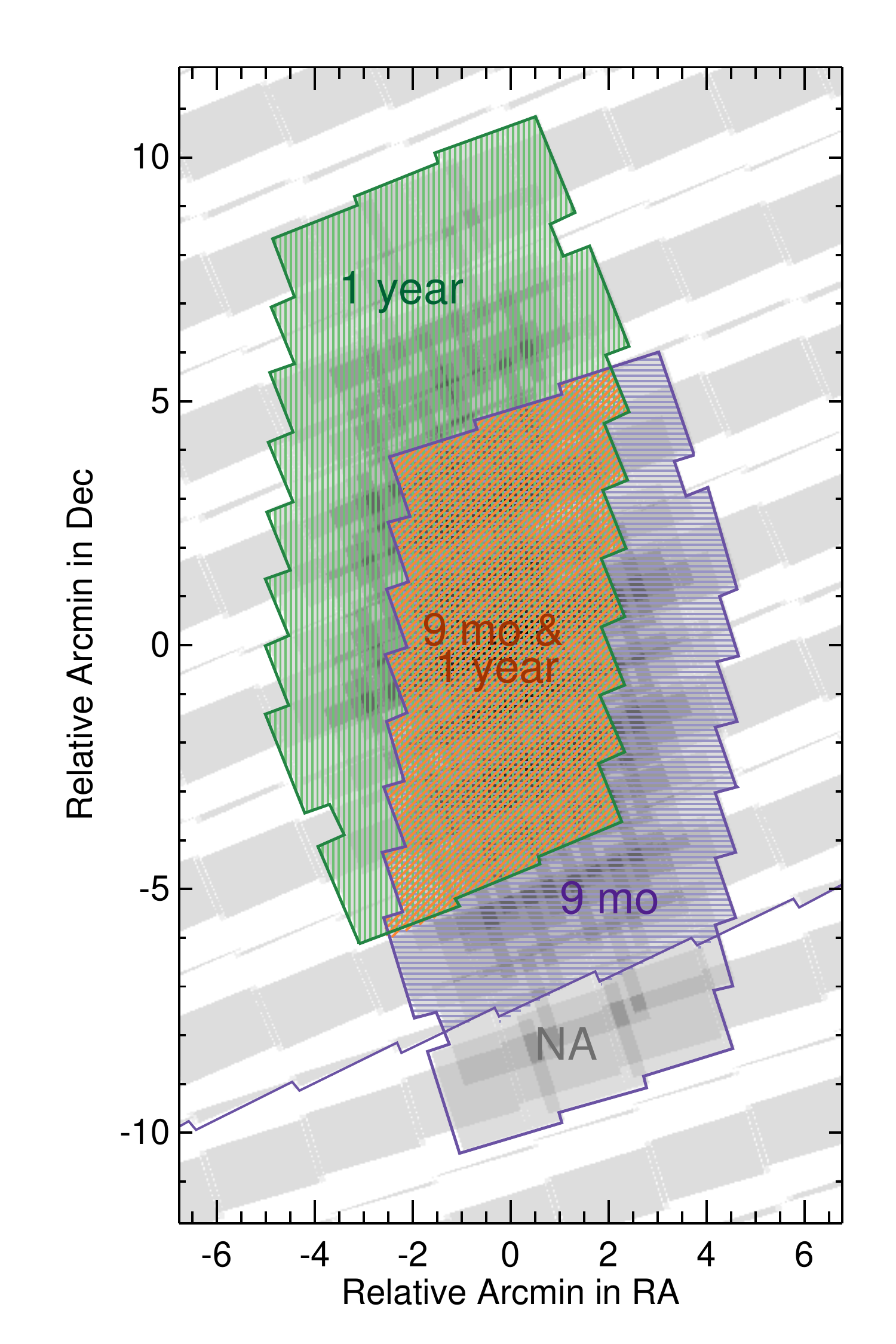}
 \caption{ An illustration of the part of the COSMOS-Web area that
   will be covered by multi-epoch NIRCam observations, thanks to the
   PRIMER survey (GO \#1837). Due to the scheduling of the PRIMER
   program primarily in Cycle 1, and this region of the COSMOS-Web
   mosaic during Cycle 2, a total of 133\,arcmin$^2$ will see multiple
   visits; the first of these 
   was observed in January 2023 (the PRIMER area
   covering both green and orange highlighted regions). The second
   will occur in $\sim$April 2023 (PRIMER covering the purple and
   orange regions). The last epoch will occur in $\sim$December
   2023/January 2024 thanks to COSMOS-Web. Thus the
   purple region will have two epochs separated by $\sim$9 months
   (this corresponds to 36.5\,arcmin$^2$), the green region will have
   two epochs separated by $\sim$1 year (this corresponds to
   53.8\,arcmin$^2$), and the orange area will have three epochs of
   separation $\sim$3 months followed by another $\sim$9 months,
   spanning a year in total (this corresponds to 42.6\,arcmin$^2$).  }
 \label{fig:transient_area}
\end{figure}

COSMOS-Web sits in unique parameter space, able to search for
intrinsically rare phenomena at sensitivities beyond most wide-field
surveys. Very high-redshift ($z>5$) supernovae (SNe) in particular may
provide a unique lens on the formation of the first massive stars by
constraining the high-mass end of the Population III initial mass
function. Such a first generation of stars is indeed thought to be
top-heavy \citep{bromm99a,bromm02a}. Ranging in mass from
100--260\,\msun, such stars are most likely to die as pair-instability
supernovae \citep[PISNe;][]{heger02a}, which release up to 100 times
more energy than Type Ia or Type II SNe (with intrinsic luminosities
$\sim10^{47-48}$\,erg\,s$^{-1}$). Their extreme energy release is a
result of electron-positron pair creation via thermal heating after
the cessation of carbon burning in the core, leading to collapse and
thermonuclear burning of O and Si; the energy released unbinds the
star without leaving a remnant. Thanks to their luminosity and
relative longevity (lengthened by $1+z$ time dilation) it is plausible
to detect and identify PISNe brighter than NIR magnitudes $\sim$28\,
from $z=10-30$ in JWST NIRCam imaging surveys, particularly in the
long wavelength channels. \citet{whalen13a} present near-infrared
light curves of PISNe from radiation hydrodynamical simulations,
suggesting PISNe with $\simgt$200\,\msun\ progenitors at $z\sim15-30$
may remain detectable with F444W$\,<28$ (possibly as bright as
$\sim$26$^{th}$ magnitude) for 1-3\,years, varying by
$\sim$0.3\,mags\,yr$^{-1}$. \citet{Hummel12a} present calculations
of the expected number density of such {\it JWST}-detectable
explosions at or below $\sim$0.02\,arcmin$^{-2}$ at any given time at
$z\simgt10$.  This could result in $\sim$40 such events sitting within
the COSMOS-Web NIRCam footprint.

The primary challenge in identifying PISNe candidates in COSMOS-Web
will come from distinguishing them from high-redshift galaxies; thus,
multi-epoch observations (conducted on roughly a yearly timescale)
become critical to measuring the time-variable fading of the
explosion. While most prior near-infrared datasets in COSMOS reach
depths of only 25-26 mags (and are limited by the poor spatial
resolution of {\it Spitzer} or ground-based UltraVISTA data), there
may be a handful of exceptionally bright PISNe at $z>5$ whose
transient nature can be constrained using existing observations on a
$\sim$10 year cadence. Alternatively, over a smaller area, the CANDELS
survey \citep{grogin11a,koekemoer11a} conducted deep $\sim$28 imaging
out to 1.6\,\um\ covering 200\,arcmin$^2$ in late 2011 and early 2012;
this provides a $\sim$10\,year time baseline for potential $z\sim5-12$
PISNe relative to COSMOS-Web observations detected in F150W (the
redshift range limited directly by wavelength of deep imaging and the
opacity of the IGM in absorbing photons shortward of 1216\AA).

Out to higher redshifts, it may be possible to detect PISNe candidates
out to $z\sim30$ across a $\Delta t=1$\,year timescale using imaging
from the PRIMER {\it JWST} survey in conjunction with COSMOS-Web, as
shown in Figure~\ref{fig:transient_area}. Assuming there are no
significant changes to the JWST long-range plan as of this writing,
the PRIMER survey (GO \#1837) will obtain half of their COSMOS NIRCam
imaging in late 2022 covering an area $\sim$96\,arcmin$^2$ out to
F444W, and the other half in April 2023. Both PRIMER regions of the
field will then be covered in late 2023 by COSMOS-Web, allowing a
careful comparison of differential photometry for a potential handful
of PISNe candidates brighter than $\sim$28. The total area with deep,
$\sim$28$^{th}$ magnitude $\sim$4.5\,\um\ multi-epoch {\it JWST}
observations is $\sim$133\,arcmin$^2$. Even with only a few
detections, such PISNe candidates could potentially be extremely
useful for constraining the nature of Population III stars formed
shortly after the Big Bang.

\subsection{Ultracool Halo Sub-Dwarf Stars}

\begin{figure}
 \includegraphics[width=0.95\columnwidth]{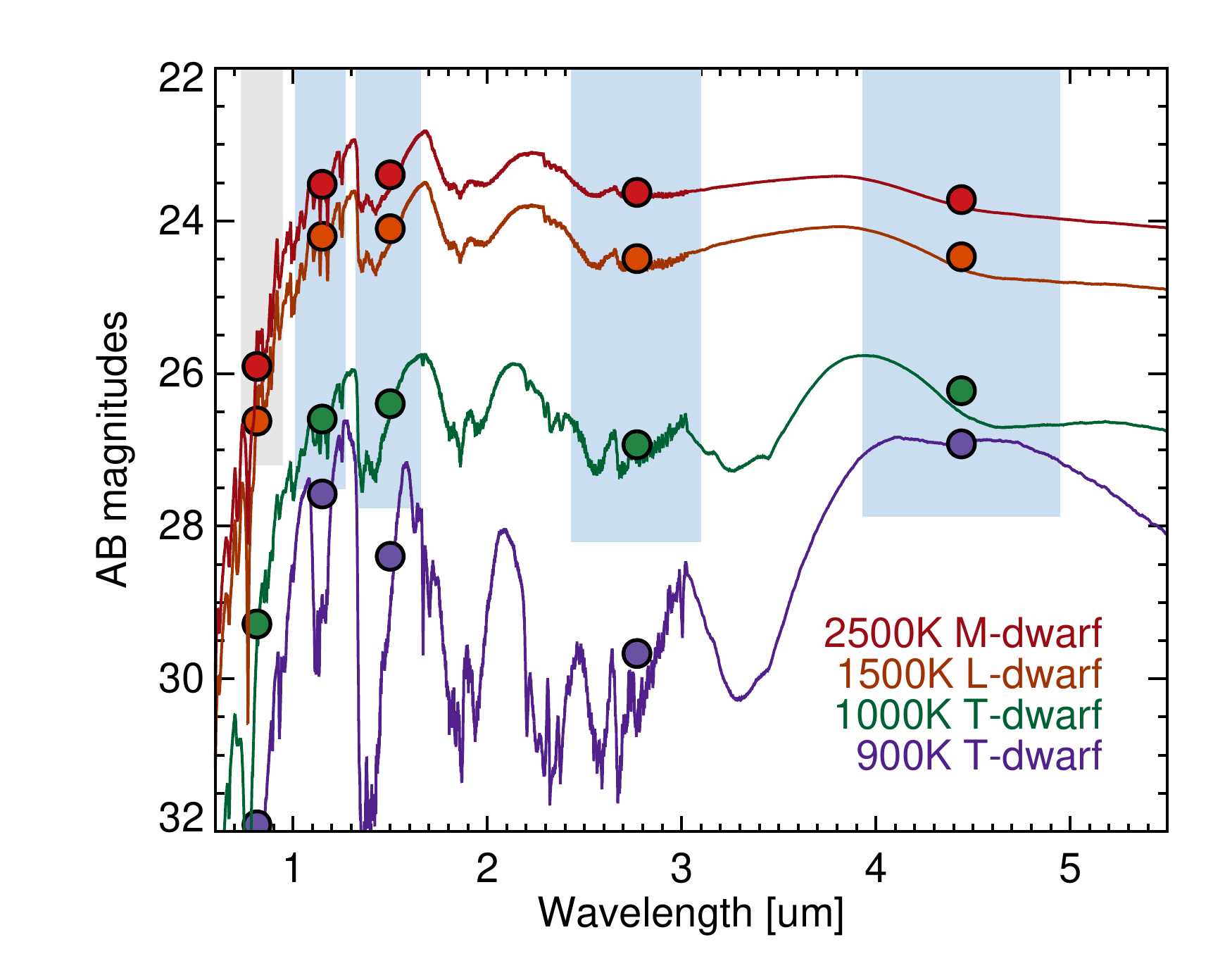}
 \caption{Ultracool halo sub-dwarf templates projected in AB
  magnitudes at a distance of 2\,kpc. Templates are taken from
  \citet{saumon08a} spanning effective surface temperatures of
  900--2500K (purple to red). Synthetic photometry is calculated in
  the COSMOS-Web bands, whose depths are shown in a similar fashion
  as in Figure~\ref{fig:sedlimits}. Halo M-dwarfs would be
  detectable out to $\sim$10\,kpc (with potential confusion with
  $z\sim6-7$ galaxies between 3--10\,kpc), while L-, T- and Y-dwarfs
  will be detectable to $\sim$2\,kpc.}
  \label{fig:ucd}
\end{figure}

Ultracool dwarfs (late M-dwarfs through Y-dwarfs) are the most
abundant stellar population by number and their prominent emission in
the near-infrared implies that deep field surveys from {\it JWST} are
prone to detect them to significant distances in the Galactic halo
\citep{ryan16a}. Indeed, mapping their number density to different
distances in the outer halo may give unique constraints on the
metal-poor initial mass function as well as the scale height of the
Milky Way for low-mass objects
\citep{burgasser03a,carnero-rosell19a}. Such discoveries are only
enabled by deep near-infrared imaging, and given the wide area of
COSMOS-Web, we anticipate finding of order $\sim$1000 such dwarfs
across the field at various distances.

Figure~\ref{fig:ucd} shows four ultracool dwarf templates from
\citet{saumon08a} with varying effective temperature from 900\,K,
through the L-T transition at $\sim$1000\,K up to late M-dwarfs at
2500\,K. Models assume a 1000\,m\,s$^{-2}$ surface gravity, consistent
with expectation for older halo stars that would likely be found in
extragalactic fields like COSMOS; no cloud cover is assumed below
1000\,K, above which models with a moderate amount of cloud cover are
adopted \citep{bowler16a}. Significant absorption bands in ultracool
T- and Y-dwarfs between 1.5--3.5\,\um, particularly at low
$\simlt$1000\,K temperatures, lead to very distinct near-infrared
colors from galaxies in NIRCam bands provided they are located at
distances $\le$\,1\,kpc.  For example, a recent late T-dwarf candidate
was identified in {\it JWST} imaging of Abell\,2744 at a distance of
$\sim$600\,pc by \citet{nonino22a}.

In addition to the science questions addressed by the detection of
ultracool dwarfs, the population has also been a dominant
contaminating source for searches of high-redshift galaxies,
particularly samples of $z\sim6-7$ sources due to their lack of
emission shortward of $\sim$1\,\um. While {\it Hubble} imaging was
limited to the shorter wavelengths, the long wavelength channels of
NIRCam are of particular use in breaking the color degeneracies for
distinguishing ultracool dwarfs from compact high-redshift galaxies.
In addition, those ultracool dwarfs that would be more consistent with
high-redshift galaxy colors are expected to be significantly brighter
\citep[peaking in density around $J\sim24$;][]{ryan16a}. Dwarfs at
considerable distances $>$1\,kpc, with apparent magnitudes fainter
than $\sim$26 have the potential to contaminate $z\sim6-7$ samples;
however, their number density is expected to be relatively low
relative to galaxies at similar magnitudes (fewer than $\sim$50 are
expected across the COSMOS-Web mosaic fainter than $J\sim26$).

\section{Summary}\label{sec:summary}

We have presented the observational design and scientific goals of
COSMOS-Web, the largest prime General Observer program in {\it
 JWST}'s Cycle 1 of observations.
COSMOS-Web is a 0.54\,deg$^2$ contiguous NIRCam survey imaged in four
filters (F115W, F150W, F277W, and F444W) to depths of $\sim$27.5--28.2
magnitudes. In parallel, COSMOS-Web also includes 0.19\,deg$^2$
non-contiguous MIRI imaging in one filter (F770W) to a depth of
$\sim$25.3--26.0 magnitudes. COSMOS-Web
is approximately 2.7$\times$ larger than all other Cycle 1 JWST NIRCam
deep field efforts combined and 3.5$\times$ larger than the combined
MIRI deep field coverage.
The improvement in photometric redshift precision in COSMOS-Web will
be substantial compared to the most recent catalogs compiled in the
COSMOS field \citep{weaver22a}, with $<$5\%\ errors on photometric
redshifts down to magnitudes $\sim$27 in F277W.

The primary science goals of COSMOS-Web are threefold. First,
COSMOS-Web will detect thousands of new galaxies within the Epoch of
Reionization (EoR, $6\simlt z\simlt 11$) and generate the largest
number of galaxies at or above the knee of the UV luminosity function.
Such intrinsically bright galaxies likely trace massive halos at early
times at the nodes of the cosmic web. COSMOS-Web's large area will
allow a detailed mapping of the galaxy density field within the EoR on
physical scales $\sim$150\,Mpc across, sufficiently large to minimize
cosmic variance by exceeding the size of the largest cosmic structures
at these redshifts.

Second, COSMOS-Web aims to detect the Universe's first massive
quiescent galaxies that were likely in place between redshifts
$4<z<6$; such galaxies mark the extreme limits of galaxy evolution at
early times by building their stellar reservoirs at extraordinary
rates (exceeding $\sim10^{10}-10^{11}$\,M$_\odot$ at $z>4$). We will
be able to distinguish them from their dust star forming counterparts,
study their morphologies and star formation histories, and thus place
constraints on their progenitors.

Lastly, COSMOS-Web will measure the evolution in the stellar mass to
halo mass relation (SMHR) from $0<z<2.5$ using weak gravitational
lensing.  The SMHR forms an essential anchor of cosmological
simulations on large scales, and these data will extend its
measurement from $z\sim1$ to $z\sim2.5$ in addition to allowing a
detailed look at the SMHR by galaxy type and star-formation history
(as probed by rest-frame optical colors and color gradients).

Beyond these core science goals, COSMOS-Web's legacy value will extend
to many subfields of extragalactic astronomy and beyond. We have
summarized the potential impact of the survey on measuring galaxy
morphologies, using spatially resolved SEDs to measure galaxy
properties, placing constraints on the dust attenuation law,
identifying and characterizing galaxy protoclusters, finding strong
gravitational lenses, identifying direct collapse black hole
candidates, studying the co-evolution of supermassive black holes and
their host galaxies, searching for $z>10$ pair instability supernovae,
and identifying ultracool sub-dwarf stars in the Milky Way's halo.  We
hope the value of this survey continues to grow with time, as have
many other deep-field observations before COSMOS-Web and {\it JWST}.

\begin{acknowledgments}

We thank the anonymous reviewer for helpful suggestions which greatly
improved the manuscript. We thank the entire JWST team, including
scientists, engineers, software developers, and the instrument and commissioning
teams for making this amazing telescope a reality. We thank our 
program coordinator Christian Soto, our NIRCam reviewer Dan Coe,
and our MIRI reviewer Stacey Bright for helping us to optimize our program
and ensuring that the entire program is schedulable. We also thank the CEERS
team for their quick release of simulations and observed data products,
early testing and modification of the data reduction pipeline, and 
assisting with preparation for COSMOS-Web data reduction.

Support for this work was provided by NASA through grant JWST-GO-01727 
and HST-AR-15802 awarded by the Space Telescope Science Institute, 
which is operated by the Association of Universities for Research in 
Astronomy, Inc., under NASA contract NAS 5-26555.

CMC thanks the National Science Foundation for support through grants
AST-1814034 and AST-2009577 as well as the University of Texas at
Austin College of Natural Sciences for support; CMC also acknowledges
support from the Research Corporation for Science Advancement from a
2019 Cottrell Scholar Award sponsored by IF/THEN, an initiative of
Lyda Hill Philanthropies. JSK acknowledges support from the College of
Science and the Laboratory for Multiwavelength Astrophysics at the
Rochester Institute of Technology. JSK acknowledges the important 
contributions to this paper, and the COSMOS-Web proposal, made by 
Shran and T'Pol, who attended every planning telecon and 
kept everyone's spirits up in the early days of the COVID-19 pandemic.

 The Cosmic Dawn Center (DAWN) is funded by the Danish
  National Research Foundation under grant No. 140. JDR was supported
by JPL, which is under a contract for NASA by Caltech. This research
is also partly supported by the Centre National d'Etudes Spatiales
(CNES). OI, CL, HHMCC acknowledge the funding of the French Agence
Nationale de la Recherche for the project iMAGE (grant
ANR-22-CE31-0007). BER was supported in part by NASA grant
80NSSC22K0814. MT and FG acknowledge the support from grant PRIN MIUR
2017 20173ML3WW\_001. BT acknowledges support from the European
Research Council (ERC) under the European Union’s Horizon 2020
research and innovation program (grant agreement 950533) and from the
Israel Science Foundation (grant 1849/19)

The authors acknowledge Research Computing at the \cite{RC} for providing 
computational resources and support for the work reported in this publication. 
 This work used the CANDIDE computer system at the IAP 
supported by grants from the PNCG, CNES and the DIM-ACAV and 
maintained by S. Rouberol.
We acknowledge use of the {\it lux} supercomputer at UC
Santa Cruz, funded by NSF MRI grant AST1828315.  

\end{acknowledgments}

\appendix
\twocolumngrid

\section{Details of the COSMOS-Web Mosaic Visits}  \label{appendix}

Here we provide detailed information on the 152 visits that comprise
the COSMOS-Web mosaic. Table~\ref{tab:visits} lists all of the
individual visits, their reference positions, and their position
angles. The observation number is given as in the COSMOS-Web Proposal
(\#1727) opened in the Astronomer's Proposal Tool (APT), and the visit
name mirrors the target name in APT. The listed position angles are
relative to the NIRCam frame (and differ from the V3 angle by
0.09$^\circ$). The three visits that have position angles differing
from the other visits in the mosaic are CWEBTILE-0-4, CWEBTILE-5-18,
and CWEBTILE-7-15. Their angles are different due to availability of
guide stars visible in the fine guidance sensor (FGS); no modification
of their positions were required to keep the NIRCam mosaic
contiguous. The positions as listed correspond to the reference
position of NRCALL\_FULL and sit at the reference point (V2,V3) =
($-$0.32,\,$-$492.59) with 4TIGHT dither offsets taken
$\pm$24$\farcs$7 along V2 and $\pm$3$\farcs$00 along V3.  The relative
positions of single visit coverage with respect to this reference
point are shown in Figure~\ref{fig:4tight}.

  \startlongtable
\begin{deluxetable}{c@{ }cccc}
  \tabletypesize{\footnotesize}
  \tablecolumns{5}
  \tablecaption{COSMOS-Web Visit Positions}
  \tablehead{
    \colhead{Obs No.} & \colhead{Visit Name} & \colhead{R.A.} & \colhead{Dec.} & \colhead{P.A.} 
  }
  \startdata
1   & CWEBTILE-0-0 & 09:59:42.539  &  $+$02:38:15.90 & 293 \\
2   & CWEBTILE-1-0 & 09:59:34.622  &  $+$02:32:49.78 & 293 \\
39  & CWEBTILE-2-0 & 09:59:26.708  &  $+$02:27:23.67 & 293 \\
40  & CWEBTILE-3-0 & 09:59:18.790  &  $+$02:21:57.55 & 293 \\
77  & CWEBTILE-4-0 & 09:59:10.876  &  $+$02:16:31.44 & 107 \\
78  & CWEBTILE-5-0 & 09:59:02.963  &  $+$02:11:05.33 & 107 \\
115 & CWEBTILE-6-0 & 09:58:55.049  &  $+$02:05:39.21 & 107 \\
116 & CWEBTILE-7-0 & 09:58:47.135  &  $+$02:00:13.10 & 107 \\
3   & CWEBTILE-0-1 & 09:59:50.742  &  $+$02:37:31.18 & 293 \\
4   & CWEBTILE-1-1 & 09:59:42.825  &  $+$02:32:05.06 & 293 \\
41  & CWEBTILE-2-1 & 09:59:34.907  &  $+$02:26:38.95 & 293 \\
42  & CWEBTILE-3-1 & 09:59:26.990  &  $+$02:21:12.83 & 293 \\
79  & CWEBTILE-4-1 & 09:59:19.076  &  $+$02:15:46.72 & 107 \\
80  & CWEBTILE-5-1 & 09:59:11.158  &  $+$02:10:20.60 & 107 \\
117 & CWEBTILE-6-1 & 09:59:03.245  &  $+$02:04:54.49 & 107 \\
118 & CWEBTILE-7-1 & 09:58:55.334  &  $+$01:59:28.37 & 107 \\
5   & CWEBTILE-0-2 & 09:59:58.942  &  $+$02:36:46.45 & 293 \\
6   & CWEBTILE-1-2 & 09:59:51.024  &  $+$02:31:20.34 & 293 \\
43  & CWEBTILE-2-2 & 09:59:43.107  &  $+$02:25:54.22 & 293 \\
44  & CWEBTILE-3-2 & 09:59:35.189  &  $+$02:20:28.11 & 293 \\
81  & CWEBTILE-4-2 & 09:59:27.272  &  $+$02:15:02.00 & 107 \\
82  & CWEBTILE-5-2 & 09:59:19.358  &  $+$02:09:35.88 & 107 \\
119 & CWEBTILE-6-2 & 09:59:11.444  &  $+$02:04:09.77 & 107 \\
120 & CWEBTILE-7-2 & 09:59:03.530  &  $+$01:58:43.65 & 107 \\
7   & CWEBTILE-0-3 & 10:00:07.141  &  $+$02:36:01.73 & 293 \\
8   & CWEBTILE-1-3 & 09:59:59.224  &  $+$02:30:35.62 & 293 \\
45  & CWEBTILE-2-3 & 09:59:51.306  &  $+$02:25:09.50 & 293 \\
46  & CWEBTILE-3-3 & 09:59:43.389  &  $+$02:19:43.39 & 293 \\
83  & CWEBTILE-4-3 & 09:59:35.471  &  $+$02:14:17.27 & 107 \\
84  & CWEBTILE-5-3 & 09:59:27.557  &  $+$02:08:51.16 & 107 \\
121 & CWEBTILE-6-3 & 09:59:19.640  &  $+$02:03:25.04 & 107 \\
122 & CWEBTILE-7-3 & 09:59:11.726  &  $+$01:57:58.93 & 107 \\
9   & CWEBTILE-0-4 & 10:00:15.341  &  $+$02:35:17.01 & 113$\dagger$ \\
10  & CWEBTILE-1-4 & 10:00:07.423  &  $+$02:29:50.90 & 293 \\
47  & CWEBTILE-2-4 & 09:59:59.502  &  $+$02:24:24.78 & 293 \\
48  & CWEBTILE-3-4 & 09:59:51.584  &  $+$02:18:58.67 & 293 \\
85  & CWEBTILE-4-4 & 09:59:43.671  &  $+$02:13:32.55 & 107 \\
86  & CWEBTILE-5-4 & 09:59:35.753  &  $+$02:08:06.44 & 107 \\
123 & CWEBTILE-6-4 & 09:59:27.839  &  $+$02:02:40.32 & 107 \\
124 & CWEBTILE-7-4 & 09:59:19.926  &  $+$01:57:14.21 & 107 \\
11  & CWEBTILE-0-5 & 10:00:23.540  &  $+$02:34:32.29 & 293 \\
12  & CWEBTILE-1-5 & 10:00:15.623  &  $+$02:29:06.17 & 293 \\
49  & CWEBTILE-2-5 & 10:00:07.701  &  $+$02:23:40.059 & 293 \\
50  & CWEBTILE-3-5 & 09:59:59.784  &  $+$02:18:13.94 & 293 \\
87  & CWEBTILE-4-5 & 09:59:51.867  &  $+$02:12:47.83 & 107 \\
88  & CWEBTILE-5-5 & 09:59:43.949  &  $+$02:07:21.72 & 107 \\
125 & CWEBTILE-6-5 & 09:59:36.035  &  $+$02:01:55.60 & 107 \\
126 & CWEBTILE-7-5 & 09:59:28.121  &  $+$01:56:29.49 & 107 \\
13  & CWEBTILE-0-6 & 10:00:31.743  &  $+$02:33:47.57 & 293 \\
14  & CWEBTILE-1-6 & 10:00:23.822  &  $+$02:28:21.45 & 293 \\
51  & CWEBTILE-2-6 & 10:00:15.901  &  $+$02:22:55.34 & 293 \\
52  & CWEBTILE-3-6 & 10:00:07.983  &  $+$02:17:29.22 & 293 \\
89  & CWEBTILE-4-6 & 10:00:00.066  &  $+$02:12:03.11 & 107 \\
90  & CWEBTILE-5-6 & 09:59:52.148  &  $+$02:06:36.99 & 107 \\
127 & CWEBTILE-6-6 & 09:59:44.231  &  $+$02:01:10.88 & 107 \\
128 & CWEBTILE-7-6 & 09:59:36.317  &  $+$01:55:44.76 & 107 \\
15  & CWEBTILE-0-7 & 10:00:39.943  &  $+$02:33:02.85 & 293 \\
16  & CWEBTILE-1-7 & 10:00:32.021  &  $+$02:27:36.73 & 293 \\
53  & CWEBTILE-2-7 & 10:00:24.100  &  $+$02:22:10.62 & 293 \\
54  & CWEBTILE-3-7 & 10:00:16.179  &  $+$02:16:44.50 & 293 \\
91  & CWEBTILE-4-7 & 10:00:08.262  &  $+$02:11:18.39 & 107 \\
92  & CWEBTILE-5-7 & 10:00:00.344  &  $+$02:05:52.27 & 107 \\
129 & CWEBTILE-6-7 & 09:59:52.430  &  $+$02:00:26.16 & 107 \\
130 & CWEBTILE-7-7 & 09:59:44.513  &  $+$01:55:00.04 & 107 \\
17  & CWEBTILE-0-8 & 10:00:48.142  &  $+$02:32:18.12 & 293 \\
18  & CWEBTILE-1-8 & 10:00:40.217  &  $+$02:26:52.01 & 293 \\
55  & CWEBTILE-2-8 & 10:00:32.300  &  $+$02:21:25.89 & 293 \\
56  & CWEBTILE-3-8 & 10:00:24.379  &  $+$02:15:59.78 & 293 \\
93  & CWEBTILE-4-8 & 10:00:16.461  &  $+$02:10:33.66 & 107 \\
94  & CWEBTILE-5-8 & 10:00:08.544  &  $+$02:05:07.55 & 107 \\
131 & CWEBTILE-6-8 & 10:00:00.626  &  $+$01:59:41.44 & 107 \\
132 & CWEBTILE-7-8 & 09:59:52.709  &  $+$01:54:15.32 & 107 \\
19  & CWEBTILE-0-9 & 10:00:56.338  &  $+$02:31:33.40 & 293 \\
20  & CWEBTILE-1-9 & 10:00:48.417  &  $+$02:26:07.29 & 293 \\
57  & CWEBTILE-2-9 & 10:00:40.496  &  $+$02:20:41.17 & 293 \\
58  & CWEBTILE-3-9 & 10:00:32.578  &  $+$02:15:15.06 & 293 \\
95  & CWEBTILE-4-9 & 10:00:24.657  &  $+$02:09:48.94 & 107 \\
96  & CWEBTILE-5-9 & 10:00:16.740  &  $+$02:04:22.83 & 107 \\
133 & CWEBTILE-6-9 & 10:00:08.822  &  $+$01:58:56.71 & 107 \\
134 & CWEBTILE-7-9 & 10:00:00.905  &  $+$01:53:30.60 & 107 \\
21  & CWEBTILE-0-10 & 10:01:04.537 &  $+$02:30:48.68 & 293 \\
22  & CWEBTILE-1-10 & 10:00:56.616 &  $+$02:25:22.57 & 293 \\
59  & CWEBTILE-2-10 & 10:00:48.695 &  $+$02:19:56.45 & 293 \\
60  & CWEBTILE-3-10 & 10:00:40.774 &  $+$02:14:30.34 & 293 \\
97  & CWEBTILE-4-10 & 10:00:32.856 &  $+$02:09:04.22 & 107 \\
98  & CWEBTILE-5-10 & 10:00:24.935 &  $+$02:03:38.11 & 107 \\
135 & CWEBTILE-6-10 & 10:00:17.018 &  $+$01:58:11.99 & 107 \\
136 & CWEBTILE-7-10 & 10:00:09.104 &  $+$01:52:45.88 & 107 \\
23  & CWEBTILE-0-11 & 10:01:12.737 &  $+$02:30:03.96 & 293 \\
24  & CWEBTILE-1-11 & 10:01:04.816 &  $+$02:24:37.84 & 293 \\
61  & CWEBTILE-2-11 & 10:00:56.895 &  $+$02:19:11.73 & 293 \\
62  & CWEBTILE-3-11 & 10:00:48.973 &  $+$02:13:45.61 & 293 \\
99  & CWEBTILE-4-11 & 10:00:41.052 &  $+$02:08:19.50 & 107 \\
100 & CWEBTILE-5-11 & 10:00:33.135 &  $+$02:02:53.38 & 107 \\
137 & CWEBTILE-6-11 & 10:00:25.214 &  $+$01:57:27.27 & 107 \\
138 & CWEBTILE-7-11 & 10:00:17.300 &  $+$01:52:01.15 & 107 \\
25  & CWEBTILE-0-12 & 10:01:20.936 &  $+$02:29:19.24 & 293 \\
26  & CWEBTILE-1-12 & 10:01:13.011 &  $+$02:23:53.12 & 293 \\
63  & CWEBTILE-2-12 & 10:01:05.090 &  $+$02:18:27.01 & 293 \\
64  & CWEBTILE-3-12 & 10:00:57.169 &  $+$02:13:00.89 & 293 \\
101 & CWEBTILE-4-12 & 10:00:49.248 &  $+$02:07:34.78 & 107 \\
102 & CWEBTILE-5-12 & 10:00:41.331 &  $+$02:02:08.66 & 107 \\
139 & CWEBTILE-6-12 & 10:00:33.413 &  $+$01:56:42.55 & 107 \\
140 & CWEBTILE-7-12 & 10:00:25.496 &  $+$01:51:16.43 & 107 \\
27  & CWEBTILE-0-13 & 10:01:29.136 &  $+$02:28:34.51 & 293 \\
28  & CWEBTILE-1-13 & 10:01:21.211 &  $+$02:23:08.40 & 293 \\
65  & CWEBTILE-2-13 & 10:01:13.290 &  $+$02:17:42.28 & 293 \\
66  & CWEBTILE-3-13 & 10:01:05.365 &  $+$02:12:16.17 & 293 \\
103 & CWEBTILE-4-13 & 10:00:57.448 &  $+$02:06:50.06 & 107 \\
104 & CWEBTILE-5-13 & 10:00:49.526 &  $+$02:01:23.94 & 107 \\
141 & CWEBTILE-6-13 & 10:00:41.609 &  $+$01:55:57.83 & 107 \\
142 & CWEBTILE-7-13 & 10:00:33.691 &  $+$01:50:31.71 & 107 \\
29  & CWEBTILE-0-14 & 10:01:37.335 &  $+$02:27:49.79 & 293 \\
30  & CWEBTILE-1-14 & 10:01:29.410 &  $+$02:22:23.68 & 293 \\
67  & CWEBTILE-2-14 & 10:01:21.486 &  $+$02:16:57.56 & 293 \\
68  & CWEBTILE-3-14 & 10:01:13.564 &  $+$02:11:31.45 & 293 \\
105 & CWEBTILE-4-14 & 10:01:05.643 &  $+$02:06:05.33 & 107 \\
106 & CWEBTILE-5-14 & 10:00:57.722 &  $+$02:00:39.22 & 107 \\
143 & CWEBTILE-6-14 & 10:00:49.805 &  $+$01:55:13.10 & 107 \\
144 & CWEBTILE-7-14 & 10:00:41.887 &  $+$01:49:46.99 & 107 \\
31  & CWEBTILE-0-15 & 10:01:45.531 &  $+$02:27:05.07 & 293 \\
32  & CWEBTILE-1-15 & 10:01:37.606 &  $+$02:21:38.96 & 293 \\
69  & CWEBTILE-2-15 & 10:01:29.685 &  $+$02:16:12.84 & 293 \\
70  & CWEBTILE-3-15 & 10:01:21.760 &  $+$02:10:46.73 & 293 \\
107 & CWEBTILE-4-15 & 10:01:13.839 &  $+$02:05:20.61 & 107 \\
108 & CWEBTILE-5-15 & 10:01:05.918 &  $+$01:59:54.50 & 107 \\
145 & CWEBTILE-6-15 & 10:00:58.000 &  $+$01:54:28.38 & 107 \\
146 & CWEBTILE-7-15 & 10:00:50.079 &  $+$01:49:02.27 & 105$\dagger$ \\
33  & CWEBTILE-0-16 & 10:01:53.730 &  $+$02:26:20.35 & 293 \\
34  & CWEBTILE-1-16 & 10:01:45.806 &  $+$02:20:54.23 & 293 \\
71  & CWEBTILE-2-16 & 10:01:37.881 &  $+$02:15:28.12 & 293 \\
72  & CWEBTILE-3-16 & 10:01:29.956 &  $+$02:10:02.00 & 293 \\
109 & CWEBTILE-4-16 & 10:01:22.035 &  $+$02:04:35.89 & 107 \\
110 & CWEBTILE-5-16 & 10:01:14.114 &  $+$01:59:09.77 & 107 \\
147 & CWEBTILE-6-16 & 10:01:06.196 &  $+$01:53:43.66 & 107 \\
148 & CWEBTILE-7-16 & 10:00:58.275 &  $+$01:48:17.55 & 107 \\
35  & CWEBTILE-0-17 & 10:02:01.930 &  $+$02:25:35.63 & 293 \\
36  & CWEBTILE-1-17 & 10:01:54.001 &  $+$02:20:09.51 & 293 \\
73  & CWEBTILE-2-17 & 10:01:46.077 &  $+$02:14:43.40 & 293 \\
74  & CWEBTILE-3-17 & 10:01:38.156 &  $+$02:09:17.28 & 293 \\
111 & CWEBTILE-4-17 & 10:01:30.231 &  $+$02:03:51.17 & 107 \\
112 & CWEBTILE-5-17 & 10:01:22.310 &  $+$01:58:25.05 & 107 \\
149 & CWEBTILE-6-17 & 10:01:14.392 &  $+$01:52:58.94 & 107 \\
150 & CWEBTILE-7-17 & 10:01:06.471 &  $+$01:47:32.82 & 107 \\
37  & CWEBTILE-0-18 & 10:02:10.126 &  $+$02:24:50.90 & 293 \\
38  & CWEBTILE-1-18 & 10:02:02.201 &  $+$02:19:24.79 & 293 \\
75  & CWEBTILE-2-18 & 10:01:54.276 &  $+$02:13:58.67 & 293 \\
76  & CWEBTILE-3-18 & 10:01:46.351 &  $+$02:08:32.56 & 293 \\
113 & CWEBTILE-4-18 & 10:01:38.430 &  $+$02:03:06.45 & 107 \\
114 & CWEBTILE-5-18 & 10:01:30.505 &  $+$01:57:40.33 & 104.5$\dagger$ \\
151 & CWEBTILE-6-18 & 10:01:22.588 &  $+$01:52:14.22 & 107 \\
152 & CWEBTILE-7-18 & 10:01:14.667 &  $+$01:46:48.10 & 107 \\
\enddata
\tablecomments{
  The position angle (P.A.) of the visit is specified in the last\\
  column; only three visits have non-standard position angles caused \\
  by guide star catalog limitations and they are marked with a
  $\dagger$. We\\ quote 0$\farcs$01 accuracy on tile positions.  }
\label{tab:visits}
\end{deluxetable}

\end{document}